\documentclass[journal, 10pt, onecolumn, letterpaper]{IEEEtran}

\usepackage{ifpdf}

\usepackage[noadjust]{cite}

\ifCLASSINFOpdf
  \usepackage[pdftex]{graphicx}
\else
  \usepackage[dvips]{graphicx}
\fi

\usepackage[cmex10]{amsmath}
\usepackage{array}

\ifCLASSOPTIONcompsoc
  \usepackage[caption=false,font=normalsize,labelfont=sf,textfont=sf]{subfig}
\else
  \usepackage[caption=false,font=footnotesize]{subfig}
\fi

\usepackage{fixltx2e}

\usepackage{url}

\usepackage[utf8]{inputenc} 
\usepackage[T1]{fontenc}
\usepackage{ifthen}

\usepackage{mathtools}
\usepackage{amssymb}
\usepackage{relsize}
\usepackage{bbm}
\usepackage{xfrac}
\usepackage{amsthm}
\theoremstyle{plain}
\newtheorem{definition}{Definition}
\newtheorem{lemma}{Lemma}

\newtheorem{theorem}{Theorem}
\newtheorem{corollary}{Corollary}
\newtheorem{remark}{Remark}

\newtheorem{example}{Example}
\newtheorem{proposition}{Proposition}
\usepackage[varg, cmbraces]{newtxmath}

\usepackage{xcolor}
\definecolor{burgundy}{rgb}{0.545098,0,0}
\definecolor{navyblue}{rgb}{0.0, 0.0, 0.5}
\definecolor{leafgreen}{rgb}{0.290196, 0.470588, 0.0}
\definecolor{bluegreen}{rgb}{0, 0.470588, 0.415686}
\definecolor{zuhl}{rgb}{0.1875, 0.26171875, 0.46484375}
\definecolor{orange}{rgb}{1, 0.6470588235, 0}
\definecolor{red}{rgb}{1, 0, 0}

\usepackage{color}
\usepackage{overpic}
\usepackage{pict2e}
\usepackage{booktabs}

\newcommand{\bvec}[1]{\boldsymbol{#1}}

\newcommand{\argmax}{\operatorname{arg~max}\limits}
\newcommand{\argmin}{\operatorname{arg~min}\limits}

\newcommand{\lemref}[1]{Lemma~\ref{#1}}
\newcommand{\propref}[1]{Proposition~\ref{#1}}
\newcommand{\thref}[1]{Theorem~\ref{#1}}

\newcommand{\corref}[1]{Corollary~\ref{#1}}
\newcommand{\sectref}[1]{Section~\ref{#1}}
\newcommand{\remref}[1]{Remark~\ref{#1}}

\newcommand{\appref}[1]{Appendix~\ref{#1}}

\usepackage{soul}
\setstcolor{red}

\allowdisplaybreaks[3]

\usepackage{array}
\usepackage[linesnumbered, algoruled, boxed, lined, vlined]{algorithm2e}
\SetKwRepeat{Do}{do}{while}

\usepackage{hyperref}
\hypersetup{
    colorlinks,
    linkcolor = {red!60!black},
    citecolor = {blue!60!black},
    urlcolor = {blue!50!black}
}
\usepackage{microtype}
\usepackage[nospread]{flushend}

\begin{document}

\title{Asymptotic Expansions of Smooth R\'{e}nyi Entropies and Their Applications}

\IEEEoverridecommandlockouts

\author{%
\IEEEauthorblockN{%
Yuta~Sakai,~\IEEEmembership{Member,~IEEE,} and
Vincent~Y.~F.~Tan,~\IEEEmembership{Senior~Member,~IEEE,}}%
\thanks{This work is supported by an NRF Fellowship (R-263-000-D02-281).}
\thanks{Y.~Sakai is with the Department of Electrical and Computer Engineering, National University of Singapore, Singapore, Email: \url{eleyuta@nus.edu.sg}.
V.~Y.~F.~Tan is with the Department of Electrical and Computer Engineering and the Department of Mathematics, National University of Singapore, Singapore, Email: \url{vtan@nus.edu.sg}.}}%

\maketitle

\begin{abstract}
This study considers the unconditional smooth R\'{e}nyi entropy, the smooth conditional R\'{e}nyi entropy proposed by Kuzuoka [\emph{IEEE Trans.\ Inf.\ Theory}, vol.~66, no.~3, pp.~1674--1690, 2020], and a new quantity which we term the conditional smooth R\'{e}nyi entropy.
In particular, we examine asymptotic expansions of these entropies when the underlying source with its side-information are stationary and memoryless.
Using these smooth R\'{e}nyi entropies, we establish one-shot coding theorems of several information-theoretic problems: Campbell's source coding problems, guessing problems, and task encoding problems, all allowing errors.
In each problem, we consider two error formalisms: the average and maximum error criteria, where the averaging and maximization are taken with respect to the side-information of the source.
Applying our asymptotic expansions to the one-shot coding theorems, we derive various asymptotic fundamental limits for these problems when their error probabilities are allowed to be non-vanishing.
We show that, in non-degenerate settings, the first-order fundamental limits differ under the average and maximum error criteria.
This is in contrast to a different but related setting considered by the present authors (for variable-length conditional source coding allowing errors) in which the first-order terms are identical but the second-order terms are different under these criteria.
\end{abstract}


\IEEEpeerreviewmaketitle

\section{Introduction}

R\'{e}nyi's information measures \cite{renyi_1961} admit various operational meanings in various information-theoretic problems, e.g., Campbell's source coding problems \cite{campbell_1965} which concern with the cumulant generating function of codeword lengths of a prefix-free code (see also \cite{courtade_verdu_isit2014_lossless} for a fixed-to-variable length code without prefix-free constraints), guessing problems \cite{massey_isit1994, arikan_1996}, task encoding problems \cite{bunte_lapidoth_2014}, and so on.
By proposing a new set of axioms, R\'{e}nyi in \cite{renyi_1961} generalized the Shannon entropy $H$ to the R\'{e}nyi entropy $H_{\alpha}$ which is parameterized by $\alpha \in (0, 1) \cup (1, \infty)$, a quantity which is known as the \emph{order.}

Renner and Wolf in \cite{renner_wolf_isit2004, renner_wolf_asiacrypt2005} generalized $H_{\alpha}$ by incorporating another parameter $0 \le \varepsilon < 1$ to form the smooth R\'{e}nyi entropy $H_{\alpha}^{\varepsilon}$; this parameter $\varepsilon$ is known as the smoothness parameter (cf.\ \cite{konig_renner_schaffner_2009}).
Note that the two definitions stated in \cite[Definition~I.1]{renner_wolf_isit2004} and \cite[Section~2.1]{renner_wolf_asiacrypt2005} are slightly different, and the latter is amenable to being generalized to a conditional version of the smooth R\'{e}nyi entropy defined in \cite[Definition~1]{renner_wolf_asiacrypt2005}.
Basic properties of the smooth R\'{e}nyi entropy $H_{\alpha}^{\varepsilon}$ of \cite{renner_wolf_asiacrypt2005} were investigated by Renner and Wolf \cite{renner_wolf_asiacrypt2005} and Koga \cite{koga_itw2013}.
Recently, Kuzuoka \cite{kuzuoka_2019} provided another definition of the smooth conditional R\'{e}nyi entropy based on Arimoto's conditional R\'{e}nyi entropy \cite{arimoto_1977}.
He provided a general formula for the smooth conditional R\'{e}nyi entropy.
Moreover, using the smooth conditional R\'{e}nyi entropy, he \cite{kuzuoka_2019} established one-shot converse and achievability bounds on both Campbell's source coding and guessing problems allowing errors in the presence of common side-information.

\subsection{Main Contributions}

For a stationary memoryless pair of random vectors $X^{n} = (X_{1}, \dots, X_{n})$ and $Y^{n} = (Y_{1}, \dots, Y^{n})$, this study investigates asymptotic expansions of three information measures: the unconditional version of the smooth R\'{e}nyi entropy $H_{\alpha}^{\varepsilon}( X^{n} )$ \cite{renner_wolf_asiacrypt2005}, Kuzuoka's smooth conditional R\'{e}nyi entropy $H_{\alpha}^{\varepsilon}(X^{n} \mid Y^{n})$ \cite{kuzuoka_2019}, and a new quantity that we propose which we term the conditional smooth R\'{e}nyi entropy $\check{H}_{\alpha}^{\varepsilon}(X^{n} \mid Y^{n})$ which is different from both Renner and Wolf's and Kuzuoka's proposals \cite{renner_wolf_asiacrypt2005, kuzuoka_2019}.
For $H_{\alpha}^{\varepsilon}( X^{n} )$, we derive exact first, second, and third-order terms, i.e., coefficients in the $n$, $\sqrt{n}$, and $\log n$ scales, respectively.
More precisely, for fixed real numbers $0 < \alpha < 1$ and $0 < \varepsilon < 1$, we show that 
\begin{align}
H_{\alpha}^{\varepsilon}( X^{n} )
=
n \, H(X) - \sqrt{ n \, V(X) } \, \Phi^{-1}( \varepsilon ) - \frac{ 1 }{ 2 \, (1 - \alpha) } \log n + \mathrm{O}( 1 )
\qquad (\mathrm{as} \ n \to \infty) ,
\label{eq:main-result}
\end{align}
provided that $V(X) > 0$ and $T(X) < \infty$, where these notations are standard in the second-order asymptotics literature (cf.\ \cite{tan_2014}), and will be explicitly defined later.
This third-order asymptotic result is derived by refining Polyanskiy, Poor, and Verd\'{u}'s technical lemma \cite[Lemma~47]{polyanskiy_poor_verdu_2010} whose proof employs the Berry--Esseen theorem.
For the conditional versions, we show that the first-order terms of $H_{\alpha}^{\varepsilon}(X^{n} \mid Y^{n})$ and $\check{H}_{\alpha}^{\varepsilon}(X^{n} \mid Y^{n})$ differ in most non-degenerate cases, and we show that the remainder terms scale as $+\mathrm{O}( \sqrt{n} )$ due to Chebyshev's inequality.

To apply our asymptotic expansions of the smooth R\'{e}nyi entropies to information-theoretic problems \cite{campbell_1965, courtade_verdu_isit2014_lossless, massey_isit1994, arikan_1996, bunte_lapidoth_2014, kuzuoka_2019}, we establish one-shot coding theorems on the problems but \emph{allowing errors.}
While Kumar \emph{et al.} \cite{kumar_sunny_thakre_kumar_2019} studied these problems \cite{campbell_1965, courtade_verdu_isit2014_lossless, massey_isit1994, arikan_1996, bunte_lapidoth_2014} for sources $X$ with finite alphabets $\mathcal{X}$ in the absence of side-information $Y$, we consider sources with \emph{countably infinite} alphabets $\mathcal{X}$ in the \emph{presence} of side-information $Y$.
The extension from finite to countably infinite alphabets $\mathcal{X}$ was mentioned as a direction of future research by Kumar \emph{et al.} \cite[Section~V]{kumar_sunny_thakre_kumar_2019}.
Moreover, in each problem, we consider two error formalisms: average and maximum error criteria, where the averaging and maximization are taken with respect to the side-information $Y$.
Our one-shot coding theorems under average and maximum error criteria are formulated by using Kuzuoka's proposal $H_{\alpha}^{\varepsilon}(X \mid Y)$ and our proposal $\check{H}_{\alpha}^{\varepsilon}(X \mid Y)$, respectively.
We then characterize asymptotic expansions of fundamental limits of these problems.
In the presence of the side-information $Y$, we show that the first-order terms for the average and maximum error formalisms are different in most non-degenerate cases, in contrast to the main result of \cite{sakai_tan_2019_VL}.
In the absence of the side-information $Y$, we provide third-order asymptotic expansions of the fundamental limits by applying \eqref{eq:main-result} to our one-shot coding theorems.
Specifically, for Campbell's source coding problem, our third-order asymptotic expansion is formulated by the right-hand side of \eqref{eq:main-result}.
By comparing this asymptotic result to Strassen's third-order asymptotic result for fixed-length source coding \cite{strassen_1964}, we can quantify the improvement with which ``variable-length compression'' yields over ``fixed-length compression.''

\subsection{Other Related Works}

\subsubsection{Smooth Min- and Max-Entropies}

Renner and Wolf \cite{renner_wolf_isit2004, renner_wolf_asiacrypt2005} proposed the smooth R\'{e}nyi entropy $H_{\alpha}^{\varepsilon}$ to provide operational interpretations of the smooth max-entropy $H_{0}^{\varepsilon}$ and the smooth min-entropy $H_{\infty}^{\varepsilon}$ using two information-theoretic problems, namely, fixed-length source coding and intrinsic randomness \cite{vembu_verdu_1995, han_2003}; these two entropies are special cases of $H_{\alpha}^{\varepsilon}$ by taking the limits as $\alpha \to 0^{+}$ and as $\alpha \to \infty$, respectively.
Interestingly, it is immediate from Strassen's seminal result for fixed-length source coding of independent and identically distributed (i.i.d.) sources $X^{n}$ \cite{strassen_1964} that
\begin{align}
H_{0}^{\varepsilon}( X^{n} )
=
n \, H(X) - \sqrt{ n \, V(X) } \, \Phi^{-1}( \varepsilon ) - \frac{ 1 }{ 2 } \log n + \mathrm{O}( 1 )
\qquad (\mathrm{as} \ n \to \infty)
\end{align}
for a fixed $0 < \varepsilon < 1$, provided that $V(X) > 0$ and $T(X) < \infty$.
This result is consistent with our main result stated in \eqref{eq:main-result}.

In subsequent works, several applications of the smooth max- and min-entropies $H_{0}^{\varepsilon}$ and $H_{\infty}^{\varepsilon}$, respectively, were studied.
Operational characterizations of $H_{0}^{\varepsilon}$ and $H_{\infty}^{\varepsilon}$ in various quantum information-theoretic problems were discussed by K\"{o}nig, Renner, and Schaffner \cite{konig_renner_schaffner_2009}.
Using $H_{0}^{\varepsilon}$ and $H_{\infty}^{\varepsilon}$, Tomamichel, Colbeck, and Renner \cite{tomamichel_colbeck_renner_2009} formulated a quantum version of the asymptotic equipartition property (AEP) in the presence of quantum mechanical side-information.
They \cite{tomamichel_colbeck_renner_2010} also discussed the duality between $H_{0}^{\varepsilon}$ and $H_{\infty}^{\varepsilon}$ in the context of quantum information theory.
Uyematsu \cite{uyematsu_2010, uyematsu_isit2010} provided general formulas for fixed-length source coding problems and resolvability problems in terms of $H_{0}^{\varepsilon}$.
Uyematsu and Kunimatsu provided a general formula for intrinsic randomness problems \cite{uyematsu_kunimatsu_itw2013} in terms of $H_{\infty}^{\varepsilon}$.
Finally, Saito and Matsushima \cite{saito_matsushima_2016} provided a general formula of the threshold of overflow probabilities for variable-length compressions in terms of $H_{0}^{\varepsilon}$.

\subsubsection{Smooth R\'{e}nyi Entropy of Order $0 < \alpha \le 1$}

Operational characterizations of the smooth R\'{e}nyi entropy $H_{\alpha}^{\varepsilon}$ defined in \cite[Section~2.1]{renner_wolf_asiacrypt2005} for $0 < \alpha < 1$ were initiated by Kuzuoka \cite{kuzuoka_isit2016}.
He established one-shot bounds and a general formula for Campbell's source coding problem \cite{campbell_1965} allowing errors.
Sason and Verd\'{u} \cite[Theorem~17]{sason_verdu_2018} also provided a converse bound for Campbell's source coding problem in the absence of prefix-free constraints.
Yagi and Han \cite{yagi_han_isit2017} provided a general formula for the variable-length resolvability problem in terms of the smooth R\'{e}nyi entropy of order one \cite{renner_wolf_isit2004}.

\subsubsection{Unified One-Shot Coding Theorems}

Recently, Kumar, Sunny, Thakre, and Kumar \cite{kumar_sunny_thakre_kumar_2019} proved unified one-shot coding theorems that can be specialized to various information-theoretic problems \cite{campbell_1965, courtade_verdu_isit2014_lossless, massey_isit1994, arikan_1996, bunte_lapidoth_2014, kuzuoka_2019} in the error-free regime.
We will contrast our work to that of Kumar \emph{et al.}; see \sectref{sect:unified}.
Note that in \cite[Lemma~2]{courtade_verdu_isit2014_lossless}, Courtade and Verd\'{u} provided a unified lemma which can be specialized to Campbell's source coding problem \cite{campbell_1965} in the absence of prefix-free constraints.
Sason and Verd\'{u} \cite[Lemma~7]{sason_verdu_2018} established inequalities between the guessing moment and the moment generating function of codeword lengths for a variable-length source code without prefix-free constraints; see \cite[Lemma~6]{sakai_tan_2019_VL} for a similar result to \cite[Lemma~7]{sason_verdu_2018} in the almost lossless regime.

\subsubsection{Variable-Length Compression Allowing Errors}

While this study examines the cumulant generating function of codeword lengths of a variable-length source code \cite{campbell_1965, courtade_verdu_isit2014_lossless} allowing errors as in Kuzuoka's works \cite{kuzuoka_isit2016, kuzuoka_2019}, the (ordinary) expectation of codeword lengths of a variable-length source code allowing errors has been investigated by several researchers \cite{han_2000, koga_yamamoto_2005, kostina_polyanskiy_verdu_2015, sakai_tan_2019_VL}.
Specifically, the present authors \cite{sakai_tan_2019_VL} derived second-order asymptotic expansions of the fundamental limits of variable-length conditional source coding problems under both average and maximum error criteria.
We then showed that the difference between the average and maximum error criteria is manifested in the second-order terms in these asymptotic expansions, and this difference can be quantified by the law of total variance for the information variance.
However, in this work, the difference between the two error criteria is manifested in the first-order term.

\subsection{Paper Organization}

The rest of this paper is organized as follows:
\sectref{sect:unconditional} introduces various definitions of the smooth R\'{e}nyi entropies, and establishes various asymptotic expansions of these quantities for i.i.d.\ sources.
\sectref{sect:appl} provides one-shot coding theorems for various information-theoretic problems \cite{campbell_1965, courtade_verdu_isit2014_lossless, massey_isit1994, arikan_1996, bunte_lapidoth_2014, kuzuoka_2019} in the regime in which the error probabilities are allowed to be non-vanishing.
In this section, we also derive asymptotic expansions of the fundamental limits of these problems by applying the results in \sectref{sect:unconditional}.
\sectref{sect:conclusion} concludes this study and discusses several directions for future works.
Technical proofs are relegated to the appendices.

\section{Asymptotics of Smooth R\'{e}nyi Entropies}
\label{sect:unconditional}

\subsection{Unconditional Version of Smooth R\'{e}nyi Entropy}

Let $\mathcal{X}$ be a countably infinite alphabet, and $X$ an $\mathcal{X}$-valued random variable (r.v.).
Denote by $P_{X} \coloneqq \mathbb{P} \circ X^{-1}$ the probability law of $X$.
Throughout this paper, denote by $\log$ the logarithm to the base $2$.
Given $\alpha \in (0, 1) \cup (1, \infty)$ and $0 \le \varepsilon < 1$, Renner and Wolf \cite[Section~2.1]{renner_wolf_asiacrypt2005} defined the \emph{$\varepsilon$-smooth R\'{e}nyi entropy} of $X$ by%
\footnote{In \cite[Definition~I.1]{renner_wolf_isit2004}, Renner and Wolf also proposed another definition of the $\varepsilon$-smooth R\'{e}nyi entropy.}
\begin{align}
H_{\alpha}^{\varepsilon}( X )
=
H_{\alpha}^{\varepsilon}( P_{X} )
\coloneqq
\frac{ 1 }{ 1 - \alpha } \log \left( \inf_{Q \in \mathcal{B}_{\mathcal{X}}^{\varepsilon}( P_{X} )} \sum_{x \in \mathcal{X}} Q( x )^{\alpha} \right) ,
\label{def:smooth}
\end{align}
where the infimum is taken over the collection $\mathcal{B}_{\mathcal{X}}^{\varepsilon}( P_{X} )$ of sub-probability distributions $Q$ on $\mathcal{X}$ given as
\begin{align}
\mathcal{B}_{\mathcal{X}}^{\varepsilon}( P_{X} )
\coloneqq
\left\{ Q \ \middle| \ \sum_{x \in \mathcal{X}} Q( x ) \ge 1 - \varepsilon \ \mathrm{and} \ 0 \le Q( a ) \le P_{X}( a ) \ \mathrm{for} \ \mathrm{all} \ a \in \mathcal{X} \right\} .
\label{def:eps-ball}
\end{align}
Note that $H_{\alpha}^{\varepsilon}( P )$ coincides with the R\'{e}nyi entropy $H_{\alpha}(X)$ \cite{renyi_1961} if $\varepsilon = 0$.
In other words, one has
\begin{align}
H_{\alpha}^{\varepsilon}( X ) \Big|_{\varepsilon = 0}
=
H_{\alpha}( X )
\coloneqq
\frac{ 1 }{ 1 - \alpha } \log \left( \sum_{x \in \mathcal{X}} P_{X}( x )^{\alpha} \right) .
\end{align}

Given an $\mathcal{X}$-valued r.v.\ $X$ and a real number $0 \le \varepsilon < 1$, we define $\mathcal{A}_{X}^{\varepsilon} \subsetneq \mathcal{X}$ as any proper subset that satisfies the following properties:
\begin{align}
x_{1} \in \mathcal{A}_{X}^{\varepsilon} \ \mathrm{and} \ x_{2} \in \mathcal{X} \setminus \mathcal{A}_{X}^{\varepsilon}
\quad \Longrightarrow \quad
P_{X}( x_{1} ) \ge P_{X}( x_{2} )
\label{eq:largest-A_cond1}
\end{align}
and
\begin{align}
P_{X}( \mathcal{A}_{X}^{\varepsilon} )
<
1 - \varepsilon
\le
P_{X}( \mathcal{A}_{X}^{\varepsilon} ) + \max_{x \in \mathcal{X} \setminus \mathcal{A}_{X}^{\varepsilon}} P_{X}( x ) .
\label{eq:largest-A_cond2}
\end{align}
Note that $\mathcal{A}_{X}^{\varepsilon}$ is the empty set $\emptyset$ if and only if
\begin{align}
\max_{x \in \mathcal{X}} P_{X}( x )
\ge
1 - \varepsilon .
\end{align}
The following lemma states a formula of $H_{\alpha}^{\varepsilon}( X )$ without the infimum operation used in the right-hand side of \eqref{def:smooth}.

\begin{lemma}[{Koga \cite[Theorem~2]{koga_itw2013}}]
\label{lem:majorization}
For every $0 < \alpha < 1$ and $0 \le \varepsilon < 1$, it holds that
\begin{align}
H_{\alpha}^{\varepsilon}( X )
=
\frac{ 1 }{ 1 - \alpha } \log \left( \sum_{x \in \mathcal{A}_{X}^{\varepsilon}} P_{X}( x )^{\alpha} + \Big( 1 - \varepsilon - P_{X}(\mathcal{A}_{X}^{\varepsilon}) \Big)^{\alpha} \right) .
\end{align}
\end{lemma}

\begin{remark}
The collection of all $\mathcal{A}_{X}^{\varepsilon} \subsetneq \mathcal{X}$ satisfying \eqref{eq:largest-A_cond1} and \eqref{eq:largest-A_cond2} is, in general, not a singleton.
However, this diversity of choices is irrelevant in this study, because the sub-probability distribution $Q_{X}^{\varepsilon}$ given as
\begin{align}
Q_{X}^{\varepsilon}( \mathcal{B} )
=
P_{X}( \mathcal{B} \cap \mathcal{A}_{X}^{\varepsilon} )
\qquad (\mathrm{for} \ \mathcal{B} \subset \mathcal{X})
\end{align}
is unique.
In fact, instead of $\mathcal{A}_{X}^{\varepsilon}$, the original statement \cite[Theorem~2]{koga_itw2013} of \lemref{lem:majorization} is stated in terms of a decreasing rearrangement of the probability masses of $P_{X}( \cdot )$.
Assume that $\mathcal{X} = \{ 1, 2, \dots \}$ and
\begin{align}
P_{X}( 1 ) \ge P_{X}( 2 ) \ge P_{X}( 3 ) \ge P_{X}( 4 ) \ge P_{X}( 5 ) \ge \cdots .
\end{align}
Then, the subset $\mathcal{A}_{X}^{\varepsilon}$ can be written as
\begin{align}
\mathcal{A}_{X}^{\varepsilon}
=
\begin{cases}
\emptyset
& \mathrm{if} \ P_{X}( 1 ) \ge 1 - \varepsilon ,
\\
\{ 1, 2, \dots, J \}
& \mathrm{otherwise} ,
\end{cases}
\end{align}
where the positive integer $J$ is chosen so that
\begin{align}
J
=
\sup\left\{ j \ge 0 \ \middle| \ \sum_{k = 1}^{j} P_{X}( k ) < 1 - \varepsilon \right\} .
\end{align}
In this case, it is clear that the $P_{X}$-probability of $\mathcal{A}_{X}^{\varepsilon}$ can be written as
\begin{align}
P_{X}( \mathcal{A}_{X}^{\varepsilon} )
=
\sum_{k = 1}^{J} P_{X}( k ) .
\end{align}
\end{remark}

Now, we shall investigate asymptotic expansions of $H_{\alpha}^{\varepsilon}( X^{n} )$ as $n \to \infty$ for fixed real parameters $0 < \alpha < 1$ and $0 < \varepsilon < 1$, where $X^{n} = (X_{1}, \dots, X_{n})$ denotes $n$ i.i.d.\ copies of $X$.
Define the following three information quantities:
\begin{align}
H( X )
=
H( P_{X} )
& \coloneqq
\sum_{x \in \mathcal{X}} P_{X}( x ) \log \frac{ 1 }{ P_{X}( x ) } ,
\\
V( X )
=
V( P_{X} )
& \coloneqq
\sum_{x \in \mathcal{X}} P_{X}( x ) \, \bigg( \log \frac{ 1 }{ P_{X}( x ) } - H( X ) \bigg)^{2} ,
\\
T( X )
=
T( P_{X} )
& \coloneqq
\sum_{x \in \mathcal{X}} P_{X}( x ) \, \bigg| \log \frac{ 1 }{ P_{X}( x ) } - H( X ) \bigg|^{3} .
\end{align}
In addition, denote by $\Phi^{-1} : (0, 1) \to \mathbb{R}$ the inverse of the Gaussian cumulative distribution function
\begin{align}
\Phi( u )
& \coloneqq
\int_{-\infty}^{u} \varphi( t ) \, \mathrm{d} t ,
\end{align}
where
\begin{align}
\varphi( t )
& \coloneqq
\frac{ 1 }{ \sqrt{ 2 \pi } } \mathrm{e}^{-t^{2}/2} .
\end{align}
The following theorem states an asymptotic expansion of the smooth R\'{e}nyi entropy up to the third-order term.

\begin{theorem}
\label{th:CLT}
Fix two real numbers $0 < \alpha < 1$ and $0 < \varepsilon < 1$.
If $V( X )$ is zero, then
\begin{align}
H_{\alpha}^{\varepsilon}( X^{n} )
=
n \, H( X ) + \mathrm{O}( 1 )
\qquad (\mathrm{as} \ n \to \infty) .
\label{eq:zero-variance}
\end{align}
On the other hand, if $V( X )$ is positive and finite, then
\begin{align}
H_{\alpha}^{\varepsilon}( X^{n} )
=
n \, H( X ) + \mathrm{O}( \sqrt{n} )
\qquad (\mathrm{as} \ n \to \infty) .
\label{eq:no-third}
\end{align}
In particular, if $V( X )$ is positive and $T( X )$ is finite, then
\begin{align}
H_{\alpha}^{\varepsilon}( X^{n} )
=
n \, H( X ) - \sqrt{ n \, V( X ) } \, \Phi^{-1}( \varepsilon ) - \frac{ 1 }{ 2 \, (1 - \alpha) } \log n + \mathrm{O}( 1 )
\qquad (\mathrm{as} \ n \to \infty) .
\label{eq:CLT}
\end{align}
\end{theorem}

\begin{IEEEproof}[Proof of \thref{th:CLT}]
Since \eqref{eq:zero-variance} and \eqref{eq:no-third} are special cases of \thref{th:2nd_avg} stated later when $Y$ is almost surely constant; we defer their proofs to the proof of \thref{th:2nd_avg}.
It remains to prove the asymptotic expansion stated in \eqref{eq:CLT}.

Similar to \cite[Equation~(13)]{kostina_polyanskiy_verdu_2015}, define the \emph{$\varepsilon$-cutoff random transformation action} on a real-valued r.v.\ $Z$ by
\begin{align}
\langle Z \rangle_{\varepsilon}
\coloneqq
\begin{cases}
Z
& \mathrm{if} \ Z < \eta ,
\\
B \, Z
& \mathrm{if} \ Z = \eta ,
\\
0
& \mathrm{if} \ Z > \eta ,
\end{cases}
\label{eq:cutoff}
\end{align}
where $B$ is the Bernoulli r.v.\ with parameter $1 - \beta$ in which $B \Perp Z$, and $\eta \in \mathbb{R}$ and $0 \le \beta < 1$ are chosen so that
\begin{align}
\mathbb{P}\{ Z > \eta \} + \beta \, \mathbb{P}\{ Z = \eta \}
=
\varepsilon .
\label{eq:parameters}
\end{align}
Then, we assert the following lemma.

\begin{lemma}
\label{lem:CLT_MGF}
Let $Z_{1}, Z_{2}, \dots$ be a sequence of independent and real-valued r.v.'s.
For each positive integer $n$, define the following three quantities:
\begin{align}
E_{n}
& =
\sum_{i = 1}^{n} \mathbb{E}[ Z_{i} ] ,
\label{def:En} \\
V_{n}
& =
\sum_{i = 1}^{n} \mathbb{E}[ (Z_{i} - \mathbb{E}[Z_{i}])^{2} ] ,
\label{def:Vn} \\
T_{n}
& =
\sum_{i = 1}^{n} \mathbb{E}[ |Z_{i} - \mathbb{E}[Z_{i}]|^{3} ] .
\label{def:Tn}
\end{align}
Suppose that
\begin{itemize}
\item
there exists a positive constant $c_{1}$ such that $n \, c_{1} < V_{n} < n/c_{1}$ for sufficiently large $n$, and
\item
there exists a positive constant $c_{2}$ such that $T_{n} < c_{2} \, V_{n}$ for sufficiently large $n$.
\end{itemize}
For any fixed real numbers $0 < \varepsilon < 1$ and $s > 0$, it holds that
\begin{align}
\frac{ 1 }{ s } \log \mathbb{E}\bigg[ \bigg\langle \exp\bigg( s \sum_{i = 1}^{n} Z_{i} \bigg) \bigg\rangle_{\varepsilon} \bigg]
=
E_{n} - \sqrt{ V_{n} } \, \Phi^{-1}( \varepsilon ) - \frac{ 1 }{ 2 \, s } \log n + \mathrm{O}( 1 )
\qquad (\mathrm{as} \ n \to \infty) .
\label{eq:CLT_MGF}
\end{align}
\end{lemma}

\begin{IEEEproof}[Proof of \lemref{lem:CLT_MGF}]
We prove \lemref{lem:CLT_MGF} by applying Polyanskiy, Poor, and Verd\'{u}'s upper bound \cite[Lemma~47]{polyanskiy_poor_verdu_2010} on the antilogarithm of the left-hand side of \eqref{eq:CLT_MGF}, and by establishing a corresponding inequality in the opposite direction.
See \appref{app:CLT_MGF} for details.
\end{IEEEproof}

Let $X^{n} = (X_{1}, \dots, X_{n})$ be $n$ i.i.d.\ copies of $X$.
For each $\bvec{x} \in \mathcal{X}^{n}$, define the information density $\iota_{n}( \bvec{x} )$ by
\begin{align}
\iota_{n}( \bvec{x} )
& \coloneqq
\log \frac{ 1 }{ P_{X^{n}}( \bvec{x} ) } .
\end{align}
For the sake of brevity, white $\mathcal{A}_{n} = \mathcal{A}_{X^{n}}^{\varepsilon}$.
Choose an $\bvec{x}^{\ast} \in \mathcal{X}^{n} \setminus \mathcal{A}_{n}$ so that
\begin{align}
\bvec{x}^{\ast}
\in
\argmax\limits_{\bvec{x} \in \mathcal{X}^{n} \setminus \mathcal{A}_{n}} P_{X^{n}}( \bvec{x} ) .
\end{align}
Then, it follows from \lemref{lem:majorization} that
\begin{align}
H_{\alpha}^{\varepsilon}( X^{n} )
& =
\frac{ 1 }{ 1 - \alpha } \log \left( \sum_{\bvec{x} \in \mathcal{A}_{n}} P_{X^{n}}( \bvec{x} )^{\alpha} + \Big( 1 - \varepsilon - P_{X^{n}}( \mathcal{A}_{n} ) \Big)^{\alpha} \right)
\notag \\
& =
\frac{ 1 }{ 1 - \alpha } \log \left( \sum_{\bvec{x} \in \mathcal{A}_{n}} P_{X^{n}}( \bvec{x} ) \, \exp\left( (1 - \alpha) \, \log \frac{ 1 }{ P_{X^{n}}( \bvec{x} ) } \right) + (1 - \varepsilon - P_{X^{n}}( \mathcal{A}_{n} )) \, \exp\left( (1 - \alpha) \, \log \frac{ 1 }{ 1 - \varepsilon - P_{X^{n}}( \mathcal{A}_{n} ) } \right) \right)
\notag \\
& =
\frac{ 1 }{ 1 - \alpha } \log \left( \sum_{\bvec{x} \in \mathcal{A}_{n}} P_{X^{n}}( \bvec{x} ) \, \exp\left( (1 - \alpha) \, \log \frac{ 1 }{ P_{X^{n}}( \bvec{x} ) } \right) \right.
\notag \\
& \qquad \qquad \qquad
{} + (1 - \varepsilon - P_{X^{n}}( \mathcal{A}_{n} )) \left( \exp\left( (1 - \alpha) \, \log \frac{ 1 }{ 1 - \varepsilon - P_{X^{n}}( \mathcal{A}_{n} ) } \right)
\right. \notag \\
& \left. \vphantom{\sum_{\bvec{x} \in \mathcal{A}_{n}}} \left. \qquad \qquad \qquad \qquad \qquad \qquad \qquad \qquad
{} + \exp\left( (1 - \alpha) \, \log \frac{ 1 }{ P_{X^{n}}( \bvec{x}_{n}^{\ast} ) } \right) - \exp\left( (1 - \alpha) \, \log \frac{ 1 }{ P_{X^{n}}( \bvec{x}_{n}^{\ast} ) } \right) \right) \right)
\notag \\
& =
\frac{ 1 }{ 1 - \alpha } \log \left( \mathbb{E}\left[ \left\langle \exp\left( (1 - \alpha) \, \log \frac{ 1 }{ P_{X^{n}}( X^{n} ) } \right) \right\rangle_{\varepsilon} \right] \right.
\notag \\
& \left. \qquad \qquad \qquad
{} + (1 - \varepsilon - P_{X^{n}}( \mathcal{A}_{n} )) \, \left( \exp\left( (1 - \alpha) \, \log \frac{ 1 }{ 1 - \varepsilon - P_{X^{n}}( \mathcal{A}_{n} ) } \right) - \exp\left( (1 - \alpha) \, \log \frac{ 1 }{ P_{X^{n}}( \bvec{x}_{n}^{\ast} ) } \right) \right) \right) ,
\label{eq:MGF_cutoff}
\end{align}
where the last equality follows by the definition of $\langle \cdot \rangle_{\varepsilon}$ stated in \eqref{eq:cutoff}.
Noting that
\begin{align}
0
<
1 - \varepsilon - P_{X^{n}}( \mathcal{A}_{n} )
\le
P_{X^{n}}( \bvec{x}^{\ast} ) ,
\end{align}
it follows from \eqref{eq:MGF_cutoff} that
\begin{align}
\frac{ 1 }{ 1 - \alpha } \log \left( \mathbb{E}\left[ \left\langle \exp\left( (1-\alpha) \log \frac{ 1 }{ P_{X^{n}}(X^{n}) } \right) \right\rangle_{\varepsilon} \right] \right)
\le
H_{\alpha}^{\varepsilon}( X^{n} )
\le
\frac{ 1 }{ 1 - \alpha } \log \left( \mathbb{E}\left[ \left\langle \exp\left( (1-\alpha) \log \frac{ 1 }{ P_{X^{n}}(X^{n}) } \right) \right\rangle_{\varepsilon} \right] + 1 \right)
\label{eq:bounds}
\end{align}
for every $n \ge 1$.
Since $\iota_{n}( \bvec{x} ) \ge n \, H_{\infty}( P )$, we see that
\begin{align}
\mathbb{E}\Big[ \Big\langle \mathrm{e}^{(1-\alpha) \, \iota_{n}( X^{n} )} \Big\rangle_{\varepsilon} \Big]
\ge
(1 - \varepsilon) \, \exp\Big( n \, (1 - \alpha) \, H_{\infty}(X) \Big) ,
\label{eq:cutoff-moment_max-entropy}
\end{align}
where the min-entropy $H_{\infty}( X )$ is defined as
\begin{align}
H_{\infty}( X )
\coloneqq
\lim_{\alpha \to \infty} H_{\alpha}( X )
=
\log \left( \frac{ 1 }{ \max_{x \in \mathcal{X}} P_{X}( x ) } \right) .
\end{align}
In addition, since $V(P) > 0$ implies that $H_{\infty}(P) > 0$, we can find an integer $n_{0}$ satisfying
\begin{align}
(1 - \varepsilon) \, \exp\Big( n \, (1 - \alpha) \, H_{\infty}(X) \Big)
>
1
\label{eq:max-entropy_n0}
\end{align}
for every $n \ge n_{0}$.
Hence, it follows from \eqref{eq:bounds}, \eqref{eq:cutoff-moment_max-entropy}, and \eqref{eq:max-entropy_n0} that
\begin{align}
\frac{ 1 }{ 1 - \alpha } \log \mathbb{E}\Big[ \Big\langle \mathrm{e}^{(1-\alpha) \, \iota_{n}( X^{n} )} \Big\rangle_{\varepsilon} \Big]
\le
H_{\alpha}^{\varepsilon}( P^{n} )
\le
\frac{ 1 }{ 1 - \alpha } \log \mathbb{E}\Big[ \Big\langle \mathrm{e}^{(1-\alpha) \, \iota_{n}( X^{n} )} \Big\rangle_{\varepsilon} \Big] + \frac{ 1 }{ 1 - \alpha } \log 2
\label{eq:bounds_smooth_CGF}
\end{align}
for every $n \ge n_{0}$.
Now, \thref{th:CLT} follows from \eqref{eq:bounds_smooth_CGF} and \lemref{lem:CLT_MGF} with $s = 1-\alpha$, completing the proof of \thref{th:CLT}.
\end{IEEEproof}

\begin{remark}
In the right-hand side of \eqref{eq:CLT} stated in \thref{th:CLT}, while the first-order term $+ n \, H(X)$ and the second-order term $- \sqrt{ n \, V(X) } \, \Phi^{-1}( \varepsilon )$ are independent of $\alpha$, the third-order term $+(\log n) / (2 \, (1-\alpha))$ depends on $\alpha$.
As will be seen in \eqref{eq:comparison_FF} of \sectref{sect:Campbell_prefix-free}, these dependencies characterize the difference between the fundamental limits of fixed-to-fixed length (block) source coding \cite{strassen_1964} and Campbell's source coding problems \cite{campbell_1965} in the almost lossless regime.
\end{remark}

\begin{remark}
The left-hand side of \eqref{eq:CLT_MGF} stated in \lemref{lem:CLT_MGF} is asymptotically equal to the cumulant generating function of $\langle Z_{1} + \dots + Z_{n} \rangle_{\varepsilon}$ with the normalization factor $1/s$, and we now consider the expectation of $\langle Z_{1} + \dots + Z_{n} \rangle_{\varepsilon}$.
A minor extension of Kostina, Polyanskiy, and Verd\'{u}'s result \cite[Lemma~1]{kostina_polyanskiy_verdu_2015} shows that
\begin{align}
\mathbb{E}\left[ \left\langle \sum_{i = 1}^{n} Z_{i} \right\rangle_{\varepsilon} \right]
=
(1 - \varepsilon) \, E_{n} - \sqrt{ V_{n} } \, f_{\mathrm{G}}( \varepsilon ) + \mathrm{O}( 1 )
\qquad (n \to \infty) ,
\label{eq:cutoff_asympt}
\end{align}
where the function $f_{\mathrm{G}} : [0, 1] \to [0, 1/\sqrt{2\pi}]$ is defined by
\begin{align}
f_{\mathrm{G}}( s )
\coloneqq
\begin{cases}
\varphi( \Phi^{-1}(s) )
& \mathrm{if} \ 0 < s < 1 ,
\\
0
& \mathrm{if} \ s = 0 \ \mathrm{or} \ s = 1 .
\end{cases}
\end{align}
Refer to \cite[Appendix~C]{sakai_tan_2019_VL} for a proof of \eqref{eq:cutoff_asympt}.
\end{remark}

\begin{remark}
For a general source $\mathbf{X} = \{ X^{n} = (X_{1}^{(n)}, \dots, X_{n}^{(n)}) \}_{n = 1}^{\infty}$ satisfying the strong converse property, Koga \cite[Theorem~3]{koga_itw2013} showed the following asymptotic expansion:
\begin{align}
H_{\alpha}^{\varepsilon}( X^{n} )
=
n \, H( \mathbf{X} ) + \mathrm{o}( n )
\qquad (\mathrm{as} \ n \to \infty) ,
\label{eq:Koga_general-formula}
\end{align}
where the spectral entropy rate $H( \mathbf{X} )$ is defined as the limit in probability of the sequence
\begin{align}
\left\{ \frac{ 1 }{ n } \log \frac{ 1 }{ P_{X^{n}}( X^{n} ) } \right\}_{n = 1}^{\infty} ,
\end{align}
provided that $\mathbf{X}$ satisfies the strong converse property.
If $X^{n}$ consists of $n$ i.i.d.\ copies of $X$, then \eqref{eq:Koga_general-formula} can be specialized to
\begin{align}
H_{\alpha}^{\varepsilon}( X^{n} )
=
n \, H( X ) + \mathrm{o}( n )
\qquad (\mathrm{as} \ n \to \infty) ,
\label{eq:Koga_iid}
\end{align}
which is a more general result than the following asymptotic result
\begin{align}
\lim_{\varepsilon \to 0^{+}} \lim_{n \to \infty} \frac{ 1 }{ n } H_{\alpha}^{\varepsilon}( X^{n} )
=
H( X )
\end{align}
shown by Renner and Wolf \cite[Lemma~3]{renner_wolf_asiacrypt2005}.
These prior results are consistent with the related results in \thref{th:CLT}.
\end{remark}

Note that $H_{\alpha}^{\varepsilon}( X )$ can be a negative number.
In fact, it is easy to see that
\begin{align}
\lim_{\varepsilon \to 1^{-}} H_{\alpha}^{\varepsilon}( X )
=
- \infty
\end{align}
for every discrete r.v.\ $X$ and every $0 < \alpha < 1$.
\thref{th:CLT} or \eqref{eq:Koga_iid} implies that $H_{\alpha}^{\varepsilon}( X^{n} )$ is positive for sufficiently large $n$, provided that $H( X ) > 0$.

While the R\'{e}nyi entropy satisfies the additivity property for independent r.v.'s, i.e.,
\begin{align}
H_{\alpha}( X^{n} )
=
n \, H_{\alpha}( X )
\label{eq:additivity-Renyi}
\end{align}
for i.i.d.\ r.v.'s $X^{n} = (X_{1}, \dots, X_{n})$, the $\varepsilon$-smooth R\'{e}nyi entropy does not satisfy the additivity property in general.
Thus, we see from \thref{th:CLT} and \eqref{eq:additivity-Renyi} that
\begin{align}
\lim_{n \to \infty} \frac{ H_{\alpha}^{\varepsilon}( X^{n} ) }{ H_{\alpha}( X^{n} ) } 
=
\frac{ H(X) }{ H_{\alpha}(X) }
\le
1
\end{align}
for every $0 < \alpha < 1$ and $0 < \varepsilon < 1$, provided that $0 < H_{\alpha}(X) < \infty$,
where note that $H(X) \le H_{\alpha}(X)$ with equality if and only if $V(X) = 0$.
Namely, the $\varepsilon$-smoothing reduces the R\'{e}nyi entropy in the first-order term.

\subsection{Smooth Conditional R\'{e}nyi Entropy---Kuzuoka's Proposal}

Recently, Kuzuoka \cite{kuzuoka_2019} introduced the smooth conditional R\'{e}nyi entropy to characterize fundamental limits of several information-theoretic problems subject to constraints on \emph{average} error probabilities, where the \emph{averaging} is taken with respect to the common side-information $Y$ in this study.
Let $\mathcal{Y}$ be a countable alphabet.
Consider a $\mathcal{Y}$-valued r.v.\ $Y$ playing the role of \emph{side-information} of $X$.
Denote by $P_{X, Y} \coloneqq \mathbb{P} \circ (X, Y)^{-1}$ (resp.\ $P_{Y} \coloneqq \mathbb{P} \circ Y^{-1}$) the joint (resp.\ marginal) probability distribution induced by $(X, Y)$ (resp.\ $Y$).
Assume without loss of generality that $P_{Y}( y ) > 0$ for every $y \in \mathcal{Y}$.
Then, the conditional probability distribution $P_{X|Y}$ of $X$ given $Y$ is defined as
\begin{align}
P_{X|Y = y}( x )
=
P_{X|Y}(x \mid y)
\coloneqq
\frac{ P_{X, Y}(x, y) }{ P_{Y}( y ) } ,
\end{align}
for each $(x, y) \in \mathcal{X} \times \mathcal{Y}$.
Given $\alpha \in (0, 1) \cup (1, \infty)$ and $0 \le \varepsilon < 1$, Kuzuoka \cite{kuzuoka_2019} defined the $\varepsilon$-smooth conditional R\'{e}nyi entropy of $X$ given $Y$ by
\begin{align}
H_{\alpha}^{\varepsilon}(X \mid Y)
\coloneqq
\frac{ \alpha }{ 1 - \alpha } \log \left( \inf_{Q \in \mathcal{B}_{\mathcal{X} \times \mathcal{Y}}^{\varepsilon}( P_{X, Y} )} \sum_{y \in \mathcal{Y}} \left( \sum_{x \in \mathcal{X}} Q(x, y)^{\alpha} \right)^{1/\alpha} \right) ,
\label{def:conditional_smooth}
\end{align}
where the infimum is taken over the collection $\mathcal{B}_{\mathcal{X} \times \mathcal{Y}}^{\varepsilon}( P_{X, Y} )$ of sub-probability distributions $Q$ on $\mathcal{X} \times \mathcal{Y}$ given as
\begin{align}
\mathcal{B}_{\mathcal{X} \times \mathcal{Y}}^{\varepsilon}( P_{X, Y} )
\coloneqq
\left\{ Q \ \middle| \ \sum_{(x, y) \in \mathcal{X} \times \mathcal{Y}} Q(x, y) \ge 1 - \varepsilon \ \mathrm{and} \ 0 \le Q(a, b) \le P_{X, Y}(a, b) \ \mathrm{for} \ \mathrm{all} \ (a, b) \in \mathcal{X} \times \mathcal{Y} \right\} .
\label{def:eps-ball_joint}
\end{align}
Note that $H_{\alpha}^{\varepsilon}(X \mid Y)$ coincides with Arimoto's conditional R\'{e}nyi entropy $H_{\alpha}(X \mid Y)$ \cite{arimoto_1977} if $\varepsilon = 0$.
In other words, we see that
\begin{align}
H_{\alpha}^{\varepsilon}(X \mid Y) \Big|_{\varepsilon = 0}
& =
H_{\alpha}(X \mid Y)
\coloneqq
\frac{ \alpha }{ 1 - \alpha } \log \left( \sum_{y \in \mathcal{Y}} \left( \sum_{x \in \mathcal{X}} P_{X, Y}(x, y)^{\alpha} \right)^{1/\alpha} \right) .
\end{align}
Moreover, it is clear that $H_{\alpha}^{\varepsilon}(X \mid Y)$ coincides with $H_{\alpha}^{\varepsilon}(X)$ defined in \eqref{def:smooth} if $\mathcal{Y}$ is a singleton.

\begin{remark}
In \cite{kuzuoka_2019}, Kuzuoka called $H_{\alpha}^{\varepsilon}(X \mid Y)$ the conditional smooth R\'{e}nyi entropy.
On the other hand, we see that it is defined by applying the smoothing operation on Arimoto's conditional R\'{e}nyi entropy $H_{\alpha}(X \mid Y)$, where the smoothing operation is taken with respect to the joint distribution $P_{X, Y}$ with the smoothness parameter $\varepsilon$.
From this perspective, in this paper, we call $H_{\alpha}^{\varepsilon}(X \mid Y)$ the smooth conditional R\'{e}nyi entropy.
\end{remark}

Given a real-valued r.v.\ $Z$, define the $\alpha$-expectation operator $\mathbb{E}^{(\alpha)}$ as
\begin{align}
\mathbb{E}^{(\alpha)}[ Z ]
\coloneqq
\frac{ \alpha }{ 1 - \alpha } \log \mathbb{E}\left[ \exp\left( \frac{ 1 - \alpha }{ \alpha } \, Z \right) \right]
\label{def:KN}
\end{align}
for each $\alpha \in (0, 1) \cup (1, \infty)$, where $\mathbb{E}[ \cdot ]$ denotes the expectation operator, and
$
\exp( u )
\coloneqq
2^{u}
$
denotes the exponential function of $u \in \mathbb{R}$ with base-$2$.
After some algebra, Arimoto's conditional R\'{e}nyi entropy $H_{\alpha}(X \mid Y)$ can be written as%
\footnote{In \eqref{eq:AKN}, assume that $\exp( - \infty ) = 0$.}
\begin{align}
H_{\alpha}(X \mid Y)
=
\mathbb{E}^{(\alpha)}[ H_{\alpha}( P_{X|Y} ) ]
=
\frac{ \alpha }{ 1 - \alpha } \log \left( \sum_{y \in \mathcal{Y}} P_{Y}( y ) \, \exp\left( \frac{ 1 - \alpha }{ \alpha } H_{\alpha}( P_{X|Y = y} ) \right) \right) .
\label{eq:AKN}
\end{align}
Inspired by \eqref{eq:AKN}, given a function $\delta : \mathcal{Y} \to [0, 1]$ and a real number $0 < \alpha < 1$, we define
\begin{align}
\bar{\mathsf{H}}_{\alpha}^{\delta( \cdot )}(X \mid Y)
\coloneqq
\mathbb{E}^{(\alpha)}\big[ H_{\alpha}^{\delta(Y)}( P_{X|Y} ) \big]
=
\frac{ \alpha }{ 1 - \alpha } \log \left( \sum_{y \in \mathcal{Y}} P_{Y}( y ) \, \exp\left( \frac{ 1 - \alpha }{ \alpha } \, H_{\alpha}^{\delta(y)}( P_{X|Y = y} ) \right) \right) ,
\label{def:cond-smooth_KN-avg}
\end{align}
where $H_{\alpha}^{\delta(y)}( P_{X|Y = y} )$ is given as
\begin{align}
H_{\alpha}^{\delta(y)}( P_{X|Y = y} )
=
\begin{dcases}
\frac{ 1 }{ 1 - \alpha } \log \left( \inf_{Q \in \mathcal{B}_{\mathcal{X}}^{\delta(y)}( P_{X|Y=y} )} \sum_{x \in \mathcal{X}} Q( x )^{\alpha} \right)
& \mathrm{if} \ 0 \le \delta(y) < 1 ,
\\
-\infty
& \mathrm{if} \ \delta(y) = 1
\end{dcases}
\end{align}
for each $y \in \mathcal{Y}$, and $\mathcal{B}_{\mathcal{X}}^{\delta(y)}( P_{X|Y=y} )$ is defined as in \eqref{def:eps-ball} for each $y \in \mathcal{Y}$.
In particular, if $\delta( y ) = \varepsilon$ for every $y \in \mathcal{Y}$, then we write
\begin{align}
\bar{\mathsf{H}}_{\alpha}^{\varepsilon}(X \mid Y)
=
\bar{\mathsf{H}}_{\alpha}^{\delta( \cdot )}(X \mid Y)
\end{align}
for the sake of brevity.
Kuzuoka \cite[Theorem~1]{kuzuoka_2019} derived a formula of $H_{\alpha}^{\varepsilon}(X \mid Y)$ as a generalization of \lemref{lem:majorization};
Kuzuoka's formula can be restated as follows:

\begin{lemma}[{Kuzuoka \cite[Theorem~1]{kuzuoka_2019}}]
\label{lem:Kuzuoka-formula}
For any $0 < \alpha < 1$ and $0 \le \varepsilon < 1$, it holds that
\begin{align}
H_{\alpha}^{\varepsilon}(X \mid Y)
& =
\inf_{\delta(\cdot) \in \mathcal{E}_{0}( \varepsilon )} \bar{\mathsf{H}}_{\alpha}^{\delta( \cdot )}(X \mid Y) ,
\label{eq:Kuzuoka-formula}
\end{align}
where the infimum is taken over the collection $\mathcal{E}_{0}( \varepsilon )$ of functions $\delta : \mathcal{Y} \to [0, 1]$ satisfying $\mathbb{E}[ \delta( Y ) ] = \varepsilon$.
\end{lemma}

Let $\{ (X_{i}, Y_{i}) \}_{i = 1}^{\infty}$ be i.i.d.\ copies of $(X, Y)$.
Defining two quantities
\begin{align}
H(X \mid Y)
& \coloneqq
\mathbb{E}\left[ \log \frac{ 1 }{ P_{X|Y}(X \mid Y) } \right]
\notag \\
& \:=
\sum_{y \in \mathcal{Y}} \sum_{x \in \mathcal{X}} P_{X, Y}(x, y) \log \frac{ 1 }{ P_{X|Y}(x \mid y) } ,
\\
U(X \mid Y)
& \coloneqq
\mathbb{E}\left[ \left( \log \frac{ 1 }{ P_{X|Y}(X \mid Y) } - H(X \mid Y) \right)^{2} \right]
\notag \\
& \:=
\sum_{y \in \mathcal{Y}} \sum_{x \in \mathcal{X}} P_{X, Y}(x, y) \left( \log \frac{ 1 }{ P_{X|Y}(X \mid Y) } - \sum_{b \in \mathcal{Y}} \sum_{a \in \mathcal{X}} P_{X, Y}(a, b) \log \frac{ 1 }{ P_{X|Y}(a \mid b) } \right)^{2} ,
\end{align}
we state asymptotic expansions of $H_{\alpha}^{\varepsilon}(X^{n} \mid Y^{n})$ as $n \to \infty$ for fixed real parameters $0 < \alpha < 1$ and $0 < \varepsilon < 1$ as follows:

\begin{theorem}
\label{th:2nd_avg}
Fix two real numbers $0 < \alpha < 1$ and $0 < \varepsilon < 1$.
If $U(X \mid Y) = 0$, then
\begin{align}
H_{\alpha}^{\varepsilon}(X^{n} \mid Y^{n})
=
n \, H(X \mid Y) + \mathrm{O}( 1 )
\qquad (\mathrm{as} \ n \to \infty) .
\label{eq:zero-variance_2nd_avg}
\end{align}
On the other hand, if $0 < U(X \mid Y) < \infty$, then
\begin{align}
H_{\alpha}^{\varepsilon}(X^{n} \mid Y^{n})
=
n \, H(X \mid Y) + \mathrm{O}( \sqrt{ n } )
\qquad (\mathrm{as} \ n \to \infty) .
\label{eq:2nd_avg}
\end{align}
\end{theorem}

\begin{IEEEproof}[Proof of \thref{th:2nd_avg}]
Let $\delta : \mathcal{Y} \to [0, 1]$ be a map.
For each $y \in \mathcal{Y}$ satisfying $0 \le \delta( y ) < 1$, denote by $\mathcal{A}_{X|Y}^{\delta(y)}(y) \subsetneq \mathcal{X}$ the proper subset defined as in $\mathcal{A}_{X}^{\varepsilon}$ stated in \eqref{eq:largest-A_cond1} and \eqref{eq:largest-A_cond2} so that the parameter $\varepsilon$ and the probability distribution $P_{X}( \cdot )$ are replaced by $\delta(y)$ and $P_{X|Y}(\cdot \mid y)$, respectively.
If $\delta( y ) = 1$, then suppose that $\mathcal{A}_{X|Y}^{\delta(y)}(y) = \emptyset$.
It follows from Lemmas~\ref{lem:majorization} and~\ref{lem:Kuzuoka-formula} that
\begin{align}
H_{\alpha}^{\varepsilon}(X \mid Y)
& =
\frac{ \alpha }{ 1 - \alpha } \log \left( \inf_{\delta(\cdot) \in \mathcal{E}_{0}( \varepsilon )} \sum_{y \in \mathcal{Y}} P_{Y}( y ) \left( \sum_{x \in \mathcal{A}_{X|Y}^{\delta(y)}(y)} P_{X|Y}(x \mid y)^{\alpha} + \Big( 1 - \delta(y) - P_{X|Y}(\mathcal{A}_{X|Y}^{\delta(y)}(y) \mid y) \Big)^{\alpha} \right)^{1/\alpha} \right)
\label{eq:Kuzuoka_identity}
\end{align}
for every $0 < \alpha < 1$.

Suppose that $U(X \mid Y) = 0$.
We firstly aim to now prove the asymptotic expansion in \eqref{eq:zero-variance_2nd_avg}.
If $H(X \mid Y) = 0$, it is immediate from the definition that $H_{\alpha}^{\varepsilon}(X^{n} \mid Y^{n})$ is constant for $n$, and it remains to consider the case where $H(X \mid Y) > 0$.
In this case, the upper bound part of \eqref{eq:zero-variance_2nd_avg} can also be directly proven by the definition of $H_{\alpha}^{\varepsilon}(X^{n} \mid Y^{n})$ stated in \eqref{def:conditional_smooth}.
On the other hand, the lower bound part of \eqref{eq:zero-variance_2nd_avg} can be proven by employing \lemref{lem:Kuzuoka-formula} and the reverse Markov inequality (cf.\ \cite[Lemma~5.6.1]{han_2003}).
See \appref{app:2nd_avg_zero-variance} for proofs of these asymptotic bounds.

Now suppose that $0 < U(X \mid Y) < \infty$, and we secondly prove the asymptotic expansion in \eqref{eq:2nd_avg}.
In this case, the upper bound part%
\footnote{Given an asymptotic expansion $f( n ) = g( n ) + \mathrm{O}( h(n) )$ as $n \to \infty$, its \emph{upper bound part} means that there exist two positive constants $c_{0}$ and $n_{0}$ such that $f( n ) \le g( n ) + c_{0} \, |h( n )|$ for all $n \ge n_{0}$.
Similarly, its \emph{lower bound part} means that there exist two positive constants $c_{1}$ and $n_{1}$ such that $f( n ) \ge g( n ) - c_{1} \, |h( n )|$ for all $n \ge n_{1}$.}
of \eqref{eq:2nd_avg} can be proven by applying Chebyshev's inequality.
On the other hand, the lower bound part of \eqref{eq:2nd_avg} can be proven by employing \lemref{lem:Kuzuoka-formula}, Chebyshev's inequality, and the reverse Markov inequality.
See \appref{app:2nd_avg} for proofs of these asymptotic bounds.
This completes the proof of \thref{th:2nd_avg}.
\end{IEEEproof}

As will be shown in \sectref{sect:appl}, together with certain one-shot coding theorems formulated by the smooth conditional R\'{e}nyi entropy $H_{\alpha}^{\varepsilon}(X^{n} \mid Y^{n})$, \thref{th:2nd_avg} characterizes the exact first-order result and the order of the residual term (the scaling of $\sqrt{n}$) for various information-theoretic problems \cite{campbell_1965, courtade_verdu_isit2014_lossless, massey_isit1994, arikan_1996, bunte_lapidoth_2014, kuzuoka_2019} under the average error criterion.

\begin{remark}
\label{rem:mixed-source}
Kuzuoka \cite[Theorem~2]{kuzuoka_2019} showed an asymptotic expansion of the $\varepsilon$-smooth conditional R\'{e}nyi entropy for a mixture of i.i.d.\ sources.
When $(X^{n}, Y^{n})$ consists of $n$ i.i.d.\ copies of $(X, Y)$, his result can be specialized to
\begin{align}
H_{\alpha}^{\varepsilon}(X^{n} \mid Y^{n})
=
n \, H(X \mid Y) + \mathrm{o}( n )
\qquad (\mathrm{as} \ n \to \infty) ,
\end{align}
and this first-order term is consistent with the results in \thref{th:2nd_avg}.
\end{remark}

\subsection{A Novel Definition: Conditional Smooth R\'{e}nyi Entropy}

Whereas Kuzuoka \cite{kuzuoka_2019} proposed $H_{\alpha}^{\varepsilon}(X \mid Y)$ to handle the average error criterion in several information-theoretic problems, we now introduce the conditional $\varepsilon$-smooth R\'{e}nyi entropy $\check{H}_{\alpha}^{\varepsilon}(X \mid Y)$ to handle the \emph{maximum} error criterion, where the \emph{maximum} (more precisely, the \emph{supremum}) is taken with respect to all realizations $y$ of the side-information $Y$ in this study.
For each $\alpha \in (0, 1) \cup (1, \infty)$ and $0 \le \varepsilon < 1$, define
\begin{align}
\check{H}_{\alpha}^{\varepsilon}(X \mid Y)
\coloneqq
\frac{ \alpha }{ 1 - \alpha } \log \left( \sum_{y \in \mathcal{Y}} P_{Y}( y ) \inf_{Q \in \mathcal{B}_{\mathcal{X}}^{\varepsilon}( P_{X|Y=y} )} \left( \sum_{x \in \mathcal{X}} Q( x )^{\alpha} \right)^{1/\alpha} \right) ,
\label{def:check}
\end{align}
where for each $y \in \mathcal{Y}$, the infimum is taken over the collection $\mathcal{B}_{\mathcal{X}}^{\varepsilon}( P_{X|Y=y} )$ of sub-probability distributions $Q$ on $\mathcal{X}$; see \eqref{def:eps-ball} for the definition of $\mathcal{B}_{\mathcal{X}}^{\varepsilon}( \cdot )$.
Similar to Kuzuoka's proposal $H_{\alpha}^{\varepsilon}(X \mid Y)$ defined in \eqref{def:conditional_smooth}, note that $\check{H}_{\alpha}^{\varepsilon}(X \mid Y)$ coincides with Arimoto's conditional R\'{e}nyi entropy $H_{\alpha}(X \mid Y)$ if $\varepsilon = 0$, and $\check{H}_{\alpha}^{\varepsilon}(X \mid Y)$ coincides with $H_{\alpha}^{\varepsilon}(X)$ defined in \eqref{def:smooth} if $X$ and $Y$ are independent.
In contrast to \lemref{lem:Kuzuoka-formula}, it can be verified that
\begin{align}
\check{H}_{\alpha}^{\varepsilon}(X \mid Y)
=
\bar{\mathsf{H}}_{\alpha}^{\varepsilon}(X \mid Y)
=
\frac{ \alpha }{ 1 - \alpha } \log \left( \sum_{y \in \mathcal{Y}} P_{Y}( y ) \, \exp\left( \frac{ 1 - \alpha }{ \alpha } \, H_{\alpha}^{\varepsilon}( P_{X|Y=y} ) \right) \right) .
\label{eq:check_identity1}
\end{align}

\begin{remark}
In contrast to the smooth conditional R\'{e}nyi entropy $H_{\alpha}^{\varepsilon}(X \mid Y)$, we call $\check{H}_{\alpha}^{\varepsilon}(X \mid Y)$ the conditional smooth R\'{e}nyi entropy in this paper.
This terminology comes from the observation that $\check{H}_{\alpha}^{\varepsilon}(X \mid Y)$ is defined by conditioning the smooth R\'{e}nyi entropy with the smoothness parameter $\varepsilon$; see \eqref{eq:AKN} and \eqref{eq:check_identity1}.
\end{remark}

Now, we shall provide asymptotic expansions of $\check{H}_{\alpha}^{\varepsilon}(X^{n} \mid Y^{n})$ as $n \to \infty$.
Define
\begin{align}
H( P_{X|Y} )
& \coloneqq
\mathbb{E}\left[ \log \frac{ 1 }{ P_{X|Y}(X \mid Y) } \ \middle| \ Y \right] ,
\\
V(X \mid Y)
& \coloneqq
\mathbb{E}\left[ \left( \log \frac{ 1 }{ P_{X|Y}(X \mid Y) } - H( P_{X|Y} ) \right)^{2} \right]
\notag \\
& \:=
\sum_{y \in \mathcal{Y}} \sum_{x \in \mathcal{X}} P_{X, Y}(x, y) \left( \log \frac{ 1 }{ P_{X|Y}(x \mid y) } - \sum_{a \in \mathcal{X}} P_{X|Y}(a \mid y) \log \frac{ 1 }{ P_{X|Y}(a \mid y) } \right)^{2} ,
\\
H^{(\alpha)}(X \mid Y)
& \coloneqq
\mathbb{E}^{(\alpha)}[ H( P_{X|Y} ) ]
\notag \\
& \:=
\frac{ \alpha }{ 1 - \alpha } \log \left( \sum_{y \in \mathcal{Y}} P_{Y}( y ) \, \exp\left( \frac{ 1 - \alpha }{ \alpha } \sum_{x \in \mathcal{X}} P_{X|Y}(x \mid y) \log \frac{ 1 }{ P_{X|Y}(x \mid y) } \right) \right) .
\label{eq:cond-Shannon_KN}
\end{align}

\begin{theorem}
\label{th:2nd_max}
Fix two real numbers $0 < \alpha < 1$ and $0 < \varepsilon < 1$.
If $V(X \mid Y) = 0$, then
\begin{align}
\check{H}_{\alpha}^{\varepsilon}(X^{n} \mid Y^{n})
=
n \, H^{(\alpha)}(X \mid Y) + \mathrm{O}( 1 )
\quad (\mathrm{as} \ n \to \infty) .
\label{eq:zero-variance_2nd_max}
\end{align}
If $V(X \mid Y) > 0$ and $\sup_{y \in \mathcal{Y}} V( P_{X|Y = y} ) < \infty$, then
\begin{align}
\check{H}_{\alpha}^{\varepsilon}(X^{n} \mid Y^{n})
=
n \, H^{(\alpha)}(X \mid Y) + \mathrm{O}( \sqrt{ n } )
\quad (\mathrm{as} \ n \to \infty) .
\label{eq:2nd_max}
\end{align}
\end{theorem}

\begin{IEEEproof}[Proof of \thref{th:2nd_max}]
It follows from \lemref{lem:majorization} and \eqref{eq:check_identity1} that
\begin{align}
\check{H}_{\alpha}^{\varepsilon}(X \mid Y)
=
\frac{ \alpha }{ 1 - \alpha } \log \left( \sum_{y \in \mathcal{Y}} P_{Y}( y ) \, \left( \sum_{x \in \mathcal{A}_{X|Y=y}^{\varepsilon}} P_{X|Y}(x \mid y)^{\alpha} + \Big( 1 - \varepsilon - P_{X|Y}(\mathcal{A}_{X|Y=y}^{\varepsilon} \mid y) \Big)^{\alpha} \right)^{1/\alpha} \right)
\label{eq:check_identity}
\end{align}
for every $0 < \alpha < 1$ and $0 < \varepsilon < 1$, where the proper subset $\mathcal{A}_{X|Y=y}^{\varepsilon}$ of $\mathcal{X}$ is given by \eqref{eq:largest-A_cond1} and \eqref{eq:largest-A_cond2} with $\delta(y) = \varepsilon$ for each $y \in \mathcal{Y}$.

Suppose that $V(X \mid Y) = 0$, and we firstly verify the asymptotic expansion in \eqref{eq:zero-variance_2nd_max}.
It is clear that
\begin{align}
P_{X|Y}(x \mid y)
=
\begin{dcases}
\exp\Big( - H( P_{X|Y=y} ) \Big)
& \mathrm{if} \ P_{X|Y}(x \mid y) > 0 ,
\\
0
& \mathrm{if} \ P_{X|Y}(x \mid y) = 0 .
\end{dcases}
\label{eq:zero-variance_cond-probab}
\end{align}
In this case, the upper bound part of \eqref{eq:zero-variance_2nd_max} can be directly proven by the definition of $\check{H}_{\alpha}^{\varepsilon}(X \mid Y)$ stated in \eqref{def:check}.
On the other hand, we prove the lower bound part of \eqref{eq:zero-variance_2nd_max} by employing \eqref{eq:check_identity} and the one-sided Chebyshev inequality, and by dividing into two cases: either $U(X \mid Y) = 0$ or $U(X \mid Y) > 0$.
See \appref{app:2nd_max_zero-variance} for proofs of these asymptotic bounds.

Next, suppose that $V(X \mid Y) > 0$ and $\sup_{y \in \mathcal{Y}} V( P_{X|Y = y} ) < \infty$.
We then verify the asymptotic expansion in \eqref{eq:2nd_max}.
The upper bound part of \eqref{eq:2nd_max} can be proven by applying Chebyshev's inequality.
On the other hand, we prove the lower bound part of \eqref{eq:2nd_max} by employing \eqref{eq:check_identity} and the one-sided Chebyshev's inequality, and by dividing into two cases: either $U(X \mid Y) = V(X \mid Y)$ or $U(X \mid Y) > V(X \mid Y)$.
See \appref{app:2nd_max_direct} for proofs of these asymptotic bounds.
This completes the proof of \thref{th:2nd_max}.
\end{IEEEproof}

As will be shown in \sectref{sect:appl}, together with certain one-shot coding theorems formulated by the smooth conditional R\'{e}nyi entropy $\check{H}_{\alpha}^{\varepsilon}(X^{n} \mid Y^{n})$, \thref{th:2nd_max} characterizes the exact first-order result and the order of the residual term (the scaling of $\sqrt{n}$) for various information-theoretic problems \cite{campbell_1965, courtade_verdu_isit2014_lossless, massey_isit1994, arikan_1996, bunte_lapidoth_2014, kuzuoka_2019} under the maximum error criterion.

The following proposition delineates the difference between the first-order terms stated in Theorems~\ref{th:2nd_avg} and~\ref{th:2nd_max}.

\begin{proposition}
\label{prop:KN-avg}
For any $0 < \alpha < 1$, it holds that
\begin{align}
H(X \mid Y)
\le
H^{(\alpha)}(X \mid Y)
\le
H_{\alpha}(X \mid Y) .
\label{eq:KN-avg}
\end{align}
More precisely, we observe the following equality conditions:
\begin{itemize}
\item
the left-hand inequality of \eqref{eq:KN-avg} holds with equality if and only if $U(X \mid Y) = V(X \mid Y)$, and
\item
the right-hand inequality of \eqref{eq:KN-avg} holds with equality if and only if $V(X \mid Y) = 0$.
\end{itemize}
\end{proposition}

\begin{IEEEproof}[Proof of \propref{prop:KN-avg}]
The left-hand inequality of \eqref{eq:KN-avg} follows from Jensen's inequality, and it follows from the equality condition of Jensen's inequality that it holds with equality if and only if $H( P_{X|Y = y} )$ is constant for every $y \in \mathcal{Y}$.
On the other hand, it follows by the law of total variance that
\begin{align}
U(X \mid Y)
=
V(X \mid Y) + \sum_{y \in \mathcal{Y}} P_{Y}( y ) \, \Big( H(X \mid Y) - H( P_{X|Y=y} ) \Big)^{2} ,
\label{eq:law_of_total_variance}
\end{align}
which implies that $U(X \mid Y) = V(X \mid Y)$ if and only if $H( P_{X|Y = y} )$ is constant for every $y \in \mathcal{Y}$.
This is indeed the equality condition of the left-hand inequality of \eqref{eq:KN-avg}.

On the other hand, it is known that the R\'{e}nyi entropy $\alpha \mapsto H_{\alpha}(X)$ is nonincreasing in $\alpha \ge 0$.
More precisely, we readily see that $\alpha \mapsto H_{\alpha}(X)$ is strictly decreasing in $\alpha \ge 0$ if and only if $V(X) > 0$.
Therefore, for each $y \in \mathcal{Y}$ and each $0 < \alpha < 1$, we observe that
\begin{align}
H( P_{X|Y=y} )
\le
H_{\alpha}( P_{X|Y=y} )
\label{eq:Shannon_vs_Renyi_each-y}
\end{align}
with equality if and only if $V( P_{X|Y=y} ) = 0$.
Applying \eqref{eq:Shannon_vs_Renyi_each-y} straightforwardly to the definition of $H^{(\alpha)}(X \mid Y)$ stated in \eqref{eq:cond-Shannon_KN}, we obtain the right-hand inequality of \eqref{eq:KN-avg} and the condition for equality.
This completes the proof of \propref{prop:KN-avg}.
\end{IEEEproof}

\begin{example}[binary symmetric source]
Let $\mathcal{X} = \mathcal{Y} = \{ 0, 1 \}$.
Given $0 < \delta < 1$, consider a pair $(X, Y)$ satisfying
\begin{align}
P_{X, Y}(x, y)
=
\begin{cases}
(1 - \delta)/2
& \mathrm{if} \ x = y ,
\\
\delta / 2
& \mathrm{if} \ x \neq y .
\end{cases}
\end{align}
It is easy to see that
\begin{align}
0
<
U(X \mid Y)
=
V(X \mid Y)
<
\infty ;
\end{align}
thus, it follows from \propref{prop:KN-avg} that
\begin{align}
0
<
H(X \mid Y)
=
H^{(\alpha)}(X \mid Y)
<
H_{\alpha}(X \mid Y) .
\end{align}
\end{example}

\begin{example}[binary erasure source]
Let $\mathcal{X} = \{ 0, 1 \}$ and $\mathcal{Y} = \{ 0, 1, ? \}$.
Given $0 < \delta < 1$, consider a pair $(X, Y)$ satisfying
\begin{align}
P_{X, Y}(x, y)
=
\begin{dcases}
(1 - \delta) / 2
& \mathrm{if} \ x = y ,
\\
\delta / 2
& \mathrm{if} \ y = \; ? ,
\\
0
& \mathrm{if} \ x \neq y \ \mathrm{and} \ y \neq \; ? .
\end{dcases}
\end{align}
It is easy to see that
\begin{align}
0
=
V(X \mid Y)
<
U(X \mid Y)
<
\infty ;
\end{align}
thus, it follows from \propref{prop:KN-avg} that
\begin{align}
0
<
H(X \mid Y)
<
H^{(\alpha)}(X \mid Y)
=
H_{\alpha}(X \mid Y) .
\end{align}
\end{example}

\begin{example}[binary symmetric erasure source]
Let $\mathcal{X} = \{ 0, 1 \}$ and $\mathcal{Y} = \{ 0, 1, ? \}$.
Given two real numbers $0 < \delta_{\mathrm{c}} < 1$ and $0 < \delta_{\mathrm{e}} < 1$, consider a pair $(X, Y)$ satisfying
\begin{align}
P_{X, Y}(x, y)
=
\begin{dcases}
(1 - \delta_{\mathrm{c}} - \delta_{\mathrm{e}}) / 2
& \mathrm{if} \ x = y ,
\\
\delta_{\mathrm{e}} / 2
& \mathrm{if} \ y = \; ? ,
\\
\delta_{\mathrm{c}} / 2
& \mathrm{if} \ x \neq y \ \mathrm{and} \ y \neq \; ? .
\end{dcases}
\end{align}
It is easy to see that
\begin{align}
0
<
V(X \mid Y)
<
U(X \mid Y)
<
\infty ;
\end{align}
thus, it follows from \propref{prop:KN-avg} that
\begin{align}
0
<
H(X \mid Y)
<
H^{(\alpha)}(X \mid Y)
<
H_{\alpha}(X \mid Y) .
\end{align}
\end{example}

\section{Applications}
\label{sect:appl}

This section provides applications of the results in \sectref{sect:unconditional} to Campbell's source coding problem \cite{campbell_1965, kuzuoka_2019}, the guessing problem \cite{massey_isit1994, arikan_1996, kuzuoka_2019}, and the task partition problem \cite{bunte_lapidoth_2014}, all allowing errors.
In this section, suppose that the order $\alpha$ is given as
\begin{align}
\alpha
=
\frac{ 1 }{ 1 + \rho }
\end{align}
for a given positive parameter $\rho$.
Namely, note that $0 < \alpha < 1$.

\subsection{Unified Approach---Converse and Achievability Bounds}
\label{sect:unified}

In this subsection, we establish unified converse and achievability bounds that are applicable to the three above mentioned information-theoretic problems \cite{campbell_1965, courtade_verdu_isit2014_lossless, massey_isit1994, arikan_1996, bunte_lapidoth_2014, kuzuoka_2019}.
Let $\rho > 0$ and $0 \le \varepsilon < 1$ be fixed.
Given two deterministic maps $\epsilon : \mathcal{X} \times \mathcal{Y} \to [0, 1]$ and $\kappa : \mathcal{X} \times \mathcal{Y} \to (0, \infty)$, construct a stochastic map $K : \mathcal{X} \times \mathcal{Y} \to [0, \infty)$ given as
\begin{align}
K(x, y)
=
\begin{cases}
\kappa(x, y)
& \mathrm{with} \ \mathrm{probability} \ 1 - \epsilon(x, y) ,
\\
0
& \mathrm{with} \ \mathrm{probability} \ \epsilon(x, y)
\end{cases}
\label{def:stochastic-K}
\end{align}
for each $(x, y) \in \mathcal{X} \times \mathcal{Y}$.
In this study, all fundamental limits can be formulated by the $\rho$-th moment of $K(X, Y)$ with properly chosen functions $\epsilon : \mathcal{X} \times \mathcal{Y} \to [0, 1]$ and $\kappa : \mathcal{X} \times \mathcal{Y} \to (0, \infty)$.
Therefore, we shall give certain lower and upper bounds on the $\rho$-th moment of $K(X, Y)$ under certain constraints.

We first provide unified converse bounds under the average and maximum error formalisms.
Define
\begin{align}
R(\epsilon, \kappa)
\coloneqq
\sup_{y \in \mathcal{Y}} \sum_{\substack{ x \in \mathcal{X} : \\ \epsilon(x, y) < 1 }} \frac{ 1 }{ \kappa( x, y ) } ,
\label{def:redundancy-R}
\end{align}
which represents a certain redundancy term in the converse bounds.
It is worth mentioning that the sum in the right-hand side of \eqref{def:redundancy-R} is taken over all $x \in \mathcal{X}$ satisfying the constraint that $\epsilon(x, y) < 1$, and this constraint aids to establish valid converse bounds in the guessing and the task encoding problems over the countably infinite alphabet $\mathcal{X}$.
The following two lemmas are generalizations of Courtade and Verd\'{u}'s unified converse bound \cite[Lemma~2]{courtade_verdu_isit2014_lossless} and Kumar, Sunny, Thakre, and Kumar's unified converse bound \cite[Theorem~18]{kumar_sunny_thakre_kumar_2019} from error-free settings (i.e., $\varepsilon = 0$) to almost lossless settings in which the error probability is at most $\varepsilon > 0$.

\begin{lemma}[unified converse bound---average error criterion]
\label{lem:unified_converse_avg}
If
\begin{align}
\mathbb{E}[ \epsilon(X, Y) ]
\le
\varepsilon ,
\label{eq:average-error}
\end{align}
then it holds that
\begin{align}
\frac{ 1 }{ \rho } \log \mathbb{E}[ K(X, Y)^{\rho} ]
& \ge
H_{\alpha}^{\varepsilon}(X \mid Y) - \log R(\epsilon, \kappa) .
\end{align}
\end{lemma}

\begin{IEEEproof}[Proof of \lemref{lem:unified_converse_avg}]
See \appref{app:unified_converse_avg}.
\end{IEEEproof}

\begin{lemma}[unified converse bound---maximum error criterion]
\label{lem:unified_converse_max}
If
\begin{align}
\sup_{y \in \mathcal{Y}} \mathbb{E}[ \epsilon(X, Y) \mid Y = y ]
\le
\varepsilon ,
\label{eq:maximum-error}
\end{align}
then it holds that
\begin{align}
\frac{ 1 }{ \rho } \log \mathbb{E}[ K(X, Y)^{\rho} ]
& \ge
\check{H}_{\alpha}^{\varepsilon}(X \mid Y) - \log R(\epsilon, \kappa) .
\end{align}
\end{lemma}

\begin{IEEEproof}[Proof of \lemref{lem:unified_converse_max}]
See \appref{app:unified_converse_max}.
\end{IEEEproof}

Next, we provide a unified achievability bound that is applicable to both maximum and average error formalisms.
Let $\delta : \mathcal{Y} \to [0, 1]$ be a deterministic map.
Recall that for each $y \in \mathcal{Y}$, the proper subset $\mathcal{A}_{X|Y}^{\delta(y)}(y) \subsetneq \mathcal{X}$ is defined to satisfy \eqref{eq:largest-A_cond1} and \eqref{eq:largest-A_cond2}, where $\varepsilon$ and $P_{X}(\cdot)$ are replaced by $\delta(y)$ and $P_{X|Y}(\cdot \mid y)$, respectively.
For each $y \in \mathcal{Y}$, choose an element $x^{\ast}(y) \in \mathcal{X} \setminus \mathcal{A}_{X|Y}^{\delta(y)}(y)$ so that
\begin{align}
x^{\ast}( y )
\in
\argmax\limits_{x \in \mathcal{X} \setminus \mathcal{A}_{X|Y}^{\delta(y)}(y)} P_{X|Y}(x \mid y) .
\label{def:x-ast}
\end{align}
and choose the number $M(y)$ as
\begin{align}
M(y)
=
1 - \delta( y ) - P_{X|Y}(\mathcal{A}_{X|Y}^{\delta(y)}(y) \mid y) .
\end{align}
Moreover, define the conditional probability distribution $Q_{X|Y}^{(\alpha, \delta(\cdot))}$ as
\begin{align}
\hspace{-0.5em}
Q_{X|Y}^{(\alpha, \delta(\cdot))}(x \mid y)
\coloneqq
\begin{dcases}
\frac{ P_{X|Y}(x \mid y)^{\alpha} }{ \sum_{a \in \mathcal{A}_{X|Y}^{\delta(y)}(y)} P_{X|Y}(a \mid y)^{\alpha} + M(y)^{\alpha} }
& \mathrm{if} \ x \in \mathcal{A}_{X|Y}^{\delta(y)}(y) ,
\\
\frac{ M(y)^{\alpha} }{ \sum_{a \in \mathcal{A}_{X|Y}^{\delta(y)}(y)} P_{X|Y}(a \mid y)^{\alpha} + M(y)^{\alpha} }
& \mathrm{if} \ x = x^{\ast}(y) ,
\\
0
& \mathrm{otherwise}
\end{dcases}\hspace{-0.5em}
\label{def:tilted-R}
\end{align}
for each $(x, y) \in \mathcal{X} \times \mathcal{Y}$.
Here, note that $Q_{X|Y}^{(\alpha, \delta(\cdot))}$ depends only on $P_{X|Y}$, $\alpha$, and $\delta( \cdot )$.

\begin{lemma}[unified achievability bound]
\label{lem:unified_direct}
Suppose that there exists a positive constant $c$ such that
\begin{align}
\kappa(x, y) \, Q_{X|Y}^{(\alpha, \delta(\cdot))}(x \mid y)
\le
c
\label{eq:kappa-tilted}
\end{align}
for every $(x, y) \in \mathcal{X} \times \mathcal{Y}$.
Then, there exists a deterministic map $\epsilon : \mathcal{X} \times \mathcal{Y} \to [0, 1]$ satisfying
\begin{align}
\mathbb{E}[ \epsilon(X, Y) \mid Y = y ]
& =
\delta( y )
\qquad (\mathrm{for} \ y \in \mathcal{Y}) ,
\label{eq:cond-error_delta} \\
\frac{ 1 }{ \rho } \log \mathbb{E}[ K(X, Y)^{\rho} ]
& \le
\bar{\mathsf{H}}_{\alpha}^{\delta( \cdot )}(X \mid Y) + \log c ,
\label{eq:direct-bound_KN}
\end{align}
where $\bar{\mathsf{H}}_{\alpha}^{\delta(\cdot)}(X \mid Y)$ is defined in \eqref{def:cond-smooth_KN-avg}.
\end{lemma}

\begin{IEEEproof}[Proof of \lemref{lem:unified_direct}]
See \appref{app:unified_direct}.
\end{IEEEproof}

To apply \lemref{lem:unified_direct} for the average and maximum error formalisms to various information-theoretic problems, we choose an appropriate map $\delta : \mathcal{Y} \to [0, 1]$ by referring to the identities of \eqref{eq:Kuzuoka-formula} in \lemref{lem:Kuzuoka-formula} and \eqref{eq:check_identity1}, respectively.
More precisely, for the average error formalism, we find a map $\delta^{\ast}( \cdot ) \in \mathcal{E}_{0}$ so that
\begin{align}
H_{\alpha}^{\varepsilon}(X \mid Y)
\ge
\bar{H}_{\alpha}^{\delta^{\ast}(\cdot)}(X \mid Y) - \zeta
\end{align}
for an arbitrarily small $\zeta > 0$.
On the other hand, for the maximum error formalism, we choose a constant function $\delta : \mathcal{Y} \to [0, 1]$ as $\delta( y ) = \varepsilon$ for every $y \in \mathcal{Y}$.
In Campbell's source coding problems \cite{campbell_1965, kuzuoka_2019} stated in the next subsection, our achievability bounds are proven by constructing \emph{Shannon codes} with the conditional distribution $Q_{X|Y}^{(\alpha, \delta(\cdot))}$ defined in \eqref{def:tilted-R}, where the 1-bit redundancy terms of the Shannon codes can be obtained by choosing the constant $c = 2$.

\subsection{Campbell's Source Coding Problem}
\label{sect:Campbell_prefix-free}

Given a correlated source $(X, Y)$, we consider compressing the source $X$ into a variable-length binary string when the side-information $Y$ is available at both encoder and decoder.
Denote by
\begin{align}
\{ 0, 1 \}^{\ast}
\coloneqq
\{ \varnothing \} \cup \bigcup_{n = 1}^{\infty} \{ 0, 1 \}^{n}
\end{align}
the set of finite-length binary strings containing the empty string $\varnothing$.
Let $F : \mathcal{X} \times \mathcal{Y} \to \{ 0, 1 \}^{\ast}$ and $G : \{ 0, 1 \}^{\ast} \times \mathcal{Y} \to \mathcal{X}$ be two random maps playing the roles of a \emph{stochastic encoder} and a \emph{stochastic decoder,} respectively.
We call this pair $(F, G)$ a \emph{variable-length stochastic code.}
For each $y \in \mathcal{Y}$, we say that the codeword set
\begin{align}
\mathcal{C}_{y}(X, Y, F)
\coloneqq
\{ \bvec{b} \in \{ 0, 1 \}^{\ast} \mid \mathbb{P}\{ F(X, Y) = \bvec{b} \mid Y = y \} > 0 \}
\end{align}
is \emph{prefix-free} if for every distinct $\bvec{b}_{1}, \bvec{b}_{2} \in \mathcal{C}_{y}(X, Y, F)$, a codeword $\bvec{b}_{1}$ is not a prefix of another codeword $\bvec{b}_{2}$.

Given a $\{ 0, 1 \}^{\ast}$-valued r.v.\ $B$, consider the cumulant generating function of codeword lengths \cite{campbell_1965, courtade_verdu_isit2014_lossless} defined as
\begin{align}
\Lambda(B \, \| \, \rho)
\coloneqq
\log \mathbb{E}[ 2^{\rho \ell( B )} ]
\label{def:CGF}
\end{align}
for a positive parameter $\rho$, where $\ell : \{ 0, 1 \}^{\ast} \to \mathbb{N} \cup \{ 0 \}$ stands for the length function of a binary string, i.e.,
\begin{align}
\ell( \varnothing )
& =
0 ,
\\
\ell( 0 ) = \ell( 1 )
& =
1 ,
\\
\ell( 00 ) = \ell( 01 ) = \ell( 10 ) = \ell( 11 )
& =
2 ,
\\
\ell( 000 ) = \ell( 001 ) = \ell( 010 ) = \ell( 011 ) = \ell( 100 ) = \ell( 101 ) = \ell( 110 ) = \ell( 111 )
& =
3 ,
\end{align}
and so on.
Then, we are interested to characterize fundamental limits defined as the infimum of positive real numbers $L$ such that there exists a variable-length stochastic code $(F, G)$ satisfying
\begin{align}
\Lambda(F(X, Y) \, \| \, \rho)
& \le
\rho L
\label{eq:CGF_Capmbell}
\end{align}
under a certain constraint on error probabilities.
We now introduce two error formalisms as follows:

\begin{definition}[average error criterion]
A \emph{$(\rho, L, \varepsilon)_{\mathrm{avg}}$-code} for a correlated source $(X, Y)$ is a variable-length stochastic code $(F, G)$ such that \eqref{eq:CGF_Capmbell} holds, the codeword set $\mathcal{C}_{y}(X, Y, F)$ is prefix-free for every $y \in \mathcal{Y}$, and
\begin{align}
\mathbb{P}\{ X \neq G(F(X, Y), Y) \}
& \le
\varepsilon .
\label{eq:average-error_Campbell}
\end{align}
\end{definition}

\begin{definition}[maximum error criterion]
A \emph{$(\rho, L, \varepsilon)_{\max}$-code} for a correlated source $(X, Y)$ is a variable-length stochastic code $(F, G)$ such that \eqref{eq:CGF_Capmbell} holds, the codeword set $\mathcal{C}_{y}(X, Y, F)$ is prefix-free for every $y \in \mathcal{Y}$, and
\begin{align}
\sup_{y \in \mathcal{Y}} \mathbb{P}\{ X \neq G(F(X, Y), Y) \mid Y = y \}
& \le
\varepsilon .
\end{align}
\end{definition}

Given two real numbers $\rho > 0$ and $0 \le \varepsilon < 1$, define the following two fundamental limits:
\begin{align}
\Lambda_{\mathrm{avg}}^{\ast}(X, Y \, \| \, \rho, \varepsilon )
& \coloneqq
\inf\{ L > 0 \mid \text{there exists a $(\rho, L, \varepsilon)_{\mathrm{avg}}$-code for $(X, Y)$} \} ,
\label{def:Lambda-ast_avg} \\
\Lambda_{\max}^{\ast}(X, Y \, \| \, \rho, \varepsilon )
& \coloneqq
\inf\{ L > 0 \mid \text{there exists a $(\rho, L, \varepsilon)_{\max}$-code for $(X, Y)$} \} .
\label{def:Lambda-ast_max}
\end{align}
We now state the following one-shot coding theorems.

\begin{theorem}[average error criterion]
\label{lem:one-shot_prefix-free_avg}
For every $\rho > 0$ and $0 \le \varepsilon < 1$, it holds that
\begin{align}
H_{\alpha}^{\varepsilon}(X \mid Y)
\le
\Lambda_{\mathrm{avg}}^{\ast}(X, Y \, \| \, \rho, \varepsilon )
<
H_{\alpha}^{\varepsilon}(X \mid Y) + 1 + \frac{ 1 }{ \rho } \log \left( \frac{ 1 }{ 1 - \varepsilon } \right) .
\end{align}
\end{theorem}

\begin{IEEEproof}[Proof of \thref{lem:one-shot_prefix-free_avg}]
It is clear that for any stochastic code $(F, G)$ and a correlated source $(X, Y)$, there exists a deterministic decoder $g : \{ 0, 1 \}^{\ast} \times \mathcal{Y} \to \mathcal{X}$ satisfying
\begin{align}
\mathbb{P}\{ X \neq g(F(X, Y), Y) \}
\le
\mathbb{P}\{ X \neq G(F(X, Y), Y) \} .
\end{align}
Thus, it suffices to consider deterministic decoders $g : \{ 0, 1 \}^{\ast} \times \mathcal{Y} \to \mathcal{X}$.
In this paper, a variable-length stochastic code $(F, g)$ is called a \emph{variable-length semi-stochastic code} if $g$ is deterministic.
Instead of $\Lambda(B \, \| \, \rho)$ defined in \eqref{def:CGF}, we now consider a cutoff version of the cumulant generating function of codeword lengths as follows:
\begin{align}
\tilde{\Lambda}(X, Y, F, g \, \| \, \rho)
\coloneqq
\log \mathbb{E}\big[ 2^{\rho \ell( F(X, Y) )} \, \bvec{1}_{\{ X = g(F(X, Y), Y) \}} \big] .
\end{align}
Namely, instead of \eqref{eq:CGF_Capmbell}, we are interested in the infimum of positive real numbers $L$ such that there exists a variable-length semi-stochastic code $(F, g)$ satisfying
\begin{align}
\tilde{\Lambda}(X, Y, F, g \, \| \, \rho)
& \le
\rho L
\label{eq:cutoff-CGF_Capmbell}
\end{align}
under the average error criterion.

\begin{definition}
Let $\rho > 0$, $L \ge 0$, and $0 \le \varepsilon < 1$ be real numbers.
Given a source $X$, a \emph{$(\rho, L, \varepsilon)_{\mathrm{avg}}$-weak-code} for the source $X$ is a variable-length semi-stochastic code $(F, g)$ such that \eqref{eq:cutoff-CGF_Capmbell} holds, the codeword set $\mathcal{C}_{y}(X, Y, F)$ is prefix-free for every $y \in \mathcal{Y}$, and
\begin{align}
\mathbb{P}\{ X \neq g(F(X, Y), Y) \}
& \le
\varepsilon .
\end{align}
\end{definition}

Similar to $\Lambda_{\mathrm{avg}}^{\ast}(X, Y \, \| \, \rho, \varepsilon)$ defined in \eqref{def:Lambda-ast_avg}, define
\begin{align}
\tilde{\Lambda}_{\mathrm{avg}}^{\ast}(X, Y \, \| \, \rho, \varepsilon )
\coloneqq
\inf\{ L > 0 \mid \text{there exists a $(\rho, L, \varepsilon)_{\mathrm{avg}}$-weak-code for $(X, Y)$} \} .
\label{eq:tilde-Lambda}
\end{align}
Instead of $\Lambda_{\mathrm{avg}}^{\ast}(X, Y \, \| \, \rho, \varepsilon)$, the following lemma establishes one-shot bounds on $\tilde{\Lambda}_{\mathrm{avg}}^{\ast}(X, Y \, \| \, \rho, \varepsilon )$.

\begin{lemma}
\label{lem:tilde_one-shot}
For any $\rho > 0$ and $0 \le \varepsilon < 1$, it holds that
\begin{align}
H_{\alpha}^{\varepsilon}(X \mid Y)
\le
\tilde{\Lambda}_{\mathrm{avg}}^{\ast}(X, Y \, \| \, \rho, \varepsilon)
<
H_{\alpha}^{\varepsilon}(X \mid Y) + 1 .
\label{eq:kuzuoka_one-shot}
\end{align}
\end{lemma}

\begin{IEEEproof}[Proof of \lemref{lem:tilde_one-shot}]
The converse and achievability bounds can be proven via the unified approaches as stated in Lemmas~\ref{lem:unified_converse_avg} and~\ref{lem:unified_direct}, respectively; see \appref{app:tilde_one-shot} for details.
\end{IEEEproof}

The following lemma provides inequalities between $\Lambda_{\mathrm{avg}}^{\ast}(X, Y \, \| \, \rho, \varepsilon)$ and $\tilde{\Lambda}_{\mathrm{avg}}^{\ast}(X, Y \, \| \, \rho, \varepsilon)$.

\begin{lemma}
\label{lem:no-tilde_tilde}
For any $\rho > 0$ and $0 < \varepsilon < 1$, it holds that
\begin{align}
\tilde{\Lambda}_{\mathrm{avg}}^{\ast}(X, Y \, \| \, \rho, \varepsilon)
\le
\Lambda_{\mathrm{avg}}^{\ast}(X, Y \, \| \, \rho, \varepsilon)
\le
\tilde{\Lambda}_{\mathrm{avg}}^{\ast}(X, Y \, \| \, \rho, \varepsilon) + \frac{ 1 }{ \rho } \log \left( \frac{ 1 }{ 1 - \varepsilon } \right) .
\label{eq:no-tilde_tilde}
\end{align}
\end{lemma}

\begin{IEEEproof}[Proof of \lemref{lem:no-tilde_tilde}]
See \appref{app:no-tilde_tilde}.
\end{IEEEproof}

The proof of \thref{lem:one-shot_prefix-free_avg} is immediately completed by combining Lemmas~\ref{lem:tilde_one-shot} and~\ref{lem:no-tilde_tilde}.
\end{IEEEproof}

\begin{theorem}[maximum error criterion]
\label{lem:one-shot_prefix-free_max}
For every $\rho > 0$ and $0 \le \varepsilon < 1$, it holds that
\begin{align}
\check{H}_{\alpha}^{\varepsilon}(X \mid Y)
\le
\Lambda_{\max}^{\ast}(X, Y \, \| \, \rho, \varepsilon )
<
\check{H}_{\alpha}^{\varepsilon}(X \mid Y) + 1 + \frac{ 1 }{ \rho } \log \left( \frac{ 1 }{ 1 - \varepsilon } \right) .
\end{align}
\end{theorem}

\begin{IEEEproof}[Proof of \thref{lem:one-shot_prefix-free_max}]
Similar to the previous subsection, we introduce a weaker version of semi-stochastic codes as follows:

\begin{definition}
Let $\rho > 0$, $L \ge 0$, and $0 \le \varepsilon < 1$ be real numbers.
Given a source $X$, a \emph{$(\rho, L, \varepsilon)_{\max}$-weak-code} for the source $X$ is a variable-length semi-stochastic code $(F, g)$ such that \eqref{eq:cutoff-CGF_Capmbell} holds, the codeword set $\mathcal{C}_{y}(X, Y, F)$ is prefix-free for every $y \in \mathcal{Y}$, and
\begin{align}
\sup_{y \in \mathcal{Y}} \mathbb{P}\{ X \neq g(F(X, Y), Y) \mid Y = y \}
& \le
\varepsilon .
\end{align}
\end{definition}

Similar to $\Lambda_{\max}^{\ast}(X, Y \, \| \, \rho, \varepsilon)$ defined in \eqref{def:Lambda-ast_max}, define
\begin{align}
\tilde{\Lambda}_{\max}^{\ast}(X, Y \, \| \, \rho, \varepsilon )
\coloneqq
\inf\{ L > 0 \mid \text{there exists a $(\rho, L, \varepsilon)_{\max}$-weak-code for the correlated source $(X, Y)$} \} .
\label{eq:tilde-Lambda_max}
\end{align}
Instead of $\Lambda_{\max}^{\ast}(X, Y \, \| \, \rho, \varepsilon)$, the following lemma establishes one-shot bounds on $\tilde{\Lambda}_{\max}^{\ast}(X, Y \, \| \, \rho, \varepsilon )$.

\begin{lemma}
\label{lem:tilde_one-shot_max}
For any $\rho > 0$ and $0 \le \varepsilon < 1$, it holds that
\begin{align}
\check{H}_{\alpha}^{\varepsilon}(X \mid Y)
\le
\tilde{\Lambda}_{\max}^{\ast}(X, Y \, \| \, \rho, \varepsilon)
<
\check{H}_{\alpha}^{\varepsilon}(X \mid Y) + 1 .
\label{eq:kuzuoka_one-shot_max}
\end{align}
\end{lemma}

\begin{IEEEproof}[Proof of \lemref{lem:tilde_one-shot_max}]
See \appref{app:tilde_one-shot_max}.
\end{IEEEproof}

The following lemma provides inequalities between $\Lambda_{\max}^{\ast}(X, Y \, \| \, \rho, \varepsilon)$ and $\tilde{\Lambda}_{\max}^{\ast}(X, Y \, \| \, \rho, \varepsilon)$.

\begin{lemma}
\label{lem:no-tilde_tilde_max}
For any $\rho > 0$ and $0 < \varepsilon < 1$, it holds that
\begin{align}
\tilde{\Lambda}_{\max}^{\ast}(X, Y \, \| \, \rho, \varepsilon)
\le
\Lambda_{\max}^{\ast}(X, Y \, \| \, \rho, \varepsilon)
\le
\tilde{\Lambda}_{\max}^{\ast}(X, Y \, \| \, \rho, \varepsilon) + \frac{ 1 }{ \rho } \log \left( \frac{ 1 }{ 1 - \varepsilon } \right) .
\label{eq:no-tilde_tilde_max}
\end{align}
\end{lemma}

\begin{IEEEproof}[Proof of \lemref{lem:no-tilde_tilde_max}]
See \appref{app:no-tilde_tilde_max}.
\end{IEEEproof}

The proof of \thref{lem:one-shot_prefix-free_max} is immediately completed by combining Lemmas~\ref{lem:tilde_one-shot_max} and~\ref{lem:no-tilde_tilde_max}.
\end{IEEEproof}

\begin{remark}
The converse bound of \thref{lem:one-shot_prefix-free_avg} is the same as Kuzuoka's converse bound \cite[Theorem~6]{kuzuoka_2019}, and the achievability bound differs slightly compared to \cite[Theorem~7]{kuzuoka_2019}.
\end{remark}

Now, we can obtain the following asymptotic results.

\begin{corollary}[average error criterion]
\label{th:asympt_prefix-free_avg}
Let $\rho > 0$ and $0 < \varepsilon < 1$ be fixed.
If $U(X \mid Y) = 0$, then
\begin{align}
\Lambda_{\mathrm{avg}}^{\ast}(X^{n}, Y^{n} \, \| \, \rho, \varepsilon)
=
n \, H(X \mid Y) + \mathrm{O}( 1 )
\qquad (\mathrm{as} \ n \to \infty) .
\end{align}
On the other hand, if $0 < U(X \mid Y) < \infty$, then
\begin{align}
\Lambda_{\mathrm{avg}}^{\ast}(X^{n}, Y^{n} \, \| \, \rho, \varepsilon)
=
n \, H(X \mid Y) + \mathrm{O}( \sqrt{n} )
\qquad (\mathrm{as} \ n \to \infty) .
\end{align}
\end{corollary}

\begin{IEEEproof}[Proof of \corref{th:asympt_prefix-free_avg}]
\corref{th:asympt_prefix-free_avg} follows from Theorems~\ref{th:2nd_avg} and~\ref{lem:one-shot_prefix-free_avg}.
\end{IEEEproof}

\begin{corollary}[maximum error criterion]
\label{th:asympt_prefix-free_max}
Let $\rho > 0$ and $0 < \varepsilon < 1$ be fixed.
If $V(X \mid Y) = 0$, then
\begin{align}
\Lambda_{\max}^{\ast}(X^{n}, Y^{n} \, \| \, \rho, \varepsilon)
=
n \, H^{(\alpha)}(X \mid Y) + \mathrm{O}( 1 )
\qquad (\mathrm{as} \ n \to \infty) .
\end{align}
On the other hand, if $V(X \mid Y) > 0$ and $\sup_{y \in \mathcal{Y}} V( P_{X|Y=y} ) < \infty$, then
\begin{align}
\Lambda_{\max}^{\ast}(X^{n}, Y^{n} \, \| \, \rho, \varepsilon)
=
n \, H^{(\alpha)}(X \mid Y) + \mathrm{O}( \sqrt{n} )
\qquad (\mathrm{as} \ n \to \infty) .
\end{align}
\end{corollary}

\begin{IEEEproof}[Proof of \corref{th:asympt_prefix-free_max}]
\corref{th:asympt_prefix-free_max} follows from Theorems~\ref{th:2nd_max} and~\ref{lem:one-shot_prefix-free_max}.
\end{IEEEproof}

Now, consider compressing the source $X$ into a variable-length binary string \emph{in the absence of the side-information $Y$.}
Formally, when $Y$ is almost surely constant, we denote by
\begin{align}
\Lambda^{\ast}(X \, \| \, \rho, \varepsilon)
\coloneqq
\Lambda_{\mathrm{avg}}^{\ast}(X, Y \, \| \, \rho, \varepsilon)
=
\Lambda_{\max}^{\ast}(X, Y \, \| \, \rho, \varepsilon)
\end{align}
our considered fundamental limit here.

\begin{remark}
If $\varepsilon = 0$ and $|\mathcal{Y}| = 1$, then Theorems~\ref{lem:one-shot_prefix-free_avg} and~\ref{lem:one-shot_prefix-free_max} coincide with Campbell's one-shot bounds \cite[Equation~(18)]{campbell_1965}:
\begin{align}
H_{\alpha}( X )
\le
\Lambda^{\ast}(X \, \| \, \rho, 0)
<
H_{\alpha}( X ) + 1 .
\end{align}
\end{remark}

Now, we can get the following third-order asymptotic result.

\begin{corollary}[in the absence of side-information $Y$]
\label{th:asympt_prefix-free}
Let $\rho > 0$ and $0 < \varepsilon < 1$ be fixed.
If $V(X) = 0$, then
\begin{align}
\Lambda^{\ast}(X^{n} \, \| \, \rho, \varepsilon)
=
n \, H( X ) + \mathrm{O}( 1 )
\qquad (\mathrm{as} \ n \to \infty) .
\end{align}
On the other hand, if $0 < V(X) < \infty$, then
\begin{align}
\Lambda^{\ast}(X^{n} \, \| \, \rho, \varepsilon)
=
n \, H( X ) + \mathrm{O}( \sqrt{n} )
\qquad (\mathrm{as} \ n \to \infty) .
\end{align}
In particular, if $V(X) > 0$ and $T(X) < \infty$, then
\begin{align}
\Lambda^{\ast}(X^{n} \, \| \, \rho, \varepsilon)
=
n \, H( X ) - \sqrt{ n \, V( X ) } \, \Phi^{-1}( \varepsilon ) - \frac{ 1 + \rho }{ 2 \, \rho } \log n + \mathrm{O}( 1 )
\qquad (\mathrm{as} \ n \to \infty) .
\label{eq:third-asympt_prefix-free}
\end{align}
\end{corollary}

\begin{IEEEproof}[Proof of \corref{th:asympt_prefix-free}]
\corref{th:asympt_prefix-free} follows from Theorems~\ref{th:CLT} and~\ref{lem:one-shot_prefix-free_avg}.
\end{IEEEproof}

Finally, consider compressing the source $X$ into a \emph{fixed-length} binary string in the absence of the side-information $Y$.
Namely, on the stochastic encoders $F : \mathcal{X} \to \{ 0, 1 \}^{\ast}$, we now impose the constraint that each codeword has the same length, i.e., the codeword length $\ell( F(X) )$ is almost surely constant.
This is the well-known fixed-to-fixed length (block) source coding problem.

\begin{definition}[block source coding]
\label{def:block-code}
A $(\rho, L, \varepsilon)_{\mathrm{FF}}$-code for a source $X$ is a stochastic code $(F, G)$ such that
\begin{align}
\Lambda(F(X) \, \| \, \rho)
& \le
\rho L ,
\\
\mathbb{P}\{ X \neq G(F(X)) \}
& \le
\varepsilon ,
\end{align}
and $\ell( F(X) )$ is almost surely constant.
\end{definition}

Given $\rho > 0$ and $0 \le \varepsilon < 1$, consider the following fundamental limit:
\begin{align}
\Lambda^{\ast}(X \, \| \, \rho, \varepsilon)
\coloneqq
\inf\{ L > 0 \mid \text{there exists a $(\rho, L, \varepsilon)_{\mathrm{FF}}$-code for the source $X$} \} .
\end{align}
It is clear that
\begin{align}
\Lambda(F(X) \, \| \, \rho)
=
\ell( F(X) )
\qquad (\mathrm{a.s.}) ,
\end{align}
provided that $\ell( F(X) )$ is almost surely constant.
Hence, the fundamental limit $\Lambda_{\mathrm{FF}}^{\ast}(X \, \| \, \rho, \varepsilon)$ is independent of the parameter $\rho > 0$, and it can be written as
\begin{align}
\Lambda_{\mathrm{FF}}^{\ast}(X \, \| \, \rho, \varepsilon)
=
\lceil \log (1 + | \mathcal{A}_{X}^{\varepsilon} |) \rceil ,
\end{align}
where the proper subset $\mathcal{A}_{X}^{\varepsilon} \subsetneq \mathcal{X}$ is defined in \eqref{eq:largest-A_cond1} and \eqref{eq:largest-A_cond2}.
Therefore, it follows by Strassen's seminal result \cite{strassen_1964} that
\begin{align}
\Lambda_{\mathrm{FF}}^{\ast}(X^{n} \, \| \, \rho, \varepsilon)
=
n \, H(X) - \sqrt{ n \, V(X) } \, \Phi^{-1}( \varepsilon ) - \frac{ 1 }{ 2 } \log n + \mathrm{O}( 1 ) ,
\label{eq:Strassen}
\end{align}
provided that $V(X) > 0$ and $T(X) < \infty$.

We shall compare the two fundamental limits $\Lambda^{\ast}(X \, \| \, \rho, \varepsilon)$ and $\Lambda_{\mathrm{FF}}^{\ast}(X \, \| \, \rho, \varepsilon)$.
Since every fixed-to-fixed length source code is prefix-free, it is clear that a $(\rho, L, \varepsilon)_{\mathrm{FF}}$-code for a source $X$ is a $(\rho, L, \varepsilon)$-code for the source $X$.
Namely, we readily see that
\begin{align}
\Lambda^{\ast}(X \, \| \, \rho, \varepsilon)
\le
\Lambda_{\mathrm{FF}}^{\ast}(X \, \| \, \rho, \varepsilon) .
\label{eq:3rd-order_FF}
\end{align}
Indeed, it follows from \corref{th:asympt_prefix-free} and \eqref{eq:3rd-order_FF} that
\begin{align}
\Lambda_{\mathrm{FF}}^{\ast}(X^{n} \, \| \, \rho, \varepsilon) - \Lambda^{\ast}(X^{n} \, \| \, \rho, \varepsilon)
=
\frac{ 1 }{ 2 \, \rho } \log n + \mathrm{O}( 1 )
\qquad (\mathrm{as} \ n \to \infty) ,
\label{eq:comparison_FF}
\end{align}
provided that $V(X) > 0$ and $T(X) < \infty$.
In other words, the first- and second-order terms (i.e., the $n$ and $\sqrt{n}$ scales, respectively) of $\Lambda^{\ast}(X^{n} \, \| \, \rho, \varepsilon)$ and $\Lambda_{\mathrm{FF}}^{\ast}(X^{n} \, \| \, \rho, \varepsilon)$ are the same, and the third-order term (i.e., the $\log n$ scale) of $\Lambda^{\ast}(X^{n} \, \| \, \rho, \varepsilon)$ is strictly smaller than that of $\Lambda_{\mathrm{FF}}^{\ast}(X^{n} \, \| \, \rho, \varepsilon)$.
Roughly speaking, the benefit of variable-length codewords appears only in the third-order term in Campbell's fixed-to-variable length source coding problem \cite{campbell_1965}.

\subsection{Guessing Problem}

We introduce Kuzuoka's guessing problem \cite[Section~III]{kuzuoka_2019}.
A \emph{guessing strategy with a giving-up policy} is a pair $(\mathsf{g}, \pi)$ of deterministic maps $\mathsf{g} : \mathcal{X} \times \mathcal{Y} \to \mathbb{N}$ and $\pi : \mathbb{N} \times \mathcal{Y} \to [0, 1]$ in which $\mathsf{g}(\cdot, y) : \mathcal{X} \to \mathbb{N}$ is bijective for each $y \in \mathcal{Y}$.
This pair $(\mathsf{g}, \pi)$ induces the following strategy:
Assume that the guesser knows the side-information $Y = y$.
For each guessing epoch, the guesser can stochastically give-up guessing based on the giving-up policy $\pi$.
Formally, at the $k$-th stage ($k \ge 1$), he declares an error with probability $\pi(k, y)$, or he asks the question ``Is $X = x_{k}$?'' with probability $1 - \pi(k, y)$, where the candidate $x_{k}$ is chosen by the guessing function $\mathsf{g}$ as $\mathsf{g}(x_{k}, y) = k$.
The guesser repeats these epochs until he succeeds in guessing $X$ or he declares an error.
Construct a stochastic map $\bar{\mathsf{G}} : \mathcal{X} \times \mathcal{Y} \to \mathbb{N} \cup \{ 0 \}$ so that%
\footnote{Originally, Kuzuoka \cite[Section~III]{kuzuoka_2019} introduced a positive error cost when the guesser declares an error.
This setting was argued to be more practical; see also \cite[Section~IV]{sakai_tan_2019_VL}.
On the other hand, he derived his one-shot bounds and general formula in the absence of the error cost.
Our study also focuses on the guessing problem without the error cost.}
\begin{align}
\bar{\mathsf{G}}(x, y)
=
\begin{dcases}
\mathsf{g}(x, y)
& \mathrm{with} \ \mathrm{probability} \ \prod_{k = 1}^{\mathsf{g}(x, y)} \Big( 1 - \pi(k, y) \Big)
\\
0
& \mathrm{with} \ \mathrm{probability} \ 1 - \prod_{k = 1}^{\mathsf{g}(x, y)} \Big( 1 - \pi(k, y) \Big)
\end{dcases}
\label{def:giving-up_guessing_function}
\end{align}
for each $(x, y) \in \mathcal{X} \times \mathcal{Y}$.
We call this stochastic map $\bar{\mathsf{G}} : \mathcal{X} \times \mathcal{Y} \to \mathbb{N} \cup \{ 0 \}$ the \emph{giving-up guessing function induced by the guessing strategy $(\mathsf{g}, \pi)$.}
Then, it is clear that the guesser declares some error if and only if $\bar{\mathsf{G}}(X, Y) = 0$.
We now aim to minimize the guessing $\rho$-th moment $\mathbb{E}[ \bar{\mathsf{G}}(X, Y)^{\rho} ]$ for a fixed $\rho > 0$ subject to certain error constraints.
In other words, we are interested to characterize fundamental limits defined as the infimum of positive real numbers $M$ such that there exists a guessing strategy $(\mathsf{g}, \pi)$ satisfying
\begin{align}
\log \mathbb{E}[ \bar{\mathsf{G}}(X, Y)^{\rho} ]
\le
\rho M
\label{eq:CGF_Kuzuoka}
\end{align}
under a certain constraint on error probabilities.
We now introduce two error formalisms as follows:

\begin{definition}[average error criterion]
A $(\rho, M, \varepsilon)_{\mathrm{avg}}$-strategy for a correlated source $(X, Y)$ is a guessing strategy $(\mathsf{g}, \pi)$ such that \eqref{eq:CGF_Kuzuoka} holds and
\begin{align}
\mathbb{P}\{ \bar{\mathsf{G}}(X, Y) = 0 \}
& \le
\varepsilon .
\end{align}
\end{definition}

\begin{definition}[maximum error criterion]
A $(\rho, M, \varepsilon)_{\max}$-strategy for a correlated source $(X, Y)$ is a guessing strategy $(\mathsf{g}, \pi)$ such that \eqref{eq:CGF_Kuzuoka} holds and
\begin{align}
\sup_{y \in \mathcal{Y}} \mathbb{P}\{ \bar{\mathsf{G}}(X, Y) = 0 \mid Y = y \}
& \le
\varepsilon .
\end{align}
\end{definition}

Given two real numbers $\rho > 0$ and $0 \le \varepsilon < 1$, define the following two fundamental limits
\begin{align}
\mathbf{G}_{\mathrm{avg}}^{\ast}(X, Y \, \| \, \rho, \varepsilon)
& \coloneqq
\inf\{ M > 0 \mid \text{there exists a $(\rho, M, \varepsilon)_{\mathrm{avg}}$-strategy for $(X, Y)$} \} ,
\\
\mathbf{G}_{\max}^{\ast}(X, Y \, \| \, \rho, \varepsilon)
& \coloneqq
\inf\{ M > 0 \mid \text{there exists a $(\rho, M, \varepsilon)_{\max}$-strategy for $(X, Y)$} \} .
\end{align}
Now, we state the following one-shot bounds.

\begin{theorem}[average error criterion]
\label{th:one-shot_guess_avg}
For every $\rho > 0$ and $0 < \varepsilon < 1$, it holds that
\begin{align}
H_{\alpha}^{\varepsilon}(X \mid Y) - \log \left( 1 + \frac{ H(X \mid Y) }{ \varepsilon } \right)
\le
\mathbf{G}_{\mathrm{avg}}^{\ast}(X, Y \, \| \, \rho, \varepsilon )
\le
H_{\alpha}^{\varepsilon}(X \mid Y) .
\label{eq:one-shot_guess_avg}
\end{align}
\end{theorem}

\begin{IEEEproof}[Proof of \thref{th:one-shot_guess_avg}]
For each $y \in \mathcal{Y}$, denote by $\varsigma_{y} : \mathbb{N} \to \mathcal{X}$ a bijection satisfying
\begin{align}
P_{X|Y}(\varsigma_{y}(1) \mid y) \ge P_{X|Y}(\varsigma_{y}(2) \mid y) \ge P_{X|Y}(\varsigma_{y}(3) \mid y) \ge P_{X|Y}(\varsigma_{y}(4) \mid y) \ge P_{X|Y}(\varsigma_{y}(5) \mid y) \ge \cdots .
\label{def:varsigma}
\end{align}
Define two parameters $J$ and $\xi$ by
\begin{align}
J
& \coloneqq
\sup\left\{ j \ge 0 \ \middle| \ \sum_{y \in \mathcal{Y}} P_{Y}( y ) \sum_{k = 1}^{j} P_{X|Y}(\varsigma_{y}(k) \mid y) < 1 - \varepsilon \right\} ,
\label{def:J} \\
\xi
& \coloneqq
1 - \varepsilon - \sum_{y \in \mathcal{Y}} P_{Y}( y ) \sum_{k = 1}^{J} P_{X|Y}(\varsigma_{y}(k) \mid y) ,
\label{def:xi}
\end{align}
respectively.
The following lemma characterizes an optimal guessing strategy under the average error criterion.

\begin{lemma}[optimal guessing strategy---average error criterion]
\label{lem:optimal-strategy_avg}
Consider a guessing strategy $(\mathsf{g}^{\ast}, \pi_{\mathrm{avg}}^{\ast})$ given by
\begin{align}
\mathsf{g}^{\ast}(x, y)
& =
\varsigma_{y}^{-1}( x ) ,
\label{eq:optimal-guessing} \\
\pi_{\mathrm{avg}}^{\ast}(k, y)
& =
\begin{dcases}
0
& \mathrm{if} \ 1 \le k \le J ,
\\
1 - \frac{ \xi }{ P_{X, Y}(\varsigma_{y}(k), y) }
& \mathrm{if} \ k = J + 1 ,
\\
1
& \mathrm{if} \ J + 2 \le k < \infty .
\end{dcases}
\label{eq:pi-ast_avg}
\end{align}
Denote by $\bar{\mathsf{G}}_{\mathrm{avg}}^{\ast} : \mathcal{X} \times \mathcal{Y} \to \mathbb{N} \cup \{ 0 \}$ the giving-up guessing function induced by $(\mathsf{g}^{\ast}, \pi_{\mathrm{avg}}^{\ast})$.
For any $\rho > 0$, it holds that
\begin{align}
\frac{ 1 }{ \rho } \log \mathbb{E}[ \bar{\mathsf{G}}_{\mathrm{avg}}^{\ast}(X, Y)^{\rho} ]
& =
\mathbf{G}_{\mathrm{avg}}^{\ast}(X, Y \, \| \, \rho, \varepsilon) ,
\label{eq:optimal-guessing_moment_avg} \\
\mathbb{P}\{ \bar{\mathsf{G}}_{\mathrm{avg}}^{\ast}(X, Y) = 0 \}
& =
\varepsilon .
\label{eq:optimal-guessing_error-probab_avg}
\end{align}
\end{lemma}

\begin{IEEEproof}[Proof of \lemref{lem:optimal-strategy_avg}]
See \appref{app:optimal-strategy_avg}.
\end{IEEEproof}

The left-hand inequality of \eqref{eq:one-shot_guess_avg} can be proven by combining \lemref{lem:optimal-strategy_avg} and the unified converse bound stated in \lemref{lem:unified_converse_avg}.
On the other hand, the right-hand inequality of \eqref{eq:one-shot_guess_avg} can be proven by the unified achievability bound stated in \lemref{lem:unified_direct}.
See \appref{app:one-shot_guess_avg} for proofs of these lower and upper bounds.
\end{IEEEproof}

\begin{theorem}[maximum error criterion]
\label{th:one-shot_guess_max}
For every $\rho > 0$ and $0 < \varepsilon < 1$, it holds that
\begin{align}
\check{H}_{\alpha}^{\varepsilon}(X \mid Y) - \log \left( 1 + \frac{ \sup_{y \in \mathcal{Y}} H( P_{X|Y=y} ) }{ \varepsilon } \right)
\le
\mathbf{G}_{\max}^{\ast}(X, Y \, \| \, \rho, \varepsilon )
\le
\check{H}_{\alpha}^{\varepsilon}(X \mid Y) .
\label{eq:one-shot_guess_max}
\end{align}
\end{theorem}

\begin{IEEEproof}[Proof of \thref{th:one-shot_guess_max}]
For each $y \in \mathcal{Y}$, define two parameters $J( y )$ and $\xi( y )$ by
\begin{align}
J( y )
& \coloneqq
\sup\left\{ j \ge 0 \ \middle| \ \sum_{k = 1}^{j} P_{X|Y}(\varsigma_{y}(k) \mid y) < 1 - \varepsilon \right\} ,
\label{def:J-y} \\
\xi( y )
& \coloneqq
1 - \varepsilon - \sum_{k = 1}^{J(y)} P_{X|Y}(\varsigma_{y}(k) \mid y) ,
\label{def:xi-y}
\end{align}
respectively.
In contrast to \lemref{lem:optimal-strategy_avg}, the following lemma characterizes an optimal guessing strategy under the maximum error criterion.

\begin{lemma}[optimal guessing strategy---maximum error criterion]
\label{lem:optimal-strategy_max}
Consider a guessing strategy $(\mathsf{g}^{\ast}, \pi_{\max}^{\ast})$ given by \eqref{eq:optimal-guessing} and
\begin{align}
\pi_{\mathrm{avg}}^{\ast}(k, y)
& =
\begin{dcases}
0
& \mathrm{if} \ 1 \le k \le J( y ) ,
\\
1 - \frac{ \xi( y ) }{ P_{X, Y}(\varsigma_{y}(k), y) }
& \mathrm{if} \ k = J( y ) + 1 ,
\\
1
& \mathrm{if} \ J( y ) + 2 \le k < \infty .
\end{dcases}
\label{eq:pi-ast_max}
\end{align}
Denote by $\bar{\mathsf{G}}_{\max}^{\ast} : \mathcal{X} \times \mathcal{Y} \to \mathbb{N} \cup \{ 0 \}$ the giving-up guessing function induced by $(\mathsf{g}^{\ast}, \pi_{\max}^{\ast})$.
For any $\rho > 0$, it holds that
\begin{align}
\frac{ 1 }{ \rho } \log \mathbb{E}[ \bar{\mathsf{G}}_{\max}^{\ast}(X, Y)^{\rho} ]
& =
\mathbf{G}_{\max}^{\ast}(X, Y \, \| \, \rho, \varepsilon) ,
\label{eq:optimal-guessing_moment_max} \\
\mathbb{P}\{ \bar{\mathsf{G}}_{\max}^{\ast}(X, Y) = 0 \mid Y = y \}
& =
\varepsilon
\qquad (\mathrm{for} \ \mathrm{all} \ y \in \mathcal{Y}) .
\label{eq:optimal-guessing_error-probab_max}
\end{align}
\end{lemma}

\begin{IEEEproof}[Proof of \lemref{lem:optimal-strategy_max}]
See \appref{app:optimal-strategy_max}.
\end{IEEEproof}

Now, the left-hand inequality of \eqref{eq:one-shot_guess_max} can be proven by combining \lemref{lem:optimal-strategy_max} and the unified converse bound stated in \lemref{lem:unified_converse_max}.
On the other hand, the right-hand inequality of \eqref{eq:one-shot_guess_max} can be proven by the unified achievability bound stated in \lemref{lem:unified_direct}.
See \appref{app:one-shot_guess_max} for proofs of these lower and upper bounds.
\end{IEEEproof}

\begin{remark}
Kuzuoka provided one-shot bounds \cite[Theorems~3 and~4]{kuzuoka_2019} on the same guessing problem when $X$ takes values in a finite alphabet.
His converse bound \cite[Theorem~3]{kuzuoka_2019} is proven by taking the sum in the right-hand side of \eqref{def:redundancy-R} over all $x \in \mathcal{X}$ even if $\epsilon(x, y) < 1$, and this works only if $\mathcal{X}$ is finite due to the divergence of the harmonic series (see also \cite[Section~II-A]{arikan_1996}).
In other words, his one-shot bounds \cite[Theorems~3 and~4]{kuzuoka_2019} can be written as
\begin{align}
H_{\alpha}^{\varepsilon}(X \mid Y) - \log( 1 + |\mathcal{A}| )
\le
\mathbf{G}_{\mathrm{avg}}^{\ast}(X, Y \, \| \, \rho, \varepsilon )
\le
H_{\alpha}^{\varepsilon}(X \mid Y) ,
\end{align}
provided that $X$ is supported on a finite subalphabet $\mathcal{A} \subset \mathcal{X}$.
On the other hand, our one-shot bounds stated in Theorems~\ref{th:one-shot_guess_avg} and~\ref{th:one-shot_guess_max} are also applicable to sources $X$ with countably infinite alphabets $\mathcal{X}$.
This holds because our converse bounds are proven by considering the optimal guessing strategies stated in Lemmas~\ref{lem:optimal-strategy_avg} and~\ref{lem:optimal-strategy_max} and by restricting the sum in the right-hand side of \eqref{def:redundancy-R} over all $x \in \mathcal{X}$ satisfying $\epsilon(x, y) < 1$.
\end{remark}

Using these one-shot bounds, we obtain the following asymptotic expansions of the fundamental limits.

\begin{corollary}[average error criterion]
\label{cor:guess_avg}
Let $\rho > 0$ and $0 < \varepsilon < 1$ be fixed.
If $U(X \mid Y) = 0$, then
\begin{align}
\mathbf{G}_{\mathrm{avg}}^{\ast}(X^{n}, Y^{n} \, \| \, \rho, \varepsilon)
& =
n \, H(X \mid Y) + \mathrm{O}( 1 )
\qquad (\mathrm{as} \ n \to \infty) .
\end{align}
On the other hand, if $U(X \mid Y) > 0$, then
\begin{align}
\mathbf{G}_{\mathrm{avg}}^{\ast}(X^{n}, Y^{n} \, \| \, \rho, \varepsilon)
& =
n \, H(X \mid Y) + \mathrm{O}( \sqrt{n} )
\qquad (\mathrm{as} \ n \to \infty) .
\end{align}
\end{corollary}

\begin{IEEEproof}[Proof of \corref{cor:guess_avg}]
\corref{cor:guess_avg} follows from Theorems~\ref{th:2nd_avg} and~\ref{th:one-shot_guess_avg}.
\end{IEEEproof}

\begin{corollary}[maximum error criterion]
\label{cor:guess_max}
Let $\rho > 0$ and $0 < \varepsilon < 1$ be fixed.
If $V(X \mid Y) = 0$, then
\begin{align}
\mathbf{G}_{\max}^{\ast}(X^{n}, Y^{n} \, \| \, \rho, \varepsilon)
& =
n \, H(X \mid Y) + \mathrm{O}( 1 )
\qquad (\mathrm{as} \ n \to \infty) .
\end{align}
On the other hand, if $V(X \mid Y) > 0$ and $\sup_{y \in \mathcal{Y}} V( P_{X|Y=y} ) < \infty$, then
\begin{align}
\mathbf{G}_{\max}^{\ast}(X^{n}, Y^{n} \, \| \, \rho, \varepsilon)
& =
n \, H^{(\alpha)}(X \mid Y) + \mathrm{O}( \sqrt{n} )
\qquad (\mathrm{as} \ n \to \infty) .
\end{align}
\end{corollary}

\begin{IEEEproof}[Proof of \corref{cor:guess_max}]
\corref{cor:guess_max} follows from Theorems~\ref{th:2nd_max} and~\ref{th:one-shot_guess_max}.
\end{IEEEproof}

\begin{remark}
Suppose here that $X$ is supported on some finite sub-alphabet $\mathcal{A} \subsetneq \mathcal{X}$.
Since there is no difference between the average and maximum error criteria in the error-free setting (i.e., $\varepsilon = 0$), we now define
\begin{align}
\mathbf{G}_{\text{$\mathrm{error}$-$\mathrm{free}$}}^{\ast}(X, Y \, \| \, \rho)
\coloneqq
\mathbf{G}_{\mathrm{avg}}^{\ast}(X, Y \, \| \, \rho, 0)
=
\mathbf{G}_{\max}^{\ast}(X, Y \, \| \, \rho, 0) .
\end{align}
Then, it follows by Ar{\i}kan's seminal result \cite{arikan_1996} that
\begin{align}
\mathbf{G}_{\text{$\mathrm{error}$-$\mathrm{free}$}}^{\ast}(X^{n}, Y^{n} \, \| \, \rho)
=
n \, H_{\alpha}(X \mid Y) + \mathrm{O}( 1 )
\qquad (\mathrm{as} \ n \to \infty) .
\end{align}
Therefore, the differences among the first-order terms of the asymptotic expansions of the three fundamental limits $\mathbf{G}_{\mathrm{avg}}^{\ast}(X, Y \, \| \, \rho, \varepsilon)$, $\mathbf{G}_{\max}^{\ast}(X, Y \, \| \, \rho, \varepsilon)$, and $\mathbf{G}_{\text{$\mathrm{error}$-$\mathrm{free}$}}^{\ast}(X^{n}, Y^{n} \, \| \, \rho)$ are characterized by the relations amongst the different conditional entropies as delineated by \propref{prop:KN-avg}.
\end{remark}

Now, consider guessing problems for the source $X$ \emph{in the absence of the side-information $Y$.}
Formally, when $Y$ is almost surely constant, we denote by
\begin{align}
\mathbf{G}^{\ast}(X \, \| \, \rho, \varepsilon)
\coloneqq
\mathbf{G}_{\mathrm{avg}}^{\ast}(X, Y \, \| \, \rho, \varepsilon)
=
\mathbf{G}_{\max}^{\ast}(X, Y \, \| \, \rho, \varepsilon) .
\end{align}
our considered fundamental limit here.
Then, we can get the following second-order asymptotic result.

\begin{corollary}[in the absence of side-information $Y$]
\label{cor:guess_no-side}
Let $\rho > 0$ and $0 < \varepsilon < 1$ be fixed.
If $V(X) = 0$, then
\begin{align}
\mathbf{G}^{\ast}(X^{n} \, \| \, \rho, \varepsilon)
=
n \, H( X ) + \mathrm{O}( 1 )
\qquad (\mathrm{as} \ n \to \infty) .
\end{align}
On the other hand, if $0 < V(X) < \infty$, then
\begin{align}
\mathbf{G}^{\ast}(X^{n} \, \| \, \rho, \varepsilon)
=
n \, H( X ) + \mathrm{O}( \sqrt{n} )
\qquad (\mathrm{as} \ n \to \infty) .
\end{align}
In particular, if $V(X) > 0$ and $T(X) < \infty$, then
\begin{align}
\mathbf{G}^{\ast}(X^{n} \, \| \, \rho, \varepsilon)
=
n \, H( X ) - \sqrt{ n \, V( X ) } \, \Phi^{-1}( \varepsilon ) + \mathrm{O}( \log n )
\qquad (\mathrm{as} \ n \to \infty) .
\end{align}
\end{corollary}

\begin{IEEEproof}[Proof of \corref{cor:guess_no-side}]
\corref{cor:guess_no-side} follows from Theorems~\ref{th:CLT} and~\ref{th:one-shot_guess_avg}.
\end{IEEEproof}

\subsection{Encoding Tasks}

Suppose that $\mathcal{X}$ is a countably infinite set of tasks we wish to execute.
Given a correlated source $(X, Y)$ and a positive integer $M$, let us consider assigning a randomly occurred task $X$ into $M$ messages with the help of some side-information $Y$ of $X$.
Bunte and Lapidoth \cite{bunte_lapidoth_2014} proposed this problem and assumed that none of the tasks are ignored.
In this case, we can think of such an assignment as a finite partition of $\mathcal{X}$ in which the cardinality of the partition does not exceed the desired threshold $M$.

In this study, under certain error constraints, we allow the possibility of ignoring some tasks.
Given a deterministic map $\mathsf{f} : \mathcal{X} \times \mathcal{Y} \to \{ 0, 1, 2, \dots M \}$ called an \emph{assignment function,} consider the following assignment rule:
If $\mathsf{f}(X, Y) = m$ for some $1 \le m \le M$, then a task $X$ is assigned to a message $m$.
On the other hand, a task $X$ is ignored if and only if $\mathsf{f}(X, Y) = 0$.
Define the deterministic map $\mathcal{L} : \{ 0, 1, 2, \dots, M \} \times \mathcal{Y} \to 2^{\mathcal{X}}$ by
\begin{align}
\mathcal{L}(m, y)
\coloneqq
\begin{cases}
\emptyset
& \mathrm{if} \ m = 0 ,
\\
\{ x \in \mathcal{X} \mid \mathsf{f}(x, y) = m \}
& \mathrm{if} \ m = 1, \dots, M
\end{cases}
\label{def:sub-partition}
\end{align}
for each $1 \le m \le M$ and $y \in \mathcal{Y}$.
Then, the family $\{ \mathcal{L}(m, y) \}_{m = 0}^{M}$ forms a sub-partition%
\footnote{A sub-partition of a set $\mathcal{S}$ is a subset of a partition of the set $\mathcal{S}$.}
of $\mathcal{X}$ for each $y \in \mathcal{Y}$.
When a task $X$ occurs, all tasks in $\mathcal{L}(\mathsf{f}(X, Y), Y)$ are executed.
Namely, if $\mathsf{f}(X, Y) = 0$, then no task in $\mathcal{X}$ is executed.
Furthermore, we allow the probability of not executing any task even if $\mathsf{f}(X, Y) \neq 0$, but this occurs with a certain fixed probability.
More precisely, given a stochastic map $\mathsf{E} : 2^{\mathcal{X}} \times \mathcal{Y} \to 2^{\mathcal{X}}$ satisfying
\begin{align}
\mathbb{P}\{ \mathsf{E}( \mathcal{A}, y ) = \emptyset \} + \mathbb{P}\{ \mathsf{E}( \mathcal{A}, y ) = \mathcal{A} \}
=
1
\label{eq:error-stochas_partition}
\end{align}
for each $\mathcal{A} \subset \mathcal{X}$ and $y \in \mathcal{Y}$, define the stochastic map $\mathsf{L} : \{ 0, 1, 2, \dots, M \} \times \mathcal{Y} \to 2^{\mathcal{X}}$ by
\begin{align}
\mathsf{L}(\mathsf{f}(x, y), y)
\coloneqq
\mathsf{E}( \mathcal{L}(\mathsf{f}(x, y), y), y )
\label{def:stochastic-sub-partition}
\end{align}
for each $(x, y) \in \mathcal{X} \times \mathcal{Y}$.
Then, an error occurs if and only if $X \notin \mathsf{L}(\mathsf{f}(X, Y), Y)$.
For this stochastic sub-partition $\mathsf{L}$ induced by the pair $(\mathsf{f}, \mathsf{E})$, we aim to minimize the task sub-partitioning $\rho$-th moment $\mathbb{E}[ |\mathsf{L}(\mathsf{f}(X, Y), Y)|^{\rho} ]$ for a fixed $\rho > 0$ subject to certain error constraints.
In other words, we are interested to characterize fundamental limits defined as the infimum of positive real numbers $N$ such that there exist an assignment function $\mathsf{f} : \mathcal{X} \times \mathcal{Y} \to \{ 0, 1, 2, \dots, M \}$ and a stochastic map $\mathsf{E} : 2^{\mathcal{X}} \times \mathcal{Y} \to 2^{\mathcal{X}}$ satisfying
\begin{align}
\log \mathbb{E}[ |\mathsf{L}(\mathsf{f}(X, Y), Y)|^{\rho} ]
& \le
\rho N
\label{eq:task-moment}
\end{align}
under a certain constraint on error probabilities.
We now introduce two error formalisms as follows:

\begin{definition}[average error criterion]
A $(\rho, M, N, \varepsilon)_{\mathrm{avg}}$-assignment for a correlated source $(X, Y)$ is a pair $(\mathsf{f}, \mathsf{E})$ consisting of an assignment function $\mathsf{f} : \mathcal{X} \times \mathcal{Y} \to \{ 0, 1, 2, \dots, M \}$ and a stochastic map $\mathsf{E} : 2^{\mathcal{X}} \times \mathcal{Y} \to 2^{\mathcal{X}}$ satisfying \eqref{eq:error-stochas_partition} and \eqref{eq:task-moment} hold, and
\begin{align}
\mathbb{P}\{ X \notin \mathsf{L}(\mathsf{f}(X, Y), Y) \}
& \le
\varepsilon .
\end{align}
\end{definition}

\begin{definition}[maximum error criterion]
A $(\rho, M, N, \varepsilon)_{\max}$-assignment for a correlated source $(X, Y)$ is a pair $(\mathsf{f}, \mathsf{E})$ consisting of an assignment function $\mathsf{f} : \mathcal{X} \times \mathcal{Y} \to \{ 0, 1, 2, \dots, M \}$ and a stochastic map $\mathsf{E} : 2^{\mathcal{X}} \times \mathcal{Y} \to 2^{\mathcal{X}}$ satisfying \eqref{eq:error-stochas_partition} and \eqref{eq:task-moment} hold, and
\begin{align}
\sup_{y \in \mathcal{Y}} \mathbb{P}\{ X \notin \mathsf{L}(\mathsf{f}(X, Y), Y) \mid Y = y \}
& \le
\varepsilon .
\end{align}
\end{definition}

Given a positive integer $M$ and two real numbers $\rho > 0$ and $0 < \varepsilon < 1$,
consider the following two fundamental limits:
\begin{align}
\mathbf{L}_{\mathrm{avg}}^{\ast}(X, Y, M \, \| \, \rho, \varepsilon)
& \coloneqq
\inf\{ N > 0 \mid \text{there exists a $(\rho, M, N, \varepsilon)_{\mathrm{avg}}$-assignment for $(X, Y)$} \} ,
\\
\mathbf{L}_{\max}^{\ast}(X, Y, M \, \| \, \rho, \varepsilon)
& \coloneqq
\inf\{ N > 0 \mid \text{there exists a $(\rho, M, N, \varepsilon)_{\max}$-assignment for $(X, Y)$} \} .
\end{align}

We now state the following one-shot bounds.

\begin{theorem}[average error criterion]
\label{th:one-shot_task_avg}
Let $\rho > 0$, $0 < \varepsilon < 1$, and $M \ge 1$ be fixed.
It holds that
\begin{align}
\mathbf{L}_{\mathrm{avg}}^{\ast}(X, Y, M \, \| \, \rho, \varepsilon)
\ge
H_{\alpha}^{\varepsilon}(X \mid Y) - \log M .
\label{eq:one-shot_task_avg_converse}
\end{align}
Moreover, if $M > 2 + H(X \mid Y) / \varepsilon$, then
\begin{align}
\mathbf{L}_{\mathrm{avg}}^{\ast}(X, Y, M \, \| \, \rho, \varepsilon)
\le
\left| H_{\alpha}^{\varepsilon}(X \mid Y) - \log \left( \frac{ \varepsilon (M - 2) - H(X \mid Y) }{ 4 \, \varepsilon + 4 \, H(X \mid Y) } \right) \right|_{+} + \frac{ 1 }{ \rho } \log 2 ,
\label{eq:one-shot_task_avg_direct}
\end{align}
where $| u |_{+} \coloneqq \max\{ 0, u \}$ for $u \in \mathbb{R}$.
\end{theorem}

\begin{IEEEproof}[Proof of \thref{th:one-shot_task_avg}]
The converse bound stated in \eqref{eq:one-shot_task_avg_converse} is proven in \appref{app:one-shot_task_avg_converse} via the unified converse bound stated in \lemref{lem:unified_converse_avg}.
In the following, we shall prove the achievability bound stated in \eqref{eq:one-shot_task_avg_direct}.

Recall that the numbers $J$ and $\xi$ are defined in \eqref{def:J} and \eqref{def:xi}, respectively.
In addition, define the number
\begin{align}
\upsilon
& \coloneqq
\frac{ \xi }{ \sum_{b \in \mathcal{Y}} P_{X, Y}(\varsigma_{b}( J + 1 ), b) } ,
\label{def:upsilon}
\end{align}
where the bijection $\varsigma_{y} : \mathbb{N} \to \mathcal{X}$ is defined to satisfy \eqref{def:varsigma} for each $y \in \mathcal{Y}$.
We now introduce yet another definition of the $\varepsilon$-smooth conditional R\'{e}nyi entropy as
\begin{align}
\tilde{H}_{\alpha}^{\varepsilon}(X \mid Y)
\coloneqq
\frac{ \alpha }{ 1 - \alpha } \log \left( \sum_{y \in \mathcal{Y}} \left( \sum_{k = 1}^{J} P_{X, Y}(\varsigma_{y}(k), y)^{\alpha} + \upsilon^{\alpha} \, P_{X, Y}(\varsigma_{y}(J + 1), y)^{\alpha} \right)^{1/\alpha} \right) .
\label{def:yet}
\end{align}
Instead of establishing \eqref{eq:one-shot_task_avg_direct}, we establish the following one-shot achievability bound, which serves as an intermediate result in proving \eqref{eq:one-shot_task_avg_direct}.

\begin{lemma}
\label{lem:task_one-shot_direct_avg_tilde}
Suppose that the integer $M$ is large enough so that
\begin{align}
M
>
2 + \frac{ H(X \mid Y) }{ \varepsilon } .
\label{eq:at-most_M_avg}
\end{align}
Then, it holds that
\begin{align}
\mathbf{L}_{\mathrm{avg}}^{\ast}(X, Y, M \, \| \, \rho, \varepsilon)
\le
\left| \tilde{H}_{\alpha}^{\varepsilon}(X \mid Y) - \log \left( \frac{ \varepsilon (M - 2) - H(X \mid Y) }{ 4 \, \varepsilon } \right) \right|_{+} + \frac{ 1 }{ \rho } \log 2 .
\label{eq:task_one-shot_direct_avg_tilde}
\end{align}
\end{lemma}

\begin{IEEEproof}[Proof of \lemref{lem:task_one-shot_direct_avg_tilde}]
See \appref{app:task_one-shot_direct_avg_tilde}.
\end{IEEEproof}

The following lemma provides inequalities between $H_{\alpha}^{\varepsilon}(X \mid Y)$ and $\tilde{H}_{\alpha}^{\varepsilon}(X \mid Y)$.

\begin{lemma}
\label{lem:ineq-Kuzuoka_tilde}
For any $0 < \alpha < 1$ and $0 \le \varepsilon < 1$, it holds that
\begin{align}
H_{\alpha}^{\varepsilon}(X \mid Y)
\le
\tilde{H}_{\alpha}^{\varepsilon}(X \mid Y)
\le
H_{\alpha}^{\varepsilon}(X \mid Y) + \log \left( 1 + \frac{ H(X \mid Y) }{ \varepsilon } \right) .
\label{eq:ineq-Kuzuoka_tilde}
\end{align}
\end{lemma}

\begin{IEEEproof}[Proof of \lemref{lem:ineq-Kuzuoka_tilde}]
See \appref{app:ineq-Kuzuoka_tilde}.
\end{IEEEproof}

\begin{remark}
By \lemref{lem:ineq-Kuzuoka_tilde}, we observe that the asymptotic expansions of $H_{\alpha}^{\varepsilon}(X \mid Y)$ and $\tilde{H}_{\alpha}^{\varepsilon}(X^{n} \mid Y^{n})$ are the same up to the reminder term $+ \mathrm{O}( \log n )$, provided that $H(X \mid Y) < \infty$.
\end{remark}

Combining Lemmas~\ref{lem:task_one-shot_direct_avg_tilde} and~\ref{lem:ineq-Kuzuoka_tilde}, we have the achievability bound stated in \eqref{eq:one-shot_task_avg_direct}.
This completes the proof of \thref{th:one-shot_task_avg}.
\end{IEEEproof}

\begin{theorem}[maximum error criterion]
\label{th:one-shot_task_max}
Let $\rho > 0$, $0 < \varepsilon < 1$, and $M \ge 1$ be fixed.
It holds that
\begin{align}
\mathbf{L}_{\max}^{\ast}(X, Y, M \, \| \, \rho, \varepsilon)
\ge
\check{H}_{\alpha}^{\varepsilon}(X \mid Y) - \log M .
\label{eq:one-shot_task_max_converse}
\end{align}
Moreover, if $M > 2 + \sup_{y \in \mathcal{Y}} H( P_{X|Y=y} ) / \varepsilon$, then
\begin{align}
\mathbf{L}_{\max}^{\ast}(X, Y, M \, \| \, \rho, \varepsilon)
\le
\left| \check{H}_{\alpha}^{\varepsilon}(X \mid Y) - \log \left( \frac{ \varepsilon (M - 2) - \sup_{y \in \mathcal{Y}} H( P_{X|Y=y} ) }{ 4 \, \varepsilon } \right) \right|_{+} + \frac{ 1 }{ \rho } \log 2 .
\label{eq:one-shot_task_max_direct}
\end{align}
\end{theorem}

\begin{IEEEproof}[Proof of \thref{th:one-shot_task_max}]
The converse and achievability bounds stated in \eqref{eq:one-shot_task_max_converse} and \eqref{eq:one-shot_task_max_direct}, respectively, are proven in Appendices~\ref{app:one-shot_task_max_converse} and~\ref{app:one-shot_task_max_direct}, respectively.
These utilize the unified approaches stated in Lemmas~\ref{lem:unified_converse_max} and~\ref{lem:unified_direct}, respectively.
This completes the proof of \thref{th:one-shot_task_max}.
\end{IEEEproof}

\begin{remark}
Suppose that there exists a $y \in \mathcal{Y}$ such that the support set $\{ x \in \mathcal{X} \mid P_{X|Y}( x \mid y ) > 0 \}$ is infinite.
Since every finite partition of an infinite set contains an infinite subset, it is clear that
\begin{align}
\mathbf{L}_{\mathrm{avg}}^{\ast}(X, Y, M \, \| \, \rho, 0)
=
\mathbf{L}_{\mathrm{avg}}^{\ast}(X, Y, M \, \| \, \rho, 0)
=
\infty
\end{align}
in the zero-error setting (i.e., $\varepsilon = 0$).
However, if $0 < \varepsilon < 1$, then Theorems~\ref{th:one-shot_task_avg} and~\ref{th:one-shot_task_max} state that these fundamental limits can be finite, and the task encoding problem can be considered over a countably infinite alphabet $\mathcal{X}$ when we ignore some tasks.
\end{remark}

Consider a sequence $\{ M_{n} \}_{n = 1}^{\infty}$ of positive integers satisfying
\begin{align}
\lim_{n \to \infty} \frac{ M_{n} }{ n }
=
\infty .
\end{align}
Define
\begin{align}
\tau_{n}
\coloneqq
\log M_{n}
\end{align}
for each $n \ge 1$.
Using the above one-shot bounds, we obtain the following asymptotic results.

\begin{corollary}[average error criterion]
\label{cor:task_avg}
Let $\rho > 0$ and $0 < \varepsilon < 1$ be fixed.
If $0 < U(X \mid Y) < \infty$, then
\begin{align}
\mathbf{L}_{\mathrm{avg}}^{\ast}(X^{n}, Y^{n}, M_{n} \, \| \, \rho, \varepsilon)
=
\big| n \, H(X \mid Y) - \tau_{n} \big|_{+} + \mathrm{O}( \sqrt{n} )
\qquad (\mathrm{as} \ n \to \infty) .
\end{align}
\end{corollary}

\begin{IEEEproof}[Proof of \corref{cor:task_avg}]
\corref{cor:task_avg} follows from Theorems~\ref{th:2nd_avg} and~\ref{th:one-shot_task_avg}.
\end{IEEEproof}

\begin{corollary}[maximum error criterion]
\label{cor:task_max}
Let $\rho > 0$ and $0 < \varepsilon < 1$ be fixed.
If $V(X \mid Y) > 0$ and $\sup_{y \in \mathcal{Y}} V( P_{X|Y=y} ) < \infty$, then
\begin{align}
\mathbf{L}_{\max}^{\ast}(X^{n}, Y^{n}, M_{n} \, \| \, \rho, \varepsilon)
=
\big| n \, H^{(\alpha)}(X \mid Y) - \tau_{n} \big|_{+} + \mathrm{O}( \sqrt{n} )
\qquad (\mathrm{as} \ n \to \infty) .
\end{align}
\end{corollary}

\begin{IEEEproof}[Proof of \corref{cor:task_max}]
\corref{cor:task_max} follows from Theorems~\ref{th:2nd_max} and~\ref{th:one-shot_task_max}.
\end{IEEEproof}

\begin{remark}
Suppose here that $X$ is supported on some finite sub-alphabet $\mathcal{A} \subsetneq \mathcal{X}$.
Since there is no difference between the average and maximum error criteria in the error-free setting (i.e., $\varepsilon = 0$), we now define
\begin{align}
\mathbf{L}_{\text{$\mathrm{error}$-$\mathrm{free}$}}^{\ast}(X, Y \, \| \, \rho)
\coloneqq
\mathbf{L}_{\mathrm{avg}}^{\ast}(X, Y \, \| \, \rho, 0)
=
\mathbf{L}_{\max}^{\ast}(X, Y \, \| \, \rho, 0) .
\end{align}
To this fundamental limit, Bunte and Lapidoth \cite{bunte_lapidoth_2014} proved that
\begin{align}
\mathbf{L}_{\text{$\mathrm{error}$-$\mathrm{free}$}}^{\ast}(X^{n}, Y^{n} \, \| \, \rho)
=
\Big| n \, H_{\alpha}(X \mid Y) - \tau_{n} \Big|_{+} + \mathrm{O}( 1 )
\qquad (\mathrm{as} \ n \to \infty) .
\end{align}
Therefore, the differences among the first-order terms of asymptotic expansions of the three fundamental limits $\mathbf{L}_{\mathrm{avg}}^{\ast}(X, Y \, \| \, \rho, \varepsilon)$, $\mathbf{L}_{\max}^{\ast}(X, Y \, \| \, \rho, \varepsilon)$, and $\mathbf{L}_{\text{$\mathrm{error}$-$\mathrm{free}$}}^{\ast}(X^{n}, Y^{n} \, \| \, \rho)$ can be characterized by \propref{prop:KN-avg}.
\end{remark}

Now, consider task encoding problems for the source $X$ \emph{in the absence of the side-information $Y$.}
Formally, when $Y$ is almost surely constant, we denote by
\begin{align}
\mathsf{L}^{\ast}(X, M \, \| \, \rho, \varepsilon)
\coloneqq
\mathsf{L}_{\mathrm{avg}}^{\ast}(X, Y, M \, \| \, \rho, \varepsilon)
=
\mathsf{L}_{\max}^{\ast}(X, Y, M \, \| \, \rho, \varepsilon)
\end{align}
our considered fundamental limit here.
Then, we obtain the following third-order asymptotic result.

\begin{corollary}[in the absence of side-information $Y$]
\label{cor:task_no-side}
Let $\rho > 0$ and $0 < \varepsilon < 1$ be fixed.
If $V(X) = 0$, then
\begin{align}
\mathbf{L}^{\ast}(X^{n}, M_{n} \, \| \, \rho, \varepsilon)
=
\Big| n \, H( X ) - \tau_{n} \Big|_{+} + \mathrm{O}( 1 )
\qquad (\mathrm{as} \ n \to \infty) .
\end{align}
On the other hand, if $0 < V(X) < \infty$, then
\begin{align}
\mathbf{L}^{\ast}(X^{n}, M_{n} \, \| \, \rho, \varepsilon)
=
\Big| n \, H( X ) - \tau_{n} \Big|_{+} + \mathrm{O}( \sqrt{n} )
\qquad (\mathrm{as} \ n \to \infty) .
\end{align}
In particular, if $V(X) > 0$ and $T(X) < \infty$, then
\begin{align}
\mathbf{L}^{\ast}(X^{n}, M_{n} \, \| \, \rho, \varepsilon)
=
\bigg| n \, H( X ) - \tau_{n} - \sqrt{ n \, V( X ) } \, \Phi^{-1}( \varepsilon ) - \frac{ 1 + \rho }{ 2 \, \rho } \log n \bigg|_{+} + \mathrm{O}( 1 )
\qquad (\mathrm{as} \ n \to \infty) .
\label{eq:task_no-side}
\end{align}
\end{corollary}

\begin{IEEEproof}[Proof of \corref{cor:task_no-side}]
\corref{cor:task_no-side} follows from Theorems~\ref{th:CLT} and~\ref{th:one-shot_task_max}.
\end{IEEEproof}

\section{Concluding Remarks}
\label{sect:conclusion}

We characterized asymptotic expansions of the unconditional and two versions of the smooth conditional R\'{e}nyi entropies, and derived fundamental limits of several information-theoretic problems \cite{campbell_1965, courtade_verdu_isit2014_lossless, massey_isit1994, arikan_1996, bunte_lapidoth_2014} as applications of the asymptotic expansions.
Specifically, we compared the third-order asymptotic analyses for the classical fixed-to-fixed source coding and Campbell's source coding problems allowing errors, and showed in \eqref{eq:comparison_FF} that the difference between these two asymptotic expansions are manifested in their third-order terms.
In contrast to traditional results \cite{courtade_verdu_isit2014_lossless, arikan_1996, bunte_lapidoth_2014, kuzuoka_2019, kumar_sunny_thakre_kumar_2019} requiring the assumption of finite alphabets, due to the fact that we allow errors in the various problems we study, our results on guessing and task encoding problems are applicable to sources $X$ defined over countably infinite alphabets $\mathcal{X}$.

In \cite{sakai_tan_2019_VL}, the present authors considered the following limiting case of the cumulant generating function of codeword lengths:
\begin{align}
\lim_{\rho \to 0^{+}} \frac{ 1 }{ \rho } \log \mathbb{E}[ 2^{\rho \ell(F(X, Y))} ]
=
\mathbb{E}[ \ell(F(X, Y)) ] ,
\end{align}
without prefix-free constraints.
We \cite{sakai_tan_2019_VL} then showed that the optimal first-order coding rates (i.e., the $n$ scale) are the same under both average and maximum error criteria, and the optimal second-order coding rates (i.e., the $\sqrt{n}$ scale) differ under these two error formalisms.
This difference is characterized by the law of total variance (see \eqref{eq:law_of_total_variance}).
On the other hand, in this study, Corollaries~\ref{th:asympt_prefix-free_avg} and~\ref{th:asympt_prefix-free_max} state that the optimal first-order coding rates differ under these two error formalisms, and this difference can also be characterized by the law of total variance (see \propref{prop:KN-avg}).

In Theorems~\ref{lem:one-shot_prefix-free_avg}--\ref{th:one-shot_task_max}, we provided one-shot coding theorems in various information-theoretic problems, and these are formulated by two conditional versions of smooth R\'{e}nyi entropies $H_{\alpha}^{\varepsilon}(X^{n} \mid Y^{n})$ and $\check{H}_{\alpha}^{\varepsilon}(X^{n} \mid Y^{n})$.
Hence, further asymptotic analyses of $H_{\alpha}^{\varepsilon}(X^{n} \mid Y^{n})$ and $\check{H}_{\alpha}^{\varepsilon}(X^{n} \mid Y^{n})$ would yield further asymptotic results on the problems.
While the exact second- and third-order terms of the unconditional version of the smooth R\'{e}nyi entropy $H_{\alpha}^{\varepsilon}( X^{n} )$ were derived in \thref{th:CLT}, we showed in Theorems~\ref{th:2nd_avg} and~\ref{th:2nd_max} the exact first-order terms of $H_{\alpha}^{\varepsilon}(X^{n} \mid Y^{n})$ and $\check{H}_{\alpha}^{\varepsilon}(X^{n} \mid Y^{n})$, respectively, and that both remainder terms scale as $+\mathrm{O}( \sqrt{n} )$ due to Chebyshev's inequality.
Namely, finding the coefficients of the second- and third-order terms of $H_{\alpha}^{\varepsilon}(X^{n} \mid Y^{n})$ and $\check{H}_{\alpha}^{\varepsilon}(X^{n} \mid Y^{n})$ remain open problems.
As explained in \remref{rem:mixed-source}, Kuzuoka \cite[Theorem~2]{kuzuoka_2019} provided the first-order term of $H_{\alpha}^{\varepsilon}(X^{n} \mid Y^{n})$ when $(X^{n}, Y^{n})$ is a mixture of i.i.d.\ sources, and his result can be straightforwardly extended to the source $(X^{n}, Y^{n})$ satisfying the AEP, e.g., a mixture of stationary and ergodic sources (see \cite[Remark~2]{kuzuoka_2019}).
General formulas of the two conditional versions $H_{\alpha}^{\varepsilon}(X^{n} \mid Y^{n})$ and $\check{H}_{\alpha}^{\varepsilon}(X^{n} \mid Y^{n})$ for a general source $(\mathbf{X}, \mathbf{Y}) = \{ (X^{n}, Y^{n}) \}_{n = 1}^{\infty}$ remains open problems as well.
Finally, in this study, we only considered the smooth R\'{e}nyi entropies in the case where $0 < \alpha < 1$.
Asymptotic expansions and operational interpretations of the smooth R\'{e}nyi entropies with the order $1 < \alpha < \infty$ are of interest in future works.

\appendices

\section{Proof of \lemref{lem:CLT_MGF}}
\label{app:CLT_MGF}

It follows by the definition of $\langle \cdot \rangle_{\varepsilon}$ stated in \eqref{eq:cutoff} that
\begin{align}
\mathbb{E}\bigg[ \bigg\langle \exp\bigg( s \sum_{i = 1}^{n} Z_{i} \bigg) \bigg\rangle_{\varepsilon} \bigg]
& =
\mathbb{E}\bigg[ \exp\bigg( s \sum_{i = 1}^{n} Z_{i} \bigg) \bvec{1}_{\{ s \sum_{i = 1}^{n} Z_{i} < \eta_{n} \}} \bigg] + \alpha_{n} \, \mathrm{e}^{s \eta_{n}} \, \mathbb{P}\bigg\{ s \sum_{i = 1}^{n} Z_{i} = \eta_{n} \bigg\} ,
\label{eq:expand_cutoff}
\end{align}
where two real parameters $\eta_{n} \in \mathbb{R}$ and $0 \le \alpha_{n} < 1$ are chosen so that
\begin{align}
\mathbb{P}\bigg\{ s \sum_{i = 1}^{n} Z_{i} > \eta_{n} \bigg\} + \alpha_{n} \, \mathbb{P}\bigg\{ s \sum_{i = 1}^{n} Z_{i} = \eta_{n} \bigg\}
=
\varepsilon .
\label{eq:parameters_n}
\end{align}

Since we have assumed that there exist two positive constants $c_{1}$ and $c_{2}$ satisfying $V_{n} > n \, c_{1}$ and $T_{n} < c_{2} \, V_{n}$ for sufficiently large $n \ge n_{0}$, it can be verified by the Berry--Esseen theorem (see, e.g., \cite[Theorem~2 in Chapter~XVI.5]{feller_1971}) and Taylor's theorem for the map $\Phi^{-1} : (0, 1) \to \mathbb{R}$ that there exists a positive constant $c_{3}$ depending only on $0 < \varepsilon < 1$ such that
\begin{align}
s \, E_{n} + s \sqrt{ V_{n} } \, \Phi^{-1}(1 - \varepsilon) - c_{2} \, c_{3} \, s
\le
\eta_{n}
\le
s \, E_{n} + s \sqrt{ V_{n} } \, \Phi^{-1}(1 - \varepsilon) + c_{2} \, c_{3} \, s
\label{eq:estimate_eta-n}
\end{align}
for all $n \ge n_{0}$.
Now, choose an integer $n_{1} \ge n_{0}$ so that
\begin{align}
c_{2} \, c_{3} + s
\le
s \sqrt{ n \, c_{1} } - s \log n
\label{eq:choice-n1}
\end{align}
for all $n \ge n_{1}$.
In addition, fix a real number $\gamma$ so that
\begin{align}
\gamma
\ge
s \, \left( 12 \, c_{2} + \frac{ 1 }{ \sqrt{ c_{1} } } \right) \sqrt{ 2 \pi } \, \mathrm{e}^{(|\Phi^{-1}(1 - \varepsilon)| + 1)^{2}/2} .
\label{eq:choice-delta}
\end{align}
Then, we observe that
\begin{align}
\mathbb{E}\left[ \exp\left( s \sum_{i = 1}^{n} Z_{i} \right) \, \bvec{1}_{\{ s \sum_{i = 1}^{n} Z_{i} < \eta_{n} \}} \right]
& \ge
\sum_{k = 1}^{\infty} 2^{\eta_{n} - k \gamma} \, \mathbb{P}\left\{ \eta_{n} - k \, \gamma \le s \sum_{i = 1}^{n} Z_{i} < \eta_{n} - (k-1) \, \gamma \right\}
\notag \\
& \ge
2^{\eta_{n}} \sum_{k = 1}^{\lceil \log n \rceil} 2^{- k \gamma} \, \mathbb{P}\left\{ \eta_{n} - k \, \gamma \le s \sum_{i = 1}^{n} Z_{i} < \eta_{n} - (k-1) \, \gamma \right\}
\notag \\
& =
2^{\eta_{n}} \sum_{k = 1}^{\lceil \log n \rceil} 2^{- k \gamma} \, \mathbb{P}\left\{ (\eta_{n} - s \, E_{n}) - k \, \gamma \le s \sum_{i = 1}^{n} Z_{i} - s \, E_{n} < (\eta_{n} - s \, E_{n}) - (k-1) \, \gamma \right\}
\notag \\
& \overset{\mathclap{\text{(a)}}}{\ge}
2^{\eta_{n}} \sum_{k = 1}^{\lceil \log n \rceil} 2^{- k \gamma} \, \left( \Phi\left( \frac{ \eta_{n} - s \, E_{n} - (k-1) \, \gamma }{ s \sqrt{ V_{n} } } \right) - \Phi\left( \frac{ \eta_{n} - s \, E_{n} - k \, \gamma }{ s \sqrt{ V_{n} } } \right) - \frac{ 12 \, T_{n} }{ V_{n}^{3/2} } \right)
\notag \\
& =
2^{\eta_{n}} \sum_{k = 1}^{\lceil \log n \rceil} 2^{- k \gamma} \, \left( \frac{ 1 }{ \sqrt{ 2 \pi } } \int_{[\eta_{n} - s E_{n} - k \gamma]/(s \sqrt{V_{n}})}^{[\eta_{n} - s E_{n} - (k-1) \gamma]/(s \sqrt{V_{n}})} \mathrm{e}^{-t^{2}/2} \, \mathrm{d} t - \frac{ 12 \, T_{n} }{ V_{n}^{3/2} } \right)
\notag \\
& \overset{\mathclap{\text{(b)}}}{\ge}
2^{\eta_{n}} \sum_{k = 1}^{\lceil \log n \rceil} 2^{- k \gamma} \, \left( \frac{ \mathrm{e}^{-(|\Phi^{-1}(1 - \varepsilon)| + 1)^{2}/2} }{ \sqrt{ 2 \pi } } \int_{[\eta_{n} - s E_{n} - k \gamma]/(s \sqrt{V_{n}})}^{[\eta_{n} - s E_{n} - (k-1) \gamma]/(s \sqrt{V_{n}})} \mathrm{d} t - \frac{ 12 \, T_{n} }{ V_{n}^{3/2} } \right)
\notag \\
& =
2^{\eta_{n}} \sum_{k = 1}^{\lceil \log n \rceil} 2^{- k \gamma} \, \left( \frac{ \gamma \, \mathrm{e}^{-(|\Phi^{-1}(1 - \varepsilon)| + 1)^{2}/2} }{ s \sqrt{ 2 \pi } } - \frac{ 12 \, T_{n} }{ V_{n}} \right) \frac{ 1 }{ \sqrt{ V_{n} } }
\notag \\
& \ge
2^{\eta_{n} - \gamma} \, (1 - n^{- \gamma}) \left( \frac{ \gamma \, \mathrm{e}^{-(|\Phi^{-1}(1 - \varepsilon)| + 1)^{2}/2} }{ s \sqrt{ 2 \pi } } - \frac{ 12 \, T_{n} }{ V_{n} } \right) \frac{ 1 }{ \sqrt{ V_{n} } }
\notag \\
& \overset{\mathclap{\text{(c)}}}{\ge}
2^{\eta_{n} - \gamma} \, (1 - n^{- \gamma}) \left( \frac{ \gamma \, \mathrm{e}^{-(|\Phi^{-1}(1 - \varepsilon)| + 1)^{2}/2} }{ s \sqrt{ 2 \pi } } - 12 \, c_{2} \right) \frac{ 1 }{ \sqrt{ V_{n} } }
\notag \\
& \overset{\mathclap{\text{(d)}}}{\ge}
2^{\eta_{n} - \gamma} \, (1 - n^{- \gamma}) \left( \frac{ \gamma \, \mathrm{e}^{-(|\Phi^{-1}(1 - \varepsilon)| + 1)^{2}/2} }{ s \sqrt{ 2 \pi } } - 12 \, c_{2} \right) \sqrt{ \frac{ c_{1} }{ n } }
\notag \\
& \overset{\mathclap{\text{(e)}}}{\ge}
\frac{ 2^{\eta_{n}} \, (1 - n^{- \gamma}) }{ 2^{\gamma} \sqrt{ n } }
\label{eq:CLT_MGF_LB}
\end{align}
for sufficiently large $n \ge n_{1}$, where
\begin{itemize}
\item
(a) follows from the Berry--Esseen theorem (see, e.g., \cite[Theorem~2 in Chapter~XVI.5]{feller_1971}),
\item
(b) follows from \eqref{eq:estimate_eta-n}, \eqref{eq:choice-n1}, and the fact that $t \mapsto \mathrm{e}^{-t^{2}/2}$ is quasiconcave in $t \in \mathbb{R}$,
\item
(c) follows from the hypothesis that $T_{n} < c_{2} \, V_{n}$ for sufficiently large $n \ge n_{1}$,
\item
(d) follows from the hypothesis that $V_{n} < n/c_{1}$ for sufficiently large $n \ge n_{1}$, and
\item
(e) follows from the choice of $\gamma$ stated in \eqref{eq:choice-delta}.
\end{itemize}
Therefore, we have
\begin{align}
\frac{ 1 }{ s } \log \mathbb{E}\left[ \left\langle \exp\left( s \sum_{i = 1}^{n} Z_{i} \right) \right\rangle_{\varepsilon} \right]
& \overset{\mathclap{\text{(a)}}}{\ge}
\frac{ 1 }{ s } \log \mathbb{E}\left[ \exp\left( s \sum_{i = 1}^{n} Z_{i} \right) \, \bvec{1}_{\{ s \sum_{i = 1}^{n} Z_{i} < \eta_{n} \}} \right]
\notag \\
& \overset{\mathclap{\text{(b)}}}{\ge}
\frac{ 1 }{ s } \log \left( \frac{ 2^{\eta_{n}} \, (1 - n^{- \gamma}) }{ 2^{\gamma} \sqrt{ n } } \right)
\notag \\
& =
\frac{ 1 }{ s } \left( \eta_{n} - \gamma + \log (1 - n^{-\gamma}) - \frac{ 1 }{ 2 } \log n \right)
\notag \\
& \overset{\mathclap{\text{(c)}}}{\ge}
E_{n} + \sqrt{ V_{n} } \, \Phi^{-1}(1 - \varepsilon) - \frac{ 1 }{ 2 \, s } \log n  - c_{2} \, c_{3} + \frac{ \gamma }{ s } + \frac{ 1 }{ s } \log (1 - n^{-\gamma})
\label{eq:CLT_CGF_LB}
\end{align}
for sufficiently large $n \ge n_{1}$, where
\begin{itemize}
\item
(a) follows from \eqref{eq:expand_cutoff},
\item
(b) follows from \eqref{eq:CLT_MGF_LB}, and
\item
(c) follows from \eqref{eq:estimate_eta-n}.
\end{itemize}

On the other hand, it can be verified by the same way as \cite[Lemma~47]{polyanskiy_poor_verdu_2010} that
\begin{align}
\mathbb{E}\left[ \exp\left( t \sum_{i = 1}^{n} Z_{i} \right) \, \bvec{1}_{\{ t \sum_{i = 1}^{n} Z_{i} \le \eta_{n} \}} \right]
& \le
2^{\eta_{n}+1} \, \left( \frac{ 1 }{ s \sqrt{ 2 \pi } } + 12 \, c_{2} \right) \frac{ 1 }{ \sqrt{ n \, c_{1} } }
\label{eq:CLT_MGF_UB}
\end{align}
for sufficiently large $n \ge n_{0}$.
Thus, we obtain
\begin{align}
\frac{ 1 }{ s } \log \mathbb{E}\left[ \left\langle \exp\left( s \sum_{i = 1}^{n} Z_{i} \right) \right\rangle_{\varepsilon} \right]
& \overset{\mathclap{\text{(a)}}}{\le}
\frac{ 1 }{ s } \log \mathbb{E}\left[ \exp\left( s \sum_{i = 1}^{n} Z_{i} \right) \, \bvec{1}_{\{ s \sum_{i = 1}^{n} Z_{i} \le \eta_{n} \}} \right]
\notag \\
& \overset{\mathclap{\text{(b)}}}{\le}
\frac{ 1 }{ s } \log \left( 2^{\eta_{n}+1} \, \left( \frac{ 1 }{ s \sqrt{ 2 \pi } } + 12 \, c_{2} \right) \frac{ 1 }{ \sqrt{ n \, c_{1} } } \right)
\notag \\
& =
\frac{ 1 }{ s } \left( \eta_{n} + 1 + \log \left( \frac{ 1 }{ s \sqrt{ 2 \pi } } + 12 \, c_{2} \right) - \frac{ 1 }{ 2 } \log n - \frac{ 1 }{ 2 } \log c_{1} \right)
\notag \\
& \overset{\mathclap{\text{(c)}}}{\le}
E_{n} + \sqrt{ V_{n} } \, \Phi^{-1}(1 - \varepsilon) - \frac{ 1 }{ 2 \, s } \log n + c_{2} \, c_{3} + \frac{ 1 }{ s } + \frac{ 1 }{ s } \log \left( \frac{ 1 }{ s \sqrt{ 2 \pi } } + 12 \, c_{2} \right) - \frac{ 1 }{ 2 \, s } \log c_{1}
\label{eq:CLT_CGF_UB}
\end{align}
for sufficiently large $n \ge n_{0}$, where
\begin{itemize}
\item
(a) follows from \eqref{eq:expand_cutoff},
\item
(b) follows from \eqref{eq:CLT_MGF_UB}, and
\item
(c) follows form \eqref{eq:estimate_eta-n}.
\end{itemize}
The proof of \lemref{lem:CLT_MGF} is now completed by combining \eqref{eq:CLT_CGF_LB} and \eqref{eq:CLT_CGF_UB}.
\hfill\IEEEQEDhere

\section{Proof of \thref{th:2nd_avg}---Zero Variance $U(X \mid Y) = 0$}
\label{app:2nd_avg_zero-variance}

\subsection{Proof of \eqref{eq:zero-variance_2nd_avg} When $H(X \mid Y) = 0$}

Since we have assumed that $U(X \mid Y) = 0$, it is clear that
\begin{align}
P_{X|Y}(x \mid y)
=
\begin{dcases}
\exp\Big( -H(X \mid Y) \Big)
& \mathrm{if} \ P_{X|Y}(x \mid y) > 0 ,
\\
0
& \mathrm{if} \ P_{X|Y}(x \mid y) = 0
\end{dcases}
\label{eq:cond-probab_zero-variance}
\end{align}
for every $(x, y) \in \mathcal{X} \times \mathcal{Y}$.
Consider the case in which $H(X \mid Y) = 0$.
Then, it follows from \eqref{eq:largest-A_cond1}, \eqref{eq:largest-A_cond2}, and \eqref{eq:cond-probab_zero-variance} that
\begin{align}
\mathcal{A}_{X|Y}^{\delta(y)}(y)
=
\emptyset
\label{eq:largest-A_zero-ent}
\end{align}
for every $y \in \mathcal{Y}$.
Hence, it follows from \eqref{eq:Kuzuoka_identity} and \eqref{eq:largest-A_zero-ent} that
\begin{align}
H_{\alpha}^{\varepsilon}(X^{n} \mid Y^{n})
& =
\frac{ \alpha }{ 1 - \alpha } \log \left( \inf_{\delta_{n}( \cdot )} \sum_{\bvec{y} \in \mathcal{Y}^{n}} P_{Y^{n}}( \bvec{y} ) \, (1 - \delta_{n}(\bvec{y})) \right)
\notag \\
& =
\frac{ \alpha }{ 1 - \alpha } \log (1 - \varepsilon)
\end{align}
for every $n \ge 1$, where the infimum is taken over the mappings $\delta_{n} : \mathcal{Y}^{n} \to [0, 1]$ satisfying
\begin{align}
\sum_{\bvec{y} \in \mathcal{Y}^{n}} P_{Y^{n}}( \bvec{y} ) \, \delta_{n}( \bvec{y} )
=
\varepsilon .
\label{eq:eps_n-expectation}
\end{align}
Therefore, the asymptotic expansion in \eqref{eq:zero-variance_2nd_avg} holds if $H(X \mid Y) = 0$, completing the proof.
\hfill\IEEEQEDhere

\subsection{Proof of Upper Bound Part of \eqref{eq:zero-variance_2nd_avg} When $H(X \mid Y) > 0$}
\label{app:2nd_avg_zero-variance_direct}

For each $n \ge 1$ and $\bvec{y} \in \mathcal{Y}^{n}$, define the sub-probability distribution $\tilde{Q}_{X^{n}|Y^{n}}(\cdot \mid \bvec{y})$ on $\mathcal{X}^{n}$ as
\begin{align}
\tilde{Q}_{X^{n}|Y^{n}}(\bvec{x} \mid \bvec{y})
=
(1 - \varepsilon) \, P_{X^{n}|Y^{n}}(\bvec{x} \mid \bvec{y}) .
\label{eq:cond_tildeQn}
\end{align}
Moreover, define the joint sub-probability distribution $\tilde{Q}_{X^{n}, Y^{n}}$ on $\mathcal{X}^{n} \times \mathcal{Y}^{n}$ as
\begin{align}
\tilde{Q}_{X^{n}, Y^{n}}(\bvec{x}, \bvec{y})
=
\sum_{\bvec{y} \in \mathcal{Y}^{n}} P_{Y^{n}}( \bvec{y} ) \, \tilde{Q}_{X^{n}|Y^{n}}(\bvec{x} \mid \bvec{y}) .
\label{eq:joint_tildeQn}
\end{align}
Then, we observe that
\begin{align}
H_{\alpha}^{\varepsilon}(X^{n} \mid Y^{n})
& =
\frac{ \alpha }{ 1 - \alpha } \log \left( \inf_{Q \in \mathcal{B}_{\mathcal{X}^{n} \times \mathcal{Y}^{n}}^{\varepsilon}( P_{X^{n}, Y^{n}} )} \sum_{\bvec{y} \in \mathcal{Y}^{n}} \left( \sum_{\bvec{x} \in \mathcal{X}^{n}} Q(\bvec{x}, \bvec{y})^{\alpha} \right)^{1/\alpha} \right)
\notag \\
& \overset{\mathclap{\text{(a)}}}{\le}
\frac{ \alpha }{ 1 - \alpha } \log \left( \sum_{\bvec{y} \in \mathcal{Y}^{n}} \left( \sum_{\bvec{x} \in \mathcal{X}^{n}} \tilde{Q}_{X^{n}, Y^{n}}(\bvec{x}, \bvec{y})^{\alpha} \right)^{1/\alpha} \right)
\notag \\
& \overset{\mathclap{\text{(b)}}}{=}
\frac{ \alpha }{ 1 - \alpha } \log \left( \sum_{\bvec{y} \in \mathcal{Y}^{n}} P_{Y^{n}}( \bvec{y} ) \left( \sum_{\bvec{x} \in \mathcal{X}^{n}} \tilde{Q}_{X^{n}|Y^{n}}(\bvec{x} \mid \bvec{y})^{\alpha} \right)^{1/\alpha} \right)
\notag \\
& \overset{\mathclap{\text{(c)}}}{=}
\frac{ \alpha }{ 1 - \alpha } \log \left( \sum_{\bvec{y} \in \mathcal{Y}^{n}} P_{Y^{n}}( \bvec{y} ) \left( \sum_{\bvec{x} \in \mathcal{X}^{n}} P_{X^{n}|Y^{n}}(\bvec{x} \mid \bvec{y})^{\alpha} \right)^{1/\alpha} \right) + \frac{ \alpha }{ 1 - \alpha } \log (1 - \varepsilon)
\notag \\
& \overset{\mathclap{\text{(d)}}}{=}
n \, H(X \mid Y) + \frac{ \alpha }{ 1 - \alpha } \log (1 - \varepsilon)
\end{align}
where
\begin{itemize}
\item
(a) follows from the fact that $\tilde{Q}_{X^{n}, Y^{n}}$ belongs to $\mathcal{B}_{\mathcal{X}^{n} \times \mathcal{Y}^{n}}^{\varepsilon}( P_{X^{n}, Y^{n}} )$,
\item
(b) follows from \eqref{eq:joint_tildeQn},
\item
(c) follows from \eqref{eq:cond_tildeQn}, and
\item
(d) follows from \eqref{eq:cond-probab_zero-variance}.
\end{itemize}
This completes the proof of the upper bound part of \eqref{eq:zero-variance_2nd_avg}.
\hfill\IEEEQEDhere

\subsection{Proof of Lower Bound Part of \eqref{eq:zero-variance_2nd_avg} When $H(X \mid Y) > 0$}
\label{app:2nd_avg_zero-variance_converse}

Consider a mapping $\delta_{n} : \mathcal{Y}^{n} \to [0, 1]$ satisfying \eqref{eq:eps_n-expectation}.
For the sake of brevity, we write
\begin{align}
\mathcal{A}_{n}( \bvec{y} )
=
\mathcal{A}_{X^{n}|Y^{n}}^{\delta_{n}(\bvec{y})}( \bvec{y} )
\label{def:abbrev-A-XnYn}
\end{align}
for each $\bvec{y} \in \mathcal{Y}^{n}$; see \eqref{eq:largest-A_cond1} and \eqref{eq:largest-A_cond2} for the definition of the right-hand side of \eqref{def:abbrev-A-XnYn} by replacing $\varepsilon$ and $P_{X}( \cdot )$ by $\delta_{n}( \bvec{y} )$ and $P_{X^{n}|Y^{n}}(\cdot \mid \bvec{y})$, respectively.
Moreover, define
\begin{align}
\tilde{\mathcal{A}}_{n}( \bvec{y} )
\coloneqq
\mathcal{A}_{n}( \bvec{y} ) \cup \{ \bvec{x}^{\ast}( \bvec{y} ) \} ,
\label{def:abbrev-tildeA-XnYn}
\end{align}
where $\bvec{x}^{\ast}( \bvec{y} ) \in \mathcal{X}^{n}$ is chosen so that
\begin{align}
\bvec{x}^{\ast}( \bvec{y} )
\in
\argmax\limits_{\bvec{x} \in \mathcal{X}^{n} \setminus \mathcal{A}_{n}(\bvec{y})} P_{X^{n}|Y^{n}}(\bvec{x} \mid \bvec{y}) .
\end{align}
Furthermore, define the subset $\mathcal{U}_{n}$ of $\mathcal{Y}^{n}$ by
\begin{align}
\mathcal{U}_{n}
\coloneqq
\left\{ \bvec{y} \ \middle| \ P_{X^{n}|Y^{n}}(\tilde{\mathcal{A}}_{n}(\bvec{y}) \mid \bvec{y}) \ge \frac{ 1 - \varepsilon }{ 2 } \right\} .
\label{def:Un}
\end{align}
Note that both $\mathcal{A}_{n}( \cdot )$ and $\mathcal{U}_{n}$ depend on $\delta_{n}( \cdot )$.
A direct calculation shows
\begin{align}
1 - \varepsilon
& \overset{\mathclap{\text{(a)}}}{=}
\sum_{\bvec{y} \in \mathcal{Y}^{n}} P_{Y^{n}}( \bvec{y} ) \, (1 - \delta_{n}( \bvec{y} ))
\notag \\
& \overset{\mathclap{\text{(b)}}}{\le}
\sum_{\bvec{y} \in \mathcal{Y}^{n}} P_{Y^{n}}( \bvec{y} ) \, P_{X^{n}|Y^{n}}(\tilde{\mathcal{A}}_{n}(\bvec{y}) \mid \bvec{y})
\notag \\
& \overset{\mathclap{\text{(c)}}}{\le}
\sum_{\bvec{y} \in \mathcal{U}_{n}} P_{Y^{n}}( \bvec{y} ) + \frac{ 1 - \varepsilon }{ 2 } \sum_{\bvec{y} \in \mathcal{Y}^{n} \setminus \mathcal{U}_{n}} P_{Y^{n}}( \bvec{y} )
\notag \\
& =
P_{Y^{n}}( \mathcal{U}_{n} ) \left( \frac{ 1 + \varepsilon }{ 2 } \right) + \frac{ 1 - \varepsilon }{ 2 } ,
\end{align}
where
\begin{itemize}
\item
(a) follows from \eqref{eq:eps_n-expectation},
\item
(b) follows from the right-hand inequality of \eqref{eq:largest-A_cond2}, and
\item
(c) follows by the definition of $\mathcal{U}_{n}$ stated in \eqref{def:Un}.
\end{itemize}
Therefore, we get%
\footnote{Inequality~\eqref{eq:reverse-Markov} is indeed a reverse Markov inequality (cf.\ \cite[Lemma~5.6.1]{han_2003}).}
\begin{align}
P_{Y^{n}}( \mathcal{U}_{n} )
\ge
\frac{ 1 - \varepsilon }{ 1 + \varepsilon } .
\label{eq:reverse-Markov}
\end{align}
In addition, we see that
\begin{align}
\frac{ 1 - \varepsilon }{ 2 }
& \overset{\mathclap{\text{(a)}}}{\le}
P_{X^{n}|Y^{n}}(\tilde{\mathcal{A}}_{n}(\bvec{y}) \mid \bvec{y})
\notag \\
& \overset{\mathclap{\text{(b)}}}{=}
|\tilde{\mathcal{A}}_{n}( \bvec{y} )| \, \exp\Big( - n \, H(X \mid Y) \Big)
\label{eq:card_An-yn}
\end{align}
for each $\bvec{y} \in \mathcal{U}_{n}$, where
\begin{itemize}
\item
(a) follows by the definition of $\mathcal{U}_{n}$ stated in \eqref{def:Un}, and
\item
(b) follows from \eqref{eq:cond-probab_zero-variance}.
\end{itemize}
Furthermore, we have
\begin{align}
\sum_{\bvec{x} \in \tilde{\mathcal{A}}_{n}( \bvec{y} )} P_{X^{n}|Y^{n}}(\bvec{x} \mid \bvec{y})^{\alpha}
& \overset{\mathclap{\text{(a)}}}{=}
|\tilde{\mathcal{A}}_{n}( \bvec{y} )| \, \exp\Big( - \alpha \, n \, H(X \mid Y) \Big)
\notag \\
& \overset{\mathclap{\text{(b)}}}{\ge}
\frac{ 1 - \varepsilon }{ 2 } \, \exp\Big( (1 - \alpha) \, n \, H(X \mid Y) \Big)
\label{eq:inner_cond-smooth_zero-variance}
\end{align}
for every $\bvec{y} \in \mathcal{U}_{n}$, where
\begin{itemize}
\item
(a) follows from \eqref{eq:cond-probab_zero-variance}, and
\item
(b) follows from \eqref{eq:card_An-yn}.
\end{itemize}
Now, noting that $H(X \mid Y) > 0$, choose an integer $n_{0}$ by
\begin{align}
n_{0}
=
n_{0}(\alpha, \varepsilon, H(X \mid Y))
\coloneqq
\left\lceil \frac{ 1 }{ (1 - \alpha) \, H(X \mid Y) } \log \frac{ 4 }{ 1 - \varepsilon } \right\rceil ,
\label{eq:large_n0_cond_zero-variance}
\end{align}
where $\lceil u \rceil \coloneqq \min\{ z \in \mathbb{Z} \mid z \ge u \}$ stands for the ceiling function.
Then, we observe that
\begin{align}
H_{\alpha}^{\varepsilon}(X^{n} \mid Y^{n})
& \overset{\mathclap{\text{(a)}}}{=}
\frac{ \alpha }{ 1 - \alpha } \log \left( \inf_{\delta_{n}( \cdot )} \sum_{\bvec{y} \in \mathcal{Y}_{n}} P_{Y^{n}}( \bvec{y} ) \left( \sum_{\bvec{x} \in \mathcal{A}_{n}( \bvec{y} )} P_{X^{n}|Y^{n}}(\bvec{x} \mid \bvec{y})^{\alpha} + \Big( 1 - \delta_{n}( \bvec{y} ) - P_{X^{n}|Y^{n}}(\mathcal{A}_{n}(\bvec{y}) \mid \bvec{y}) \Big)^{\alpha} \right)^{1/\alpha} \right)
\notag \\
& \overset{\mathclap{\text{(b)}}}{\ge}
\frac{ \alpha }{ 1 - \alpha } \log \left( \inf_{\delta_{n}( \cdot )} \sum_{\bvec{y} \in \mathcal{U}_{n}} P_{Y^{n}}( \bvec{y} ) \left( \sum_{\bvec{x} \in \tilde{\mathcal{A}}_{n}( \bvec{y} )} P_{X^{n}|Y^{n}}(\bvec{x} \mid \bvec{y})^{\alpha} - 1 \Big) \right)^{1/\alpha} \right)
\notag \\
& \overset{\mathclap{\text{(c)}}}{\ge}
\frac{ \alpha }{ 1 - \alpha } \log \left( \inf_{\delta_{n}( \cdot )} \sum_{\bvec{y} \in \mathcal{U}_{n}} P_{Y^{n}}( \bvec{y} ) \left( \frac{ 1 - \varepsilon }{ 2 } \, \exp\Big( (1 - \alpha) \, n \, H(X \mid Y) \Big) - 1 \right)^{1/\alpha} \right)
\notag \\
& \overset{\mathclap{\text{(d)}}}{\ge}
\frac{ \alpha }{ 1 - \alpha } \log \left( \inf_{\delta_{n}( \cdot )} \sum_{\bvec{y} \in \mathcal{U}_{n}} P_{Y^{n}}( \bvec{y} ) \left( \frac{ 1 - \varepsilon }{ 4 } \right)^{1/\alpha} \, \exp\left( \frac{ 1 - \alpha }{ \alpha } \, n \, H(X \mid Y) \right) \right)
\notag \\
& =
n \, H(X \mid Y) + \frac{ 1 }{ 1 - \alpha } \log \frac{ 1 - \varepsilon }{ 4 } + \frac{ \alpha }{ 1 - \alpha } \log \Big( \inf_{\delta_{n}( \cdot )} P_{Y^{n}}( \mathcal{U}_{n} ) \Big)
\notag \\
& \overset{\mathclap{\text{(e)}}}{\ge}
n \, H(X \mid Y) + \frac{ 1 }{ 1 - \alpha } \log \frac{ 1 - \varepsilon }{ 4 } + \frac{ \alpha }{ 1 - \alpha } \log \frac{ 1 - \varepsilon }{ 1 + \varepsilon }
\label{eq:LB_cond-smooth_zero-variance}
\end{align}
for sufficiently large $n \ge n_{0}$, where
\begin{itemize}
\item
(a) follows from \eqref{eq:Kuzuoka_identity},
\item
(b) follows from the right-hand inequality of \eqref{eq:largest-A_cond2}, i.e.,
\begin{align}
\Big( 1 - \delta_{n}( \bvec{y} ) - P_{X^{n}|Y^{n}}(\mathcal{A}_{n}(\bvec{y}) \mid \bvec{y}) \Big)^{\alpha}
\ge
P_{X^{n}|Y^{n}}(\bvec{x}^{\ast}(\bvec{y}) \mid \bvec{y})^{\alpha} - 1 ,
\label{def:largest-A_cond_cond2_alpha}
\end{align}
\item
(c) follows from \eqref{eq:inner_cond-smooth_zero-variance},
\item
(d) follows by the choice of $n_{0}$ stated in \eqref{eq:large_n0_cond_zero-variance}, and
\item
(e) follows from \eqref{eq:reverse-Markov}.
\end{itemize}
This completes the proof of the lower bound part of \eqref{eq:zero-variance_2nd_avg}.
\hfill\IEEEQEDhere

\section{Proof of \thref{th:2nd_avg}---Positive Variance $0 < U(X \mid Y) < \infty$}
\label{app:2nd_avg}

\subsection{Proof of Upper Bound Part of \eqref{eq:2nd_avg}}
\label{app:2nd_avg_direct}

For each positive integer $n$, define a subset $\mathcal{T}_{\varepsilon}^{(n)}$ of $\mathcal{X}^{n} \times \mathcal{Y}^{n}$ as
\begin{align}
\mathcal{T}_{\varepsilon}^{(n)}
& \coloneqq
\bigg\{ (\bvec{x}, \bvec{y}) \ \bigg| \ \bigg| \log \frac{ 1 }{ P_{X^{n}|Y^{n}}(\bvec{x} \mid \bvec{y}) } - n \, H(X \mid Y) \bigg| \le \sqrt{ \frac{ n \, U(X \mid Y) }{ \varepsilon } } \bigg\} .
\label{def:cond_typical1}
\end{align}
In addition, define
\begin{align}
\mathcal{T}_{\varepsilon}^{(n)}( \bvec{y} )
& \coloneqq
\{ \bvec{x} \mid (\bvec{x}, \bvec{y}) \in \mathcal{T}_{\varepsilon}^{(n)} \} .
\end{align}
Note that for every $(\bvec{x}, \bvec{y}) \in \mathcal{T}_{\varepsilon}^{(n)}$, it holds that
\begin{align}
\exp\left( - n \, H(X \mid Y) - \sqrt{ \frac{ n \, U(X \mid Y) }{ \varepsilon } } \right)
\le
P_{X^{n}|Y^{n}}(\bvec{x} \mid \bvec{y})
\le
\exp\left( - n \, H(X \mid Y) + \sqrt{ \frac{ n \, U(X \mid Y) }{ \varepsilon } } \right) .
\label{eq:bound_pmf_Teps}
\end{align}
Since $0 < U(X \mid Y) < \infty$, it follows from Chebyshev's inequality that the $P_{X^{n}, Y^{n}}$-probability of $\mathcal{T}_{\varepsilon}^{(n)}$ is bounded from below as
\begin{align}
P_{X^{n}, Y^{n}}( \mathcal{T}_{\varepsilon}^{(n)} )
& =
1 - \mathbb{P}\{ (X^{n}, Y^{n}) \notin \mathcal{T}_{\varepsilon}^{(n)} \}
\ge
1 - \varepsilon .
\label{eq:Chebyshev}
\end{align}
On the other hand, we see that
\begin{align}
1
& \ge
\sum_{\substack{ \bvec{x} \in \mathcal{X}^{n} \\ : (\bvec{x}, \bvec{y}) \in \mathcal{T}_{\varepsilon}^{(n)} }} P_{X^{n}|Y^{n}}(\bvec{x} \mid \bvec{y})
\notag \\
& \ge
\sum_{\substack{ \bvec{x} \in \mathcal{X}^{n} \\ : (\bvec{x}, \bvec{y}) \in \mathcal{T}_{\varepsilon}^{(n)} }} \exp\left( - n \, H(X \mid Y) - \sqrt{ \frac{ n \, U(X \mid Y) }{ \varepsilon } } \right)
\notag \\
& =
|\mathcal{T}_{\varepsilon}^{(n)}( \bvec{y} )| \, \exp\left( - n \, H(X \mid Y) - \sqrt{ \frac{ n \, U(X \mid Y) }{ \varepsilon } } \right)
\label{eq:bound_card-cond_typical1}
\end{align}
for every $\bvec{y} \in \mathcal{Y}^{n}$, where the second inequality follows from the left-hand inequality of \eqref{eq:bound_pmf_Teps}.
Hence, we observe that
\begin{align}
H_{\alpha}^{\varepsilon}(X^{n} \mid Y^{n})
& \overset{\mathclap{\text{(a)}}}{=}
\frac{ \alpha }{ 1 - \alpha } \log \left( \inf_{Q \in \mathcal{B}_{\mathcal{X}^{n} \times \mathcal{Y}^{n}}^{\varepsilon}( P_{X^{n}, Y^{n}} )} \sum_{\bvec{y} \in \mathcal{Y}^{n}} \left( \sum_{\bvec{x} \in \mathcal{X}^{n}} Q(\bvec{x}, \bvec{y})^{\alpha} \right)^{1/\alpha} \right)
\notag \\
& \overset{\mathclap{\text{(b)}}}{\le}
\frac{ \alpha }{ 1 - \alpha } \log \left( \sum_{\bvec{y} \in \mathcal{Y}^{n}} \left( \sum_{\substack{ \bvec{x} \in \mathcal{X}^{n} : \\ (\bvec{x}, \bvec{y}) \in \mathcal{T}_{\varepsilon}^{(n)} }} P_{X^{n}, Y^{n}}(\bvec{x}, \bvec{y})^{\alpha} \right)^{1/\alpha} \right)
\notag \\
& =
\frac{ \alpha }{ 1 - \alpha } \log \left( \sum_{\bvec{y} \in \mathcal{Y}^{n}} P_{Y^{n}}( \bvec{y} ) \, \left( \sum_{\substack{ \bvec{x} \in \mathcal{X}^{n} : \\ (\bvec{x}, \bvec{y}) \in \mathcal{T}_{\varepsilon}^{(n)} }} P_{X^{n} | Y^{n}}(\bvec{x} \mid \bvec{y})^{\alpha} \right)^{1/\alpha} \right)
\notag \\
& \overset{\mathclap{\text{(c)}}}{\le}
\frac{ \alpha }{ 1 - \alpha } \log \left(  \sum_{\bvec{y} \in \mathcal{Y}^{n}} P_{Y^{n}}( \bvec{y} ) \, \left( \sum_{\substack{ \bvec{x} \in \mathcal{X}^{n} : \\ (\bvec{x}, \bvec{y}) \in \mathcal{T}_{\varepsilon}^{(n)} }} \exp\left( - \alpha \, \Bigg( n \, H(X \mid Y) - \sqrt{ \frac{ n \, U(X \mid Y) }{ \varepsilon } } \Bigg) \right) \right)^{1/\alpha} \right)
\notag \\
& =
\frac{ \alpha }{ 1 - \alpha } \log \left(  \sum_{\bvec{y} \in \mathcal{Y}^{n}} P_{Y^{n}}( \bvec{y} ) \, \left( |\mathcal{T}_{\varepsilon}^{(n)}( \bvec{y} )| \, \exp\left( - \alpha \, \Bigg( n \, H(X \mid Y) - \sqrt{ \frac{ n \, U(X \mid Y) }{ \varepsilon } } \Bigg) \right) \right)^{1/\alpha} \right)
\notag \\
& \overset{\mathclap{\text{(d)}}}{\le}
\frac{ \alpha }{ 1 - \alpha } \log \left( \sum_{\bvec{y} \in \mathcal{Y}^{n}} P_{Y^{n}}( \bvec{y} ) \, \exp\left( \frac{ 1 - \alpha }{ \alpha } \, n \, H(X \mid Y) + \frac{ 1 + \alpha }{ \alpha } \sqrt{ \frac{ n \, U(X \mid Y) }{ \varepsilon } } \right) \right)
\notag \\
& =
n \, H(X \mid Y) + \frac{ 1 + \alpha }{ 1 - \alpha } \sqrt{ \frac{ n \, U(X \mid Y) }{ \varepsilon } }
\end{align}
where
\begin{itemize}
\item
(a) follows by the definition of $H_{\alpha}^{\varepsilon}(X^{n} \mid Y^{n})$ stated in \eqref{def:conditional_smooth},
\item
(b) follows from \eqref{eq:Chebyshev} and the definition of $\mathcal{B}_{\mathcal{X}^{n} \times \mathcal{Y}^{n}}^{\varepsilon}( P_{X^{n}, Y^{n}} )$ stated in \eqref{def:eps-ball_joint},
\item
(c) follows from the right-hand inequality of \eqref{eq:bound_pmf_Teps}, and
\item
(d) follows from \eqref{eq:bound_card-cond_typical1}.
\end{itemize}
This completes the proof of the upper bound part of \eqref{eq:2nd_avg}.
\hfill\IEEEQEDhere

\subsection{Proof of Lower Bound Part of \eqref{eq:2nd_avg}}
\label{app:2nd_avg_converse}

Consider a mapping $\delta_{n} : \mathcal{Y}^{n} \to [0, 1]$ satisfying \eqref{eq:eps_n-expectation}.
Recall that $\mathcal{A}_{n}( \bvec{y} )$ and $\tilde{\mathcal{A}}_{n}( \bvec{y} )$ are given as \eqref{def:abbrev-A-XnYn} and \eqref{def:abbrev-tildeA-XnYn}, respectively, for each $\bvec{y} \in \mathcal{Y}^{n}$.
Fix $\gamma \in (0, 1 - \varepsilon)$ arbitrarily, and consider the subset $\mathcal{T}_{\gamma}^{(n)}$ of $\mathcal{X}^{n} \times \mathcal{Y}^{n}$ defined as in \eqref{def:cond_typical1} by replacing $\varepsilon$ by $\gamma$.
Moreover, define the subset $\mathcal{V}_{n}$ of $\mathcal{Y}^{n}$ by
\begin{align}
\mathcal{V}_{n}
\coloneqq
\left\{ \bvec{y} \ \middle| \ P_{X^{n}|Y^{n}}( \tilde{\mathcal{A}}_{n}( \bvec{y} ) \cap \mathcal{T}_{\gamma}^{(n)}( \bvec{y} ) \mid \bvec{y}) \ge \frac{ 1 - \varepsilon - \gamma }{ 2 } \right\} ,
\label{def:set-Vn}
\end{align}
where note that $\mathcal{V}_{n}$ depends on $\epsilon_{n}( \cdot )$.
A simple calculation yields
\begin{align}
1 - \varepsilon - \gamma
& \overset{\mathclap{\text{(a)}}}{=}
\sum_{\bvec{y} \in \mathcal{Y}^{n}} P_{Y^{n}}( \bvec{y} ) \, (1 - \epsilon_{n}( \bvec{y} )) - \gamma
\notag \\
& \overset{\mathclap{\text{(b)}}}{\le}
\sum_{\bvec{y} \in \mathcal{Y}^{n}} P_{Y^{n}}( \bvec{y} ) \, P_{X^{n}|Y^{n}}(\tilde{\mathcal{A}}_{n}(\bvec{y}) \mid \bvec{y}) - \gamma
\notag \\
& \overset{\mathclap{\text{(c)}}}{\le}
\sum_{\bvec{y} \in \mathcal{Y}^{n}} P_{Y^{n}}( \bvec{y} ) \, P_{X^{n}|Y^{n}}(\tilde{\mathcal{A}}_{n}(\bvec{y}) \mid \bvec{y}) + \sum_{\bvec{y} \in \mathcal{Y}^{n}} P_{Y^{n}}( \bvec{y} ) \, P_{X^{n}|Y^{n}}(\mathcal{T}_{\gamma}^{(n)}(\bvec{y}) \mid \bvec{y}) - 1
\notag \\
& =
\sum_{\bvec{y} \in \mathcal{Y}^{n}} P_{Y^{n}}( \bvec{y} ) \, \Big( P_{X^{n}|Y^{n}}(\tilde{\mathcal{A}}_{n}(\bvec{y}) \mid \bvec{y}) + P_{X^{n}|Y^{n}}(\mathcal{T}_{\gamma}^{(n)}(\bvec{y}) \mid \bvec{y}) - 1 \Big)
\notag \\
& \overset{\mathclap{\text{(d)}}}{\le}
\sum_{\bvec{y} \in \mathcal{Y}^{n}} P_{Y^{n}}( \bvec{y} ) \, P_{X^{n}|Y^{n}}(\tilde{\mathcal{A}}_{n}(\bvec{y}) \cap \mathcal{T}_{\gamma}^{(n)}(\bvec{y}) \mid \bvec{y})
\notag \\
& \overset{\mathclap{\text{(e)}}}{\le}
\sum_{\bvec{y} \in \mathcal{V}_{n}} P_{Y^{n}}( \bvec{y} ) + \frac{ 1 - \varepsilon - \gamma }{ 2 } \sum_{\bvec{y} \in \mathcal{Y}^{n} \setminus \mathcal{V}_{n}} P_{Y^{n}}( \bvec{y} )
\notag \\
& =
P_{Y^{n}}( \mathcal{V}_{n} ) \, \left( \frac{ 1 + \varepsilon + \gamma }{ 2 } \right) + \frac{ 1 - \varepsilon - \gamma }{ 2 } ,
\end{align}
where
\begin{itemize}
\item
(a) follows from \eqref{eq:eps_n-expectation},
\item
(b) follows from the right-hand inequality of \eqref{eq:largest-A_cond2},
\item
(c) follows from \eqref{eq:Chebyshev},
\item
(d) follows from the fact that $\mathbb{P}( \mathcal{E}_{1} \cap \mathcal{E}_{2} ) \ge \mathbb{P}( \mathcal{E}_{1} ) + \mathbb{P}( \mathcal{E}_{2} ) - 1$ for two events $E_{1}$ and $E_{2}$, and
\item
(e) follows by the definition of $\mathcal{V}_{n}$ stated in \eqref{def:set-Vn}.
\end{itemize}
Thus, similar to \eqref{eq:reverse-Markov}, we obtain
\begin{align}
P_{Y^{n}}( \mathcal{V}_{n} )
\ge
\frac{ 1 - \varepsilon - \gamma }{ 1 + \varepsilon + \gamma } .
\label{eq:reverse-Markov_Vn}
\end{align}
Moreover, we see that
\begin{align}
\frac{ 1 - \varepsilon - \gamma }{ 2 }
& \overset{\mathclap{\text{(a)}}}{\le}
\sum_{\bvec{x} \in \tilde{\mathcal{A}}_{n}(\bvec{y}) \cap \mathcal{T}_{\gamma}^{(n)}(\bvec{y})} P_{X^{n}|Y^{n}}(\bvec{x} \mid \bvec{y})
\notag \\
& \overset{\mathclap{\text{(b)}}}{\le}
\sum_{\bvec{x} \in \tilde{\mathcal{A}}_{n}(\bvec{y}) \cap \mathcal{T}_{\gamma}^{(n)}(\bvec{y})} \exp\left( - n \, H(X \mid Y) + \sqrt{ \frac{ n \, U(X \mid Y) }{ \gamma } } \right)
\notag \\
& =
|\tilde{\mathcal{A}}_{n}(\bvec{y}) \cap \mathcal{T}_{\gamma}^{(n)}(\bvec{y})| \, \exp\left( - n \, H(X \mid Y) + \sqrt{ \frac{ n \, U(X \mid Y) }{ \gamma } } \right)
\label{eq:bound_card_An-yn_cap_Tdelta-yn}
\end{align}
for every $\bvec{y} \in \mathcal{V}_{n}$, where
\begin{itemize}
\item
(a) follows by the definition of $\mathcal{V}_{n}$ stated in \eqref{def:set-Vn}, and
\item
(b) follows from the right-hand inequality of \eqref{eq:bound_pmf_Teps}.
\end{itemize}
Furthermore, we get
\begin{align}
\sum_{\bvec{x} \in \tilde{\mathcal{A}}_{n}(\bvec{y}) \cap \mathcal{T}_{\gamma}^{(n)}(\bvec{y})} P_{X^{n}|Y^{n}}(\bvec{x} \mid \bvec{y})^{\alpha}
& \overset{\mathclap{\text{(a)}}}{\ge}
\sum_{\bvec{x} \in \tilde{\mathcal{A}}_{n}(\bvec{y}) \cap \mathcal{T}_{\gamma}^{(n)}(\bvec{y})} \exp\left( - \alpha \left( n \, H(X \mid Y) + \sqrt{ \frac{ n \, U(X \mid Y) }{ \gamma } } \right) \right)
\notag \\
& =
|\tilde{\mathcal{A}}_{n}(\bvec{y}) \cap \mathcal{T}_{\gamma}^{(n)}(\bvec{y})| \, \exp\left( - \alpha \left( n \, H(X \mid Y) + \sqrt{ \frac{ n \, U(X \mid Y) }{ \gamma } } \right) \right)
\notag \\
& \overset{\mathclap{\text{(b)}}}{\ge}
\frac{ 1 - \varepsilon - \gamma }{ 2 } \, \exp\left( (1 - \alpha) \, n \, H(X \mid Y) - (1 + \alpha) \sqrt{ \frac{ n \, U(X \mid Y) }{ \gamma } } \right)
\label{eq:inner_cond-smooth_positive-variance}
\end{align}
for every $\bvec{y} \in \mathcal{V}_{n}$, where
\begin{itemize}
\item
(a) follows from the left-hand inequality of \eqref{eq:bound_pmf_Teps}, and
\item
(b) follows from \eqref{eq:bound_card_An-yn_cap_Tdelta-yn}.
\end{itemize}
Now, since $0 < U(X \mid Y) < \infty$ implies that $0 < H(X \mid Y) < \infty$, one can choose an integer $n_{1} = n_{1}(\varepsilon, \delta, \alpha, H(X \mid Y), U(X \mid Y))$ so that
\begin{align}
(1 - \alpha) \, n \, H(X \mid Y) - (1 + \alpha) \sqrt{ \frac{ n \, U(X \mid Y) }{ \gamma } }
\ge
\log \frac{ 4 }{ 1 - \varepsilon - \gamma }
\label{eq:large_n1}
\end{align}
for every $n \ge n_{1}$.
Then, we observe that
\begin{align}
H_{\alpha}^{\varepsilon}(X^{n} \mid Y^{n})
& \overset{\mathclap{\text{(a)}}}{=}
\frac{ \alpha }{ 1 - \alpha } \log \left( \inf_{\delta_{n}( \cdot )} \sum_{\bvec{y} \in \mathcal{Y}^{n}} P_{Y^{n}}( \bvec{y} ) \left( \sum_{\bvec{x} \in \mathcal{A}_{n}(\bvec{y})} P_{X^{n}|Y^{n}}(\bvec{x} \mid \bvec{y})^{\alpha} + \Big( 1 - \delta_{n}(\bvec{y}) - P_{X^{n}|Y^{n}}(\mathcal{A}_{n}(\bvec{y}) \mid \bvec{y}) \Big)^{\alpha} \right)^{1/\alpha} \right)
\notag \\
& \ge
\frac{ \alpha }{ 1 - \alpha } \log \left( \inf_{\delta_{n}( \cdot )} \sum_{\bvec{y} \in \mathcal{V}_{n}} P_{Y^{n}}( \bvec{y} ) \left( \sum_{\bvec{x} \in \mathcal{A}_{n}(\bvec{y})} P_{X^{n}|Y^{n}}(\bvec{x} \mid \bvec{y})^{\alpha} + \Big( 1 - \delta_{n}(\bvec{y}) - P_{X^{n}|Y^{n}}(\mathcal{A}_{n}(\bvec{y}) \mid \bvec{y}) \Big)^{\alpha} \right)^{1/\alpha} \right)
\notag \\
& \overset{\mathclap{\text{(b)}}}{\ge}
\frac{ \alpha }{ 1 - \alpha } \log \left( \inf_{\delta_{n}( \cdot )} \sum_{\bvec{y} \in \mathcal{V}_{n}} P_{Y^{n}}( \bvec{y} ) \left( \sum_{x \in \tilde{\mathcal{A}}_{n}(\bvec{y})} P_{X^{n}|Y^{n}}(\bvec{x} \mid \bvec{y})^{\alpha} - 1 \right)^{1/\alpha} \right)
\notag \\
& \ge
\frac{ \alpha }{ 1 - \alpha } \log \left( \inf_{\delta_{n}( \cdot )} \sum_{\bvec{y} \in \mathcal{V}_{n}} P_{Y^{n}}( \bvec{y} ) \left( \sum_{x \in \tilde{\mathcal{A}}_{n}(\bvec{y}) \cap \mathcal{T}_{\gamma}^{(n)}(\bvec{y})} P_{X^{n}|Y^{n}}(\bvec{x} \mid \bvec{y})^{\alpha} - 1 \right)^{1/\alpha} \right)
\notag \\
& \overset{\mathclap{\text{(c)}}}{\ge}
\frac{ \alpha }{ 1 - \alpha } \log \left( \inf_{\delta_{n}( \cdot )} \sum_{\bvec{y} \in \mathcal{V}_{n}} P_{Y^{n}}( \bvec{y} ) \left( \frac{ 1 - \varepsilon - \gamma }{ 2 } \, \exp\left( (1 - \alpha) \, n \, H(X \mid Y) - (1 + \alpha) \sqrt{ \frac{ n \, U(X \mid Y) }{ \gamma } } \right) - 1 \right)^{1/\alpha} \right)
\notag \\
& \overset{\mathclap{\text{(d)}}}{\ge}
\frac{ \alpha }{ 1 - \alpha } \log \left( \inf_{\epsilon_{n}( \cdot )} \sum_{\bvec{y} \in \mathcal{V}_{n}} P_{Y^{n}}( \bvec{y} ) \left( \frac{ 1 - \varepsilon - \gamma }{ 4 } \, \exp\left( (1 - \alpha) \, n \, H(X \mid Y) - (1 + \alpha) \sqrt{ \frac{ n \, U(X \mid Y) }{ \gamma } } \right) \right)^{1/\alpha} \right)
\notag \\
& =
n \, H(X \mid Y) - \frac{ 1 + \alpha }{ 1 - \alpha } \sqrt{ \frac{ n \, U(X \mid Y) }{ \gamma } } + \frac{ 1 }{ 1 - \alpha } \log \frac{ 1 - \varepsilon - \gamma }{ 4 } + \frac{ \alpha }{ 1 - \alpha } \log \Big( \inf_{\epsilon_{n}(\cdot)} P_{Y^{n}}( \mathcal{V}_{n} ) \Big)
\notag \\
& \overset{\mathclap{\text{(e)}}}{\ge}
n \, H(X \mid Y) - \frac{ 1 + \alpha }{ 1 - \alpha } \sqrt{ \frac{ n \, U(X \mid Y) }{ \gamma } } + \frac{ 1 }{ 1 - \alpha } \log \frac{ 1 - \varepsilon - \gamma }{ 4 } + \frac{ \alpha }{ 1 - \alpha } \log \frac{ 1 - \varepsilon - \gamma }{ 1 + \varepsilon + \gamma }
\label{eq:LB_cond-smooth}
\end{align}
for sufficiently large $n \ge n_{1}$, where
\begin{itemize}
\item
(a) follows from \eqref{eq:Kuzuoka_identity},
\item
(b) follows from \eqref{def:largest-A_cond_cond2_alpha},
\item
(c) follows from \eqref{eq:inner_cond-smooth_positive-variance},
\item
(d) follows by the choice of $n_{1}$ stated in \eqref{eq:large_n1}, and
\item
(e) follows from \eqref{eq:reverse-Markov_Vn}.
\end{itemize}
This completes the proof of the lower bound part of \eqref{eq:2nd_avg}.
\hfill\IEEEQEDhere

\section{Proof of \thref{th:2nd_max}---Zero Variance $V(X \mid Y) = 0$}
\label{app:2nd_max_zero-variance}

\subsection{Proof of Upper Bound Part of \eqref{eq:zero-variance_2nd_max}}
\label{app:2nd_max_zero-variance_direct}

Consider the conditional sub-probability distribution $\tilde{Q}_{X^{n}|Y^{n}}$ given as
\begin{align}
\tilde{Q}_{X^{n}|Y^{n}}(\bvec{x} \mid \bvec{y})
=
(1 - \varepsilon) \, P_{X^{n}|Y^{n}}(\bvec{x} \mid \bvec{y})
\label{def:cond_tildeQn}
\end{align}
for each $(\bvec{x}, \bvec{y}) \in \mathcal{X}^{n} \times \mathcal{Y}^{n}$.
Then, we observe that
\begin{align}
\check{H}_{\alpha}^{\varepsilon}(X^{n} \mid Y^{n})
& =
\frac{ \alpha }{ 1 - \alpha } \log \left( \sum_{\bvec{y} \in \mathcal{Y}} P_{Y^{n}}( \bvec{y} ) \, \inf_{Q \in \mathcal{B}_{\mathcal{X}^{n}}^{\varepsilon}( P_{X^{n}|Y^{n}=\bvec{y}} )} \left( \sum_{\bvec{x} \in \mathcal{X}^{n}} Q( \bvec{x} )^{\alpha} \right)^{1/\alpha} \right)
\notag \\
& \overset{\mathclap{\text{(a)}}}{\le}
\frac{ \alpha }{ 1 - \alpha } \log \left( \sum_{\bvec{y} \in \mathcal{Y}^{n}} P_{Y^{n}}( \bvec{y} ) \, \left( \sum_{\bvec{x} \in \mathcal{X}^{n}} \tilde{Q}_{X^{n}|Y^{n}}(\bvec{x} \mid \bvec{y})^{\alpha} \right)^{1/\alpha} \right)
\notag \\
& \overset{\mathclap{\text{(b)}}}{=}
\frac{ \alpha }{ 1 - \alpha } \log \left( \sum_{\bvec{y} \in \mathcal{Y}^{n}} P_{Y^{n}}( \bvec{y} ) \, \left( \sum_{\bvec{x} \in \mathcal{X}^{n}} P_{X^{n}|Y^{n}}(\bvec{x} \mid \bvec{y})^{\alpha} \right)^{1/\alpha} \right) + \frac{ \alpha }{ 1 - \alpha } \log (1 - \varepsilon)
\notag \\
& \overset{\mathclap{\text{(c)}}}{=}
\frac{ \alpha }{ 1 - \alpha } \log \left( \sum_{\bvec{y} \in \mathcal{Y}^{n}} P_{Y^{n}}( \bvec{y} ) \, \exp\left( \frac{ 1 - \alpha }{ \alpha } \sum_{i = 1}^{n} H( P_{X_{i}|Y_{i}=y_{i}} ) \right) \right) + \frac{ \alpha }{ 1 - \alpha } \log (1 - \varepsilon)
\notag \\
& \overset{\mathclap{\text{(d)}}}{=}
\frac{ n \, \alpha }{ 1 - \alpha } \log \left( \sum_{y \in \mathcal{Y}} P_{Y}( y ) \, \exp\left( \frac{ 1 - \alpha }{ \alpha } \, H( P_{X|Y=y} ) \right) \right) + \frac{ \alpha }{ 1 - \alpha } \log (1 - \varepsilon)
\notag \\
& \overset{\mathclap{\text{(e)}}}{=}
n \, H^{(\alpha)}(X \mid Y) + \frac{ \alpha }{ 1 - \alpha } \log (1 - \varepsilon) ,
\end{align}
where
\begin{itemize}
\item
(a) follows from the fact that $\tilde{Q}_{X^{n}|Y^{n}}(\cdot \mid \bvec{y}) \in \mathcal{B}_{\mathcal{X}^{n}}^{\varepsilon}( P_{X^{n}|Y^{n} = \bvec{y}} )$ for each $\bvec{y} \in \mathcal{Y}^{n}$,
\item
(b) follows by the definition of $\tilde{Q}_{X^{n}|Y^{n}}$ stated in \eqref{def:cond_tildeQn},
\item
(c) follows from \eqref{eq:zero-variance_cond-probab},
\item
(d) follows from the fact that $(X_{1}, Y_{1}), \dots, (X_{n}, Y_{n})$ are $n$ i.i.d.\ copies of $(X, Y)$, and
\item
(e) follows by the definition of $H^{(\alpha)}(X \mid Y)$ stated in \eqref{eq:cond-Shannon_KN}.
\end{itemize}
This completes the proof of the upper bound part of \eqref{eq:zero-variance_2nd_max}.
\hfill\IEEEQEDhere

\subsection{Proof of Lower Bound Part of \eqref{eq:zero-variance_2nd_max}}
\label{app:2nd_max_zero-variance_converse}

Firstly, suppose that $U(X \mid Y) = 0$.
Then, it is clear that
\begin{align}
P_{X^{n}|Y^{n}}(\bvec{x} \mid \bvec{y})
=
\begin{dcases}
\exp\Big( - n \, H(X \mid Y) \Big)
& \mathrm{if} \ P_{X^{n}|Y^{n}}(\bvec{x} \mid \bvec{y}) > 0 ,
\\
0
& \mathrm{if} \ P_{X^{n}|Y^{n}}(\bvec{x} \mid \bvec{y}) = 0
\end{dcases}
\label{eq:zero-total-variance}
\end{align}
for every $(\bvec{x}, \bvec{y}) \in \mathcal{X}^{n} \times \mathcal{Y}^{n}$.
For the sake of brevity, denote by
\begin{align}
\mathcal{B}_{n}( \bvec{y} )
\coloneqq
\mathcal{A}_{X^{n}|Y^{n}}^{\varepsilon}( \bvec{y} )
\label{def:abbrev-B-XnYn}
\end{align}
for each $\bvec{y} \in \mathcal{Y}^{n}$; see \eqref{eq:largest-A_cond1} and \eqref{eq:largest-A_cond2} for the definition of the right-hand side of \eqref{def:abbrev-B-XnYn} by replacing $P_{X}( \cdot )$ by $P_{X^{n}|Y^{n}}(\cdot \mid \bvec{y})$.
Moreover, define
\begin{align}
\tilde{\mathcal{B}}_{n}( \bvec{y} )
\coloneqq
\mathcal{B}_{n}( \bvec{y} ) \cup \{ \check{\bvec{x}}(\bvec{y}) \} ,
\label{def:abbrev-tildeB-XnYn}
\end{align}
where $\check{\bvec{x}}( \bvec{y} ) \in \mathcal{X}^{n}$ is chosen so that
\begin{align}
\check{\bvec{x}}( \bvec{y} )
\in
\argmax\limits_{\bvec{x} \in \mathcal{X}^{n} \setminus \mathcal{B}_{n}( \bvec{y} )} P_{X^{n}|Y^{n}}(\bvec{x} \mid \bvec{y}) .
\end{align}
for each $\bvec{y} \in \mathcal{Y}^{n}$.
We get
\begin{align}
1 - \varepsilon
& \overset{\mathclap{\text{(a)}}}{\le}
\sum_{\bvec{x} \in \tilde{\mathcal{B}}_{n}(\bvec{y})} P_{X^{n}|Y^{n}}(\bvec{x} \mid \bvec{y})
\notag \\
& \overset{\mathclap{\text{(b)}}}{=}
\sum_{\bvec{x} \in \tilde{\mathcal{B}}_{n}(\bvec{y})} \exp\Big( - n \, H(X \mid Y) \Big)
\notag \\
& =
|\tilde{\mathcal{B}}_{n}( \bvec{y} )| \exp\Big( - n \, H(X \mid Y) \Big)
\label{eq:card_event_An_total}
\end{align}
for every $\bvec{y} \in \mathcal{Y}^{n}$, where
\begin{itemize}
\item
(a) follows from the right-hand inequality of \eqref{eq:largest-A_cond2}, and
\item
(b) follows from \eqref{eq:largest-A_cond1} and \eqref{eq:zero-total-variance}.
\end{itemize}
Moreover, we see that
\begin{align}
\sum_{\bvec{x} \in \tilde{\mathcal{B}}_{n}(\bvec{y})} P_{X^{n}|Y^{n}}(\bvec{x} \mid \bvec{y})^{\alpha}
& \overset{\mathclap{\text{(a)}}}{=}
\sum_{\bvec{x} \in \tilde{\mathcal{B}}_{n}(\bvec{y})} \exp\Big( - \alpha \, n \, H(X \mid Y) \Big)
\notag \\
& =
|\tilde{\mathcal{B}}_{n}(\bvec{y})| \, \exp\Big( - \alpha \, n \, H(X \mid Y) \Big)
\notag \\
& \overset{\mathclap{\text{(b)}}}{\ge}
(1 - \varepsilon) \, \exp\Big( (1 - \alpha) \, n \, H(X \mid Y) \Big)
\qquad (\mathrm{a.s.})
\label{eq:moment_LB_max_total}
\end{align}
for every $\bvec{y} \in \mathcal{Y}^{n}$, where
\begin{itemize}
\item
(a) follows from \eqref{eq:largest-A_cond1} and \eqref{eq:zero-total-variance}, and
\item
(b) follows from \eqref{eq:card_event_An_total}.
\end{itemize}
Since $H(X \mid Y) > 0$, one can choose an integer $n_{0}$ as
\begin{align}
n_{0}
\coloneqq
\left\lceil \frac{ 1 }{ (1 - \alpha) \, H(X \mid Y) } \log \frac{ 2 }{ 1 - \varepsilon } \right\rceil .
\label{eq:large-n0_zero-total-variance}
\end{align}
Then, we observe that
\begin{align}
\check{H}_{\alpha}^{\varepsilon}(X^{n} \mid Y^{n})
& \overset{\mathclap{\text{(a)}}}{=}
\frac{ \alpha }{ 1 - \alpha } \log \left( \sum_{\bvec{y} \in \mathcal{Y}^{n}} P_{Y^{n}}( \bvec{y} ) \, \left( \sum_{\bvec{x} \in \mathcal{B}_{n}(\bvec{y})} P_{X^{n}|Y^{n}}(\bvec{x} \mid \bvec{y})^{\alpha} + \Big( 1 - \varepsilon - P_{X^{n}|Y^{n}}(\mathcal{B}_{n}(\bvec{y}) \mid \bvec{y}) \Big)^{\alpha} \right)^{1/\alpha} \right)
\notag \\
& \overset{\mathclap{\text{(b)}}}{\ge}
\frac{ \alpha }{ 1 - \alpha } \log \left( \sum_{\bvec{y} \in \mathcal{Y}^{n}} P_{Y^{n}}( \bvec{y} ) \, \left( \sum_{\bvec{x} \in \tilde{\mathcal{B}}_{n}(\bvec{y})} P_{X^{n}|Y^{n}}(\bvec{x} \mid \bvec{y})^{\alpha} - 1 \right)^{1/\alpha} \right)
\notag \\
& \overset{\mathclap{\text{(c)}}}{\ge}
\frac{ \alpha }{ 1 - \alpha } \log \left( \sum_{\bvec{y} \in \mathcal{Y}^{n}} P_{Y^{n}}( \bvec{y} ) \, \left( (1 - \varepsilon) \, \exp\Big( (1 - \alpha) \, n \, H(X \mid Y) \Big) - 1 \right)^{1/\alpha} \right)
\notag \\
& \overset{\mathclap{\text{(d)}}}{\ge}
\frac{ \alpha }{ 1 - \alpha } \log \left( \sum_{\bvec{y} \in \mathcal{Y}^{n}} P_{Y^{n}}( \bvec{y} ) \, \left( \frac{ 1 - \varepsilon }{ 2 } \right)^{1/\alpha} \, \exp\left( \frac{ 1 - \alpha }{ \alpha } \, n \, H(X \mid Y) \right) \right)
\notag \\
& =
n \, H(X \mid Y) + \frac{ 1 }{ 1 - \alpha } \log \frac{ 1 - \varepsilon }{ 2 }
\notag \\
& \overset{\mathclap{\text{(e)}}}{=}
n \, H^{(a)}(X \mid Y) + \frac{ 1 }{ 1 - \alpha } \log \frac{ 1 - \varepsilon }{ 2 }
\label{eq:LB_tilde_zero-total-variance}
\end{align}
for sufficiently large $n \ge n_{0}$, where
\begin{itemize}
\item
(a) follows from \eqref{eq:check_identity},
\item
(b) follows from the right-hand inequality of \eqref{eq:largest-A_cond2}, i.e.,
\begin{align}
\Big( 1 - \varepsilon - P_{X^{n}|Y^{n}}(\mathcal{B}_{n}(\bvec{y}) \mid \bvec{y}) \Big)^{\alpha}
\ge
P_{X^{n}|Y^{n}}(\check{\bvec{x}}(\bvec{y}) \mid \bvec{y}) - 1 ,
\label{def:largest-B_cond_cond2_alpha}
\end{align}
\item
(c) follows from \eqref{eq:moment_LB_max_total},
\item
(d) follows from the choice of $n_{0}$ stated in \eqref{eq:large-n0_zero-total-variance}, and
\item
(e) follows from \propref{prop:KN-avg} and the hypothesis that $U(X \mid Y) = V(X \mid Y) = 0$.
\end{itemize}
This completes the proof of the lower bound part of \eqref{eq:zero-variance_2nd_max} in the case where $U(X \mid Y) = 0$.

Secondly, suppose that $U(X \mid Y) > 0$.
Defining a subset $\mathcal{E}_{n}$ of $\mathcal{Y}^{n}$ by
\begin{align}
\mathcal{E}_{n}
& \coloneqq
\Big\{ \bvec{y} \ \Big| \ H( P_{X^{n}|Y^{n} = \bvec{y}} ) \ge n \, H(X \mid Y) - \sqrt{ n \, (U(X \mid Y) - V(X \mid Y)) } \Big\} ,
\label{def:event_En}
\end{align}
we see that
\begin{align}
P_{Y^{n}}( \mathcal{E}_{n} )
& \overset{\mathclap{\text{(a)}}}{\ge}
1 - \frac{ \sum_{\bvec{y} \in \mathcal{Y}^{n}} P_{Y^{n}}( \bvec{y} ) \, (H( P_{X^{n}|Y^{n} = \bvec{y}} ) - H(X^{n} \mid Y^{n}))^{2} }{ \sum_{\bvec{y} \in \mathcal{Y}^{n}} P_{Y^{n}}( \bvec{y} ) \, (H( P_{X^{n}|Y^{n} = \bvec{y}} ) - H(X^{n} \mid Y^{n}))^{2} + (U(X^{n} \mid Y^{n}) - V(X^{n} \mid Y^{n})) }
\notag \\
& \overset{\mathclap{\text{(b)}}}{=}
\frac{ 1 }{ 2 } ,
\label{eq:Cantelli_En}
\end{align}
where
\begin{itemize}
\item
(a) follows from Cantelli's inequality (or the one-sided Chebyshev inequality), and
\item
(b) follows from the law of total variance.
\end{itemize}
Similar to \eqref{eq:card_event_An_total}, we see that
\begin{align}
1 - \varepsilon
& \overset{\mathclap{\text{(a)}}}{\le}
\sum_{\bvec{x} \in \tilde{\mathcal{B}}_{n}( \bvec{x} )} P_{X^{n}|Y^{n}}(\bvec{x} \mid \bvec{y})
\notag \\
& \overset{\mathclap{\text{(b)}}}{=}
\sum_{\bvec{x} \in \tilde{\mathcal{B}}_{n}( \bvec{x} )} \exp\Big( - H( P_{X^{n}|Y^{n} = \bvec{y}} ) \Big)
\notag \\
& =
|\tilde{\mathcal{B}}_{n}( \bvec{x} )| \, \exp\Big( - H( P_{X^{n}|Y^{n} = \bvec{y}} ) \Big)
\label{eq:card_event_An}
\end{align}
for every $\bvec{y} \in \mathcal{Y}^{n}$, where
\begin{itemize}
\item
(a) follows from the right-hand inequality of \eqref{eq:largest-A_cond2}, and
\item
(b) follows from \eqref{eq:largest-A_cond1} and \eqref{eq:zero-variance_cond-probab}.
\end{itemize}
Moreover, we observe that
\begin{align}
\sum_{\bvec{x} \in \tilde{\mathcal{B}}_{n}(\bvec{y})} P_{X^{n}|Y^{n}}(\bvec{x} \mid \bvec{y})^{\alpha}
& \overset{\mathclap{\text{(a)}}}{=}
\sum_{\bvec{x} \in \tilde{\mathcal{B}}_{n}(\bvec{y})} \exp\Big( - \alpha \, H( P_{X^{n}|Y^{n} = \bvec{y}} ) \Big)
\notag \\
& =
|\tilde{\mathcal{B}}_{n}(\bvec{y})| \, \exp\Big( - \alpha \, H( P_{X^{n}|Y^{n} = \bvec{y}} ) \Big)
\notag \\
& \overset{\mathclap{\text{(b)}}}{\ge}
(1 - \varepsilon) \, \exp\Big( (1 - \alpha) \, H( P_{X^{n}|Y^{n} = \bvec{y}} ) \Big)
\label{eq:moment_LB_max}
\end{align}
for every $\bvec{y} \in \mathcal{Y}^{n}$, where
\begin{itemize}
\item
(a) follows from \eqref{eq:largest-A_cond1} and \eqref{eq:zero-variance_cond-probab}, and
\item
(b) follows from \eqref{eq:card_event_An}.
\end{itemize}
Since $U(X \mid Y) > 0$ implies that $H(X \mid Y) > 0$, one can choose an integer $n_{0}$ so that
\begin{align}
n \, H(X \mid Y) - \sqrt{ n \, (U(X \mid Y) - V(X \mid Y)) }
\ge
\frac{ 1 }{ 1 - \alpha } \log \frac{ 2 }{ 1 - \varepsilon }
\label{eq:large-n0_max}
\end{align}
for every $n \ge n_{0}$.
Then, we observe that
\begin{align}
\check{H}_{\alpha}^{\varepsilon}(X^{n} \mid Y^{n})
& \overset{\mathclap{\text{(a)}}}{\ge}
\frac{ \alpha }{ 1 - \alpha } \log \left( \sum_{\bvec{y} \in \mathcal{Y}^{n}} P_{Y^{n}}( \bvec{y} ) \, \left( \sum_{\bvec{x} \in \tilde{\mathcal{B}}_{n}(\bvec{y})} P_{X^{n}|Y^{n}}(\bvec{x} \mid \bvec{y})^{\alpha} - 1 \right)^{1/\alpha} \right)
\notag \\
& \overset{\mathclap{\text{(b)}}}{\ge}
\frac{ \alpha }{ 1 - \alpha } \log \left( \sum_{\bvec{y} \in \mathcal{Y}^{n}} P_{Y^{n}}( \bvec{y} ) \, \left( (1 - \varepsilon) \, \exp\Big( (1 - \alpha) \, H( P_{X^{n}|Y^{n} = \bvec{y}} ) \Big) - 1 \right)^{1/\alpha} \right)
\notag \\
& \ge
\frac{ \alpha }{ 1 - \alpha } \log \left( \sum_{\bvec{y} \in \mathcal{E}_{n}} P_{Y^{n}}( \bvec{y} ) \, \left( (1 - \varepsilon) \, \exp\Big( (1 - \alpha) \, H( P_{X^{n}|Y^{n} = \bvec{y}} ) \Big) - 1 \right)^{1/\alpha} \right)
\notag \\
& \overset{\mathclap{\text{(c)}}}{\ge}
\frac{ \alpha }{ 1 - \alpha } \log \left( \sum_{\bvec{y} \in \mathcal{E}_{n}} P_{Y^{n}}( \bvec{y} ) \, \left( \frac{ 1 - \varepsilon }{ 2 } \right)^{1/\alpha} \, \exp\left( \frac{ 1 - \alpha }{ \alpha } \, H( P_{X^{n}|Y^{n} = \bvec{y}} ) \right) \right)
\notag \\
& =
\frac{ \alpha }{ 1 - \alpha } \log \left( \sum_{\bvec{y} \in \mathcal{E}_{n}} P_{Y^{n}}( \bvec{y} ) \, \exp\left( \frac{ 1 - \alpha }{ \alpha } \, H( P_{X^{n}|Y^{n} = \bvec{y}} ) \right) \right) + \frac{ 1 }{ 1 - \alpha } \log \frac{ 1 - \varepsilon }{ 2 }
\notag \\
& =
\frac{ \alpha }{ 1 - \alpha } \log \left( \sum_{\bvec{y} \in \mathcal{Y}^{n}} P_{Y^{n}}( \bvec{y} ) \, \exp\left( \frac{ 1 - \alpha }{ \alpha } \, H( P_{X^{n}|Y^{n} = \bvec{y}} ) \right)
\right. \notag \\
& \left. \vphantom{\sum_{y \in \mathcal{Y}}} \qquad \qquad \qquad
{} - \sum_{\bvec{y} \in \mathcal{Y}^{n} \setminus \mathcal{E}_{n}} P_{Y^{n}}( \bvec{y} ) \, \exp\left( \frac{ 1 - \alpha }{ \alpha } \, H( P_{X^{n}|Y^{n} = \bvec{y}} ) \right) \right) + \frac{ 1 }{ 1 - \alpha } \log \frac{ 1 - \varepsilon }{ 2 }
\notag \\
& \overset{\mathclap{\text{(d)}}}{=}
\frac{ \alpha }{ 1 - \alpha } \log \left( \left( \sum_{y \in \mathcal{Y}} P_{Y}( y ) \, \exp\left( \frac{ 1 - \alpha }{ \alpha } \, H( P_{X|Y = y} ) \right) \right)^{n}
\right. \notag \\
& \left. \vphantom{\sum_{y \in \mathcal{Y}}} \qquad \qquad \qquad
{} - \sum_{\bvec{y} \in \mathcal{Y}^{n} \setminus \mathcal{E}_{n}} P_{Y^{n}}( \bvec{y} ) \, \exp\left( \frac{ 1 - \alpha }{ \alpha } \, H( P_{X^{n}|Y^{n} = \bvec{y}} ) \right) \right) + \frac{ 1 }{ 1 - \alpha } \log \frac{ 1 - \varepsilon }{ 2 }
\notag \\
& \overset{\mathclap{\text{(e)}}}{\ge}
\frac{ \alpha }{ 1 - \alpha } \log \left( \left( \sum_{y \in \mathcal{Y}} P_{Y}( y ) \, \exp\left( \frac{ 1 - \alpha }{ \alpha } \, H( P_{X|Y = y} ) \right) \right)^{n}
\right. \notag \\
& \left. \vphantom{\sum_{y \in \mathcal{Y}}} \qquad \qquad \qquad
{} - \sum_{\bvec{y} \in \mathcal{Y}^{n} \setminus \mathcal{E}_{n}} P_{Y^{n}}( \bvec{y} ) \, \exp\left( \frac{ 1 - \alpha }{ \alpha } \, n \, H(X \mid Y) \right) \right) + \frac{ 1 }{ 1 - \alpha } \log \frac{ 1 - \varepsilon }{ 2 }
\notag \\
& =
\frac{ \alpha }{ 1 - \alpha } \log \left( \left( \sum_{y \in \mathcal{Y}} P_{Y}( y ) \, \exp\left( \frac{ 1 - \alpha }{ \alpha } \, H( P_{X|Y = y} ) \right) \right)^{n}
\right. \notag \\
& \left. \vphantom{\sum_{y \in \mathcal{Y}}} \qquad \qquad \qquad
{} - (1 - P_{Y^{n}}(\mathcal{E}_{n})) \, \exp\left( \frac{ 1 - \alpha }{ \alpha } \, n \, H(X \mid Y) \right) \right) + \frac{ 1 }{ 1 - \alpha } \log \frac{ 1 - \varepsilon }{ 2 }
\notag \\
& \overset{\mathclap{\text{(f)}}}{\ge}
\frac{ \alpha }{ 1 - \alpha } \log \left( \left( \sum_{y \in \mathcal{Y}} P_{Y}( y ) \, \exp\left( \frac{ 1 - \alpha }{ \alpha } \, H( P_{X|Y = y} ) \right) \right)^{n} - \frac{ 1 }{ 2 } \, \exp\left( \frac{ 1 - \alpha }{ \alpha } \, n \, H(X \mid Y) \right) \right) + \frac{ 1 }{ 1 - \alpha } \log \frac{ 1 - \varepsilon }{ 2 }
\notag \\
& \overset{\mathclap{\text{(g)}}}{\ge}
\frac{ \alpha }{ 1 - \alpha } \log \left( \frac{ 1 }{ 2 } \left( \sum_{y \in \mathcal{Y}} P_{Y}( y ) \, \exp\left( \frac{ 1 - \alpha }{ \alpha } \, H( P_{X|Y = y} ) \right) \right)^{n} \right) + \frac{ 1 }{ 1 - \alpha } \log \frac{ 1 - \varepsilon }{ 2 }
\notag \\
& =
n \, H^{(\alpha)}(X \mid Y) + \frac{ \alpha }{ 1 - \alpha } \log \frac{ 1 }{ 2 } + \frac{ 1 }{ 1 - \alpha } \log \frac{ 1 - \varepsilon }{ 2 }
\label{eq:LB_tilde_zero-cond-variance}
\end{align}
for sufficiently large $n \ge n_{0}$, where
\begin{itemize}
\item
(a) follows as in Steps~(a) and~(b) of \eqref{eq:LB_tilde_zero-total-variance},
\item
(b) follows from \eqref{eq:moment_LB_max},
\item
(c) follows by the definition of $\mathcal{E}_{n}$ stated in \eqref{def:event_En} and the choice of $n_{0}$ stated in \eqref{eq:large-n0_max},
\item
(d) follows from the fact that $(X_{1}, Y_{1}), \dots, (X_{n}, Y_{n})$ are i.i.d.\ copies of $(X, Y)$,
\item
(e) follows by the definition of $\mathcal{E}_{n}$ stated in \eqref{def:event_En},
\item
(f) follows from \eqref{eq:Cantelli_En}, and
\item
(g) follows from Jensen's inequality.
\end{itemize}
This completes the proof of the lower bound part of \eqref{eq:zero-variance_2nd_max} in the case where $U(X \mid Y) > 0$.
\hfill\IEEEQEDhere

\section{Proof of \thref{th:2nd_max}---Positive Variance $0 < V(X \mid Y) < \infty$}

\subsection{Proof of Upper Bound Part of \eqref{eq:2nd_max}}
\label{app:2nd_max_direct}

For each $\bvec{y} \in \mathcal{Y}^{n}$, consider a subset $\mathcal{D}_{\varepsilon}^{(n)}(\bvec{y})$ of $\mathcal{X}^{n}$ given by
\begin{align}
\mathcal{D}_{\varepsilon}^{(n)}( \bvec{y} )
\coloneqq
\left\{ \bvec{x} \ \middle| \ \left| \log \frac{ 1 }{ P_{X^{n}|Y^{n}}(\bvec{x} \mid \bvec{y}) } - H( P_{X^{n}|Y^{n} = \bvec{y}} ) \right| \le \sqrt{ \frac{ V( P_{X^{n}|Y^{n} = \bvec{y}} ) }{ \varepsilon } } \right\} .
\label{eq:event_Dn-xn}
\end{align}
Note that
\begin{align}
\exp\left( - H( P_{X^{n}|Y^{n} = \bvec{y}} ) - \sqrt{ \frac{ V(P_{X^{n}|Y^{n} = \bvec{y}}) }{ \varepsilon } } \right)
\le
P_{X^{n}|Y^{n}}(\bvec{x} \mid \bvec{y})
\le
\exp\left( - H( P_{X^{n}|Y^{n} = \bvec{y}} ) + \sqrt{ \frac{ V(P_{X^{n}|Y^{n} = \bvec{y}}) }{ \varepsilon } } \right)
\label{eq:bounds_probab_Deps}
\end{align}
whenever $\bvec{x} \in \mathcal{D}_{\varepsilon}^{(n)}( \bvec{y} )$.
By the left-hand inequality of \eqref{eq:bounds_probab_Deps}, it can be verified by the same way as \eqref{eq:bound_card-cond_typical1} that
\begin{align}
|\mathcal{D}_{\varepsilon}^{(n)}( \bvec{y} )|
& \le
\exp\left( H( P_{X^{n}|Y^{n} = \bvec{y}} ) + \sqrt{ \frac{ V(P_{X^{n}|Y^{n} = \bvec{y}}) }{ \varepsilon } } \right)
\label{eq:card-Deps}
\end{align}
for every $\bvec{y} \in \mathcal{Y}^{n}$.
It follows from Chebyshev's inequality that
\begin{align}
P_{X^{n}|Y^{n}}( \mathcal{D}_{\varepsilon}^{(n)}(\bvec{y}) \mid \bvec{y} )
& \ge
1 - \varepsilon
\label{eq:Chebyshev_Deps}
\end{align}
for every $\bvec{y} \in \mathcal{Y}^{n}$ in which $V( P_{X^{n}|Y^{n} = \bvec{y}} ) > 0$.
On the other hand, it follows from \eqref{eq:zero-variance_cond-probab} that
\begin{align}
P_{X^{n}|Y^{n}}( \mathcal{D}_{\varepsilon}^{(n)}(\bvec{y}) \mid \bvec{y} )
=
1
\label{eq:non-Chebyshev_Deps}
\end{align}
for every $\bvec{y} \in \mathcal{Y}^{n}$ in which $V( P_{X^{n}|Y^{n} = \bvec{y}} ) = 0$.
Hence, we have
\begin{align}
\check{H}_{\alpha}^{\varepsilon}(X^{n} \mid Y^{n})
& \overset{\mathclap{\text{(a)}}}{=}
\frac{ \alpha }{ 1 - \alpha } \log \left( \sum_{\bvec{y} \in \mathcal{Y}^{n}} P_{Y^{n}}( \bvec{y} ) \inf_{Q \in \mathcal{B}_{\mathcal{X}^{n}}^{\varepsilon}( P_{X^{n}|Y^{n}=\bvec{y}} )} \left( \sum_{\bvec{x} \in \mathcal{X}^{n}} Q( \bvec{x} )^{\alpha} \right)^{1/\alpha} \right)
\notag \\
& \overset{\mathclap{\text{(b)}}}{\le}
\frac{ \alpha }{ 1 - \alpha } \log \left( \sum_{\bvec{y} \in \mathcal{Y}^{n}} P_{Y^{n}}( \bvec{y} ) \left( \sum_{\bvec{x} \in \mathcal{D}_{\varepsilon}^{(n)}(\bvec{y})} P_{X^{n}|Y^{n}}(\bvec{x} \mid \bvec{y})^{\alpha} \right)^{1/\alpha} \right)
\notag \\
& \overset{\mathclap{\text{(c)}}}{\le}
\frac{ \alpha }{ 1 - \alpha } \log \left( \sum_{\bvec{y} \in \mathcal{Y}^{n}} P_{Y^{n}}( \bvec{y} ) \left( \sum_{\bvec{x} \in \mathcal{D}_{\varepsilon}^{(n)}(\bvec{y})} \exp\left( - \alpha \left( H( P_{X^{n}|Y^{n}=\bvec{y}} ) - \sqrt{ \frac{ V( P_{X^{n}|Y^{n} = \bvec{y}} ) }{ \varepsilon } } \right) \right) \right)^{1/\alpha} \right)
\notag \\
& \overset{\mathclap{\text{(d)}}}{\le}
\frac{ \alpha }{ 1 - \alpha } \log \left( \sum_{\bvec{y} \in \mathcal{Y}^{n}} P_{Y^{n}}( \bvec{y} ) \, \exp\left( \frac{ 1 - \alpha }{ \alpha } \, H( P_{X^{n}|Y^{n} = \bvec{y}} ) + \frac{ 1 + \alpha }{ \alpha } \sqrt{ \frac{ V( P_{X^{n}|Y^{n}} ) }{ \varepsilon } } \right) \right)
\notag \\
& \overset{\mathclap{\text{(e)}}}{\le}
\frac{ \alpha }{ 1 - \alpha } \log \left( \sum_{\bvec{y} \in \mathcal{Y}^{n}} P_{Y^{n}}( \bvec{y} ) \, \exp\left( \frac{ 1 - \alpha }{ \alpha } \, H( P_{X^{n}|Y^{n} = \bvec{y}} ) \right) \right) + \frac{ 1 + \alpha }{ 1 - \alpha } \sqrt{ \frac{ n \, \sup_{y \in \mathcal{Y}} V( P_{X|Y = y} ) }{ \varepsilon } }
\notag \\
& \overset{\mathclap{\text{(f)}}}{=}
\frac{ n \, \alpha }{ 1 - \alpha } \log \left( \sum_{y \in \mathcal{Y}} P_{Y}( y ) \, \exp\left( \frac{ 1 - \alpha }{ \alpha } \, H( P_{X|Y = y} ) \right) \right) + \frac{ 1 + \alpha }{ 1 - \alpha } \sqrt{ \frac{ n \, \sup_{y \in \mathcal{Y}} V( P_{X|Y = y} ) }{ \varepsilon } }
\notag \\
& =
n \, H^{(\alpha)}(X \mid Y) + \frac{ 1 + \alpha }{ 1 - \alpha } \sqrt{ \frac{ n \, \sup_{y \in \mathcal{Y}} V( P_{X|Y = y} ) }{ \varepsilon } }
\end{align}
where
\begin{itemize}
\item
(a) follows by the definition of $\check{H}_{\alpha}^{\varepsilon}(X^{n} \mid Y^{n})$ stated in \eqref{def:check},
\item
(b) follows from \eqref{eq:Chebyshev_Deps} and \eqref{eq:non-Chebyshev_Deps}
\item
(c) follows from the right-hand inequality of \eqref{eq:bounds_probab_Deps},
\item
(d) follows from \eqref{eq:card-Deps},
\item
(e) follows from the hypothesis that $\sup_{y \in \mathcal{Y}} V( P_{X|Y=y} ) < \infty$, and
\item
(f) follows from the fact that $(X_{1}, Y_{1}), \dots, (X_{n}, Y_{n})$ are i.i.d.\ copies of $(X, Y)$.
\end{itemize}
This completes the proof of the upper bound part of \eqref{eq:2nd_max}.
\hfill\IEEEQEDhere

\subsection{Proof of Lower Bound Part of \eqref{eq:2nd_max}}
\label{app:2nd_max_converse}

Recall that $\mathcal{B}_{n}(\bvec{y})$ and $\tilde{\mathcal{B}}_{n}(\bvec{y})$ are defined as \eqref{def:abbrev-B-XnYn} and \eqref{def:abbrev-tildeB-XnYn}, respectively, for each $\bvec{y} \in \mathcal{Y}^{n}$.
Fix $\gamma \in (0, 1 - \varepsilon)$ arbitrarily, and consider the subset $\mathcal{D}_{\gamma}^{(n)}(\bvec{y})$ of $\mathcal{X}^{n}$ defined as in \eqref{eq:event_Dn-xn} by replacing $\varepsilon$ by $\gamma$.
A simple calculation yields
\begin{align}
1 - \varepsilon - \gamma
& \overset{\mathclap{\text{(a)}}}{\le}
P_{X^{n}|Y^{n}}(\tilde{\mathcal{B}}_{n}(\bvec{y}) \mid \bvec{y}) - \gamma
\notag \\
& \overset{\mathclap{\text{(b)}}}{\le}
P_{X^{n}|Y^{n}}(\tilde{\mathcal{B}}_{n}(\bvec{y}) \mid \bvec{y}) + P_{X^{n}|Y^{n}}( \mathcal{D}_{\gamma}^{(n)}(\bvec{y}) \mid \bvec{y} ) - 1
\notag \\
& \overset{\mathclap{\text{(c)}}}{\le}
P_{X^{n}|Y^{n}}(\tilde{\mathcal{B}}_{n}(\bvec{y}) \cap \mathcal{D}_{\gamma}^{(n)}(\bvec{y}) \mid Y^{n})
\notag \\
& \overset{\mathclap{\text{(f)}}}{\le}
|\tilde{\mathcal{B}}_{n}(\bvec{y}) \cap \mathcal{D}_{\gamma}^{(n)}(\bvec{y})| \, \exp\left( - H( P_{X^{n}|Y^{n} = \bvec{y}} ) + \sqrt{ \frac{ V( P_{X^{n}|Y^{n} = \bvec{y}} ) }{ \gamma } } \right)
\label{eq:inclusion-exclusion_as}
\end{align}
for every $\bvec{y} \in \mathcal{Y}^{n}$, where
\begin{itemize}
\item
(a) follows from the right-hand inequality of \eqref{eq:largest-A_cond2},
\item
(b) follows from \eqref{eq:Chebyshev_Deps},
\item
(c) follows from the fact that $\mu( \mathcal{A} ) + \mu( \mathcal{B} ) \le \mu( \mathcal{A} \cap \mathcal{B} ) + 1$ for every probability measure $\mu$, and
\item
(f) follows from the right-hand inequality of \eqref{eq:bounds_probab_Deps}.
\end{itemize}
In addition, we have
\begin{align}
\sum_{\bvec{x} \in \tilde{\mathcal{B}}_{n}(\bvec{y}) \cap \mathcal{D}_{\gamma}^{(n)}(\bvec{y})} P_{X^{n}|Y^{n}}(\bvec{x} \mid \bvec{y})^{\alpha}
& \overset{\mathclap{\text{(a)}}}{\ge}
\sum_{\bvec{x} \in \tilde{\mathcal{B}}_{n}(\bvec{y}) \cap \mathcal{D}_{\gamma}^{(n)}(\bvec{y})} \exp\left( - \alpha \, H( P_{X^{n}|Y^{n} = \bvec{y}} ) - \alpha \sqrt{ \frac{ V( P_{X^{n}|Y^{n}} ) }{ \gamma } } \right)
\notag \\
& =
|\tilde{\mathcal{B}}_{n}(\bvec{y}) \cap \mathcal{D}_{\gamma}^{(n)}(\bvec{y})| \, \exp\left( - \alpha \, H( P_{X^{n}|Y^{n} = \bvec{y}} ) - \alpha \sqrt{ \frac{ V( P_{X^{n}|Y^{n} = \bvec{y}} ) }{ \gamma } } \right)
\notag \\
& \overset{\mathclap{\text{(b)}}}{\ge}
(1 - \varepsilon - \gamma) \, \exp\left( (1 - \alpha) \, H( P_{X^{n}|Y^{n} = \bvec{y}} ) - (1 + \alpha) \sqrt{ \frac{ V( P_{X^{n}|Y^{n} = \bvec{y}} ) }{ \gamma } } \right)
\notag \\
& \ge
(1 - \varepsilon - \gamma) \, \exp\left( (1 - \alpha) \, H( P_{X^{n}|Y^{n} = \bvec{y}} ) - (1 + \alpha) \sqrt{ \frac{ n \, \sup_{y \in \mathcal{Y}} V( P_{X|Y=y} ) }{ \gamma } } \right)
\label{eq:LB-alpha-power_as}
\end{align}
for every $\bvec{y} \in \mathcal{Y}^{n}$, where
\begin{itemize}
\item
(a) follows from the left-hand inequality of \eqref{eq:bounds_probab_Deps}, and
\item
(b) follows from \eqref{eq:inclusion-exclusion_as}.
\end{itemize}

Firstly, suppose that $U(X \mid Y) = V(X \mid Y)$.
Since $V(X \mid Y) > 0$ implies that $H(X \mid Y) > 0$, and since we have assumed that $\sup_{y \in \mathcal{Y}} V( P_{X|Y=y} ) < \infty$, one can find a positive integer $n_{2} = n_{2}(\varepsilon, \delta, \alpha, H(X \mid Y), \sup_{y \in \mathcal{Y}} V( P_{X|Y=y} ))$ such that
\begin{align}
n \, (1 - \alpha) \, H(X \mid Y) - (1 + \alpha) \sqrt{ \frac{ n \, \sup_{y \in \mathcal{Y}} V( P_{X|Y=y} ) }{ \delta } }
\ge
\log \frac{ 2 }{ 1 - \varepsilon - \delta }
\label{eq:choice_n2}
\end{align}
for every $n \ge n_{2}$.
We observe that
\begin{align}
\check{H}_{\alpha}^{\varepsilon}(X^{n} \mid Y^{n})
& \overset{\mathclap{\text{(a)}}}{\ge}
\frac{ \alpha }{ 1 - \alpha } \log \left( \sum_{\bvec{y} \in \mathcal{Y}^{n}} P_{Y^{n}}( \bvec{y} ) \, \left( \sum_{\bvec{x} \in \tilde{\mathcal{B}}_{n}(\bvec{y})} P_{X^{n}|Y^{n}}(\bvec{x} \mid \bvec{y})^{\alpha} - 1 \right)^{1/\alpha} \right)
\notag \\
& \ge
\frac{ \alpha }{ 1 - \alpha } \log \left( \sum_{\bvec{y} \in \mathcal{Y}^{n}} P_{Y^{n}}( \bvec{y} ) \, \left( \sum_{\bvec{x} \in \tilde{\mathcal{B}}_{n}(\bvec{y}) \cap \mathcal{D}_{n}(\bvec{y})} P_{X^{n}|Y^{n}}(\bvec{x} \mid \bvec{y})^{\alpha} - 1 \right)^{1/\alpha} \right)
\notag \\
& \overset{\mathclap{\text{(b)}}}{\ge}
\frac{ \alpha }{ 1 - \alpha } \log \left( \sum_{\bvec{y} \in \mathcal{Y}^{n}} P_{Y^{n}}( \bvec{y} ) \, \left( (1 - \varepsilon - \gamma) \, \exp\left( (1 - \alpha) \, H( P_{X^{n}|Y^{n} = \bvec{y}} ) - (1 + \alpha) \sqrt{ \frac{ n \, \sup_{y \in \mathcal{Y}} V( P_{X|Y=y} ) }{ \gamma } } \right) - 1 \right)^{1/\alpha} \right)
\notag \\
& \overset{\mathclap{\text{(c)}}}{=}
\frac{ \alpha }{ 1 - \alpha } \log \left( \sum_{\bvec{y} \in \mathcal{Y}^{n}} P_{Y^{n}}( \bvec{y} ) \, \left( (1 - \varepsilon - \gamma) \, \exp\left( (1 - \alpha) \, n \, H(X \mid Y) - (1 + \alpha) \sqrt{ \frac{ n \, \sup_{y \in \mathcal{Y}} V( P_{X|Y=y} ) }{ \gamma } } \right) - 1 \right)^{1/\alpha} \right)
\notag \\
& =
\frac{ \alpha }{ 1 - \alpha } \log \left( \left( (1 - \varepsilon - \gamma) \, \exp\left( (1 - \alpha) \, n \, H(X \mid Y) - (1 + \alpha) \sqrt{ \frac{ n \, \sup_{y \in \mathcal{Y}} V( P_{X|Y=y} ) }{ \gamma } } \right) - 1 \right)^{1/\alpha} \right)
\notag \\
& \overset{\mathclap{\text{(d)}}}{\ge}
\frac{ \alpha }{ 1 - \alpha } \log \left( \left( \frac{ 1 - \varepsilon - \gamma }{ 2 } \right)^{1/\alpha} \, \exp\left( \frac{ 1 - \alpha }{ \alpha } \, n \, H(X \mid Y) - \frac{ 1 + \alpha }{ \alpha } \sqrt{ \frac{ n \, \sup_{y \in \mathcal{Y}} V( P_{X|Y=y} ) }{ \gamma } } \right) \right)
\notag \\
& =
n \, H(X \mid Y) - \frac{ 1 + \alpha }{ 1 - \alpha } \sqrt{ \frac{ n \, \sup_{y \in \mathcal{Y}} V( P_{X|Y=y} ) }{ \gamma } } + \frac{ 1 }{ 1 - \alpha } \log \frac{ 1 - \varepsilon - \gamma }{ 2 }
\notag \\
& \overset{\mathclap{\text{(e)}}}{=}
n \, H^{(\alpha)}(X \mid Y) - \frac{ 1 + \alpha }{ 1 - \alpha } \sqrt{ \frac{ n \, \sup_{y \in \mathcal{Y}} V( P_{X|Y=y} ) }{ \gamma } } + \frac{ 1 }{ 1 - \alpha } \log \frac{ 1 - \varepsilon - \gamma }{ 2 }
\label{eq:LB_tilde_equal-total-variance}
\end{align}
for sufficiently large $n \ge n_{2}$, where
\begin{itemize}
\item
(a) follows as in Steps~(a) and~(b) of \eqref{eq:LB_tilde_zero-total-variance},
\item
(b) follows from \eqref{eq:LB-alpha-power_as},
\item
(c) follows by the law of total variance and the hypothesis that $U(X \mid Y) = V(X \mid Y)$,
\item
(d) follows by the choice of $n_{2}$; see \eqref{eq:choice_n2}, and
\item
(e) follows from \propref{prop:KN-avg}.
\end{itemize}
This completes the proof of the lower bound part of \eqref{eq:2nd_max} in the case where $U(X \mid Y) = V(X \mid Y)$.

Secondly, suppose that $U(X \mid Y) > V(X \mid Y)$.
Recall that the subset $\mathcal{E}_{n}$ of $\mathcal{Y}^{n}$ is defined in \eqref{def:event_En}.
Since $V(X \mid Y) > 0$ implies that $H(X \mid Y) > 0$, and since we have assumed that $\sup_{y \in \mathcal{Y}} V( P_{X|Y=y} ) < \infty$, one can find a positive integer $n_{3} = n_{3}(\varepsilon, \delta, \alpha, H(X \mid Y), U(X \mid Y), V(X \mid Y), \sup_{y \in \mathcal{Y}} V( P_{X|Y=y} ))$ such that
\begin{align}
n \, (1 - \alpha) \, H(X \mid Y) - \sqrt{ n } \left( (1 - \alpha) \sqrt{ (U(X \mid Y) - V(X \mid Y)) } + (1 + \alpha) \sqrt{ \frac{ \sup_{y \in \mathcal{Y}} V( P_{X|Y=y} ) }{ \delta } } \right)
\ge
\log \frac{ 2 }{ 1 - \varepsilon - \delta }
\label{eq:choice_n3}
\end{align}
for every $n \ge n_{3}$.
We then observe that
\begin{align}
&
\check{H}_{\alpha}^{\varepsilon}(X^{n} \mid Y^{n})
\notag \\
& \overset{\mathclap{\text{(a)}}}{\ge}
\frac{ \alpha }{ 1 - \alpha } \log \left( \sum_{\bvec{y} \in \mathcal{Y}^{n}} P_{Y^{n}}( \bvec{y} ) \, \left( \sum_{\bvec{x} \in \tilde{\mathcal{B}}_{n}(\bvec{y})} P_{X^{n}|Y^{n}}(\bvec{x} \mid \bvec{y})^{\alpha} - 1 \right)^{1/\alpha} \right)
\notag \\
& \ge
\frac{ \alpha }{ 1 - \alpha } \log \left( \sum_{\bvec{y} \in \mathcal{E}_{n}} P_{Y^{n}}( \bvec{y} ) \, \left( \sum_{\bvec{x} \in \tilde{\mathcal{B}}_{n}(\bvec{y}) \cap \mathcal{D}_{\gamma}^{(n)}} P_{X^{n}|Y^{n}}(\bvec{x} \mid \bvec{y})^{\alpha} - 1 \right)^{1/\alpha} \right)
\notag \\
& \overset{\mathclap{\text{(b)}}}{\ge}
\frac{ \alpha }{ 1 - \alpha } \log \left( \sum_{\bvec{y} \in \mathcal{E}_{n}} P_{Y^{n}}( \bvec{y} ) \, \left( (1 - \varepsilon - \gamma) \, \exp\left( (1 - \alpha) \, H( P_{X^{n}|Y^{n} = \bvec{y}} ) - (1 + \alpha) \sqrt{ \frac{ n \, \sup_{y \in \mathcal{Y}} V( P_{X|Y=y} ) }{ \gamma } } \right) - 1 \right)^{1/\alpha} \right)
\notag \\
& \overset{\mathclap{\text{(c)}}}{\ge}
\frac{ \alpha }{ 1 - \alpha } \log \left( \sum_{\bvec{y} \in \mathcal{E}_{n}} P_{Y^{n}}( \bvec{y} ) \, \left( (1 - \varepsilon - \gamma) \, \exp\left( (1 - \alpha) \, H( P_{X^{n}|Y^{n} = \bvec{y}} ) - (1 + \alpha) \sqrt{ \frac{ n \, \sup_{y \in \mathcal{Y}} V( P_{X|Y=y} ) }{ \gamma } } \right)
\right. \right. \notag \\
& \left. \left. \qquad \qquad
{} - \frac{ 1 - \varepsilon - \delta }{ 2 } \, \exp\left( n \, (1 - \alpha) \, H(X \mid Y) - (1 - \alpha) \sqrt{ n \, (U(X \mid Y) - V(X \mid Y)) } - (1 + \alpha) \sqrt{ \frac{ n \, \sup_{y \in \mathcal{Y}} V( P_{X|Y=y} ) }{ \gamma } } \right) \right)^{1/\alpha} \right)
\notag \\
& \overset{\mathclap{\text{(d)}}}{\ge}
\frac{ \alpha }{ 1 - \alpha } \log \left( \sum_{\bvec{y} \in \mathcal{E}_{n}} P_{Y^{n}}( \bvec{y} ) \, \left( \frac{ 1 - \varepsilon - \gamma }{ 2 } \right)^{1/\alpha} \, \exp\left( \frac{ 1 - \alpha }{ \alpha } \, H( P_{X^{n}|Y^{n} = \bvec{y}} ) - \frac{ 1 + \alpha }{ \alpha } \sqrt{ \frac{ n \, \sup_{y \in \mathcal{Y}} V( P_{X|Y=y} ) }{ \gamma } } \right) \right)
\notag \\
& =
\frac{ \alpha }{ 1 - \alpha } \log \left( \sum_{\bvec{y} \in \mathcal{E}_{n}} P_{Y^{n}}( \bvec{y} ) \, \exp\left( \frac{ 1 - \alpha }{ \alpha } \, H( P_{X^{n}|Y^{n} = \bvec{y}} ) \right) \right) - \frac{ 1 + \alpha }{ 1 - \alpha } \sqrt{ \frac{ n \sup_{y \in \mathcal{Y}} V( P_{X|Y=y} ) }{ \gamma } } + \frac{ 1 }{ 1 - \alpha } \log \frac{ 1 - \varepsilon - \gamma }{ 2 }
\notag \\
& =
\frac{ \alpha }{ 1 - \alpha } \log \left( \sum_{\bvec{y} \in \mathcal{Y}^{n}} P_{Y^{n}}( \bvec{y} ) \, \exp\left( \frac{ 1 - \alpha }{ \alpha } \, H( P_{X^{n}|Y^{n} = \bvec{y}} ) \right) - \sum_{\bvec{y} \in \mathcal{Y}^{n} \setminus \mathcal{E}_{n}} P_{Y^{n}}( \bvec{y} ) \, \exp\left( \frac{ 1 - \alpha }{ \alpha } \, H( P_{X^{n}|Y^{n} = \bvec{y}} ) \right) \right)
\notag \\
& \qquad 
{} - \frac{ 1 + \alpha }{ 1 - \alpha } \sqrt{ \frac{ n \sup_{y \in \mathcal{Y}} V( P_{X|Y=y} ) }{ \gamma } } + \frac{ 1 }{ 1 - \alpha } \log \frac{ 1 - \varepsilon - \gamma }{ 2 }
\notag \\
& \overset{\mathclap{\text{(e)}}}{\ge}
\frac{ \alpha }{ 1 - \alpha } \log \left( \left( \sum_{y \in \mathcal{Y}} P_{Y}( y ) \, \exp\left( \frac{ 1 - \alpha }{ \alpha } \, H( P_{X|Y = y} ) \right) \right)^{n} - \sum_{\bvec{y} \in \mathcal{Y}^{n} \setminus \mathcal{E}_{n}} P_{Y^{n}}( \bvec{y} ) \, \exp\left( \frac{ 1 - \alpha }{ \alpha } \, H( P_{X^{n}|Y^{n} = \bvec{y}} ) \right) \right)
\notag \\
& \qquad 
{} - \frac{ 1 + \alpha }{ 1 - \alpha } \sqrt{ \frac{ n \sup_{y \in \mathcal{Y}} V( P_{X|Y=y} ) }{ \gamma } } + \frac{ 1 }{ 1 - \alpha } \log \frac{ 1 - \varepsilon - \gamma }{ 2 }
\notag \\
& \overset{\mathclap{\text{(f)}}}{\ge}
\frac{ \alpha }{ 1 - \alpha } \log \left( \left( \sum_{y \in \mathcal{Y}} P_{Y}( y ) \, \exp\left( \frac{ 1 - \alpha }{ \alpha } \, H( P_{X|Y = y} ) \right) \right)^{n} - \sum_{\bvec{y} \in \mathcal{Y}^{n} \setminus \mathcal{E}_{n}} P_{Y^{n}}( \bvec{y} ) \, \exp\left( \frac{ 1 - \alpha }{ \alpha } \, n \, H(X \mid Y) \right) \right)
\notag \\
& \qquad 
{} - \frac{ 1 + \alpha }{ 1 - \alpha } \sqrt{ \frac{ n \sup_{y \in \mathcal{Y}} V( P_{X|Y=y} ) }{ \gamma } } + \frac{ 1 }{ 1 - \alpha } \log \frac{ 1 - \varepsilon - \gamma }{ 2 }
\notag \\
& =
\frac{ \alpha }{ 1 - \alpha } \log \left( \left( \sum_{y \in \mathcal{Y}} P_{Y}( y ) \, \exp\left( \frac{ 1 - \alpha }{ \alpha } \, H( P_{X|Y = y} ) \right) \right)^{n} - (1 - P_{Y^{n}}(\mathcal{E}_{n})) \, \exp\left( \frac{ 1 - \alpha }{ \alpha } \, n \, H(X \mid Y) \right) \right)
\notag \\
& \qquad 
{} - \frac{ 1 + \alpha }{ 1 - \alpha } \sqrt{ \frac{ n \sup_{y \in \mathcal{Y}} V( P_{X|Y=y} ) }{ \gamma } } + \frac{ 1 }{ 1 - \alpha } \log \frac{ 1 - \varepsilon - \gamma }{ 2 }
\notag \\
& \overset{\mathclap{\text{(g)}}}{\ge}
\frac{ \alpha }{ 1 - \alpha } \log \left( \left( \sum_{y \in \mathcal{Y}} P_{Y}( y ) \, \exp\left( \frac{ 1 - \alpha }{ \alpha } \, H( P_{X|Y = y} ) \right) \right)^{n} - \frac{ 1 }{ 2 } \, \exp\left( \frac{ 1 - \alpha }{ \alpha } \, n \, H(X \mid Y) \right) \right)
\notag \\
& \qquad
{} - \frac{ 1 + \alpha }{ 1 - \alpha } \sqrt{ \frac{ n \sup_{y \in \mathcal{Y}} V( P_{X|Y=y} ) }{ \gamma } } + \frac{ 1 }{ 1 - \alpha } \log \frac{ 1 - \varepsilon - \gamma }{ 2 }
\notag \\
& \overset{\mathclap{\text{(h)}}}{\ge}
\frac{ \alpha }{ 1 - \alpha } \log \left( \frac{ 1 }{ 2 } \left( \sum_{y \in \mathcal{Y}} P_{Y}( y ) \, \exp\left( \frac{ 1 - \alpha }{ \alpha } \, H( P_{X|Y = y} ) \right) \right)^{n} \right) - \frac{ 1 + \alpha }{ 1 - \alpha } \sqrt{ \frac{ n \sup_{y \in \mathcal{Y}} V( P_{X|Y=y} ) }{ \gamma } } + \frac{ 1 }{ 1 - \alpha } \log \frac{ 1 - \varepsilon - \gamma }{ 2 }
\notag \\
& =
n \, H^{(\alpha)}(X \mid Y) - \frac{ 1 + \alpha }{ 1 - \alpha } \sqrt{ \frac{ n \sup_{y \in \mathcal{Y}} V( P_{X|Y=y} ) }{ \gamma } } + \frac{ 1 }{ 1 - \alpha } \log \frac{ 1 - \varepsilon - \gamma }{ 2 } + \frac{ \alpha }{ 1 - \alpha } \log 2
\label{eq:LB_tilde_diff-total-variance}
\end{align}
for every $n \ge n_{3}$, where
\begin{itemize}
\item
(a) follows as in Steps~(a) and~(b) of \eqref{eq:LB_tilde_zero-total-variance},
\item
(b) follows from \eqref{eq:LB-alpha-power_as},
\item
(c) follows by the choice of $n_{3}$; see \eqref{eq:choice_n3},
\item
(d) follows by the definition of $\mathcal{E}_{n}$ stated in \eqref{def:event_En},
\item
(e) follows from the fact that $(X_{1}, Y_{1}), \dots, (X_{n}, Y_{n})$ are $n$ i.i.d.\ copies of $(X, Y)$,
\item
(f) follows by the definition of $\mathcal{E}_{n}$ stated in \eqref{def:event_En},
\item
(g) follows from \eqref{eq:Cantelli_En}, and
\item
(h) follows from Jensen's inequality.
\end{itemize}
This completes the proof of the lower bound part of \eqref{eq:2nd_max} in the case where $U(X \mid Y) > V(X \mid Y)$.
\hfill\IEEEQEDhere

\section{Proof of \lemref{lem:unified_converse_avg}}
\label{app:unified_converse_avg}

For each $(x, y) \in \mathcal{X} \times \mathcal{Y}$, define
\begin{align}
a(x, y)
& \coloneqq
\begin{cases}
\kappa(x, y)^{-\rho/(1+\rho)}
& \mathrm{if} \ 0 \le \epsilon(x, y) < 1 ,
\\
0
& \mathrm{if} \ \epsilon(x, y) = 1 ,
\end{cases}
\\
b(x, y)
& \coloneqq
\kappa(x, y)^{\rho/(1+\rho)} \, \Big( (1 - \epsilon(x, y)) \, P_{X|Y}(x \mid y) \Big)^{1/(1+\rho)} .
\end{align}
Applying H\"{o}lder's inequality (see, e.g., \cite[Problem~4.15]{gallager_1968}),
\begin{align}
\sum_{x \in \mathcal{X}} a(x, y) \, b(x, y)
\le
\left( \sum_{x \in \mathcal{X}} a(x, y)^{1/\lambda} \right)^{\lambda} \left( \sum_{x \in \mathcal{X}} b(x, y)^{1/(1-\lambda)} \right)^{1-\lambda}
\end{align}
with $\lambda = \rho/(1+\rho)$, we get
\begin{align}
\left( \sum_{x \in \mathcal{X}} \Big( (1 - \epsilon(x, y)) \, P_{X|Y}(x \mid y) \Big)^{1/(1+\rho)} \right)^{1+\rho}
\le
\left( \sum_{\substack{ x \in \mathcal{X} : \\ 0 \le \epsilon(x, y) < 1 }} \frac{ 1 }{ \kappa(x, y) } \right)^{\rho} \left( \sum_{x \in \mathcal{X}} (1 - \epsilon(x, y)) \, P_{X|Y}(x \mid y) \, \kappa(x, y)^{\rho} \right)
\label{eq:Holder_wrt-X}
\end{align}
for every $y \in \mathcal{Y}$.
Now, choose the function $\delta : \mathcal{Y} \to [0, 1]$ so that
\begin{align}
\delta( y )
& =
\mathbb{E}[ \epsilon(X, Y) \mid Y = y ]
=
\sum_{x \in \mathcal{X}} \epsilon(x, y) \, P_{X|Y}(x \mid y) .
\label{eq:delta-y}
\end{align}
Then, we have
\begin{align}
\mathbb{E}[ K(X, Y)^{\rho} \mid Y = y ]
& =
\sum_{x \in \mathcal{X}} (1 - \epsilon(x, y)) \, P_{X|Y}(x \mid y) \, \kappa(x, y)^{\rho}
\notag \\
& \overset{\mathclap{\text{(a)}}}{\ge}
\left( \sum_{x \in \mathcal{X}} \Big( (1 - \epsilon(x, y)) \, P_{X|Y}(x \mid y) \Big)^{1/(1+\rho)} \right)^{1+\rho} \left( \sum_{\substack{ x \in \mathcal{X} : \\ 0 \le \epsilon(x, y) < 1 }} \frac{ 1 }{ \kappa(x, y) } \right)^{-\rho}
\notag \\
& \overset{\mathclap{\text{(b)}}}{\ge}
\left( \inf_{\tilde{\epsilon}( \cdot ) : \mathbb{E}[ \tilde{\epsilon}(X) \mid Y = y ] = \delta(y) } \sum_{x \in \mathcal{X}} \Big( (1 - \tilde{\epsilon}(x)) \, P_{X|Y}(x \mid y) \Big)^{1/(1+\rho)} \right)^{1+\rho} \left( \sum_{\substack{ x \in \mathcal{X} : \\ 0 \le \epsilon(x, y) < 1 }} \frac{ 1 }{ \kappa(x, y) } \right)^{-\rho}
\notag \\
& \overset{\mathclap{\text{(c)}}}{=}
\exp\left( \rho \, H_{1/(1+\rho)}^{\delta(y)}( P_{X|Y=y} ) - \rho \log \sum_{\substack{ x \in \mathcal{X} : \\ 0 \le \epsilon(x, y) < 1 }} \frac{ 1 }{ \kappa(x, y) } \right)
\label{eq:unified-converse_given-Y}
\end{align}
for every $y \in \mathcal{Y}$, where
\begin{itemize}
\item
(a) follows from \eqref{eq:Holder_wrt-X},
\item
(b) follows by the choice of $\delta : \mathcal{Y} \to [0, 1]$ stated in \eqref{eq:delta-y}, and
\item
(c) follows by the definition of $H_{1/(1+\rho)}^{\delta(y)}( P_{X|Y=y} )$ stated in \eqref{def:smooth}.
\end{itemize}
Therefore, we obtain
\begin{align}
\mathbb{E}[ K(X, Y)^{\rho} ]
& =
\sum_{y \in \mathcal{Y}} P_{Y}( y ) \, \mathbb{E}[ K(X, Y)^{\rho} \mid Y = y ]
\notag \\
& \overset{\mathclap{\text{(a)}}}{\ge}
\sum_{y \in \mathcal{Y}} P_{Y}( y ) \, \exp\left( \rho \, H_{1/(1+\rho)}^{\delta(y)}( P_{X|Y=y} ) - \rho \log \sum_{\substack{ x \in \mathcal{X} : \\ 0 \le \epsilon(x, y) < 1 }} \frac{ 1 }{ \kappa(x, y) } \right)
\notag \\
& \overset{\mathclap{\text{(b)}}}{\ge}
\left( \sum_{y \in \mathcal{Y}} P_{Y}( y ) \, \exp\Big( \rho \, H_{1/(1+\rho)}^{\delta(y)}( P_{X|Y=y} ) \Big) \right) \, \exp\Big( - \rho \log R(\epsilon, \kappa) \Big)
\notag \\
& \overset{\mathclap{\text{(c)}}}{=}
\exp\left( \rho \, \bar{\mathsf{H}}_{1/(1+\rho)}^{\delta(\cdot)}(X \mid Y) \right) \, \exp\Big( - \rho \log R(\epsilon, \kappa) \Big)
\notag \\
& \overset{\mathclap{\text{(d)}}}{\ge}
\exp\left( \rho \, \inf_{\delta(\cdot) : \mathbb{E}[ \delta(Y) ] \le \varepsilon} \bar{\mathsf{H}}_{1/(1+\rho)}^{\delta(\cdot)}(X \mid Y) \right) \, \exp\Big( - \rho \log R(\epsilon, \kappa) \Big)
\notag \\
& \overset{\mathclap{\text{(e)}}}{=}
\exp\left( \rho \, \inf_{\delta(\cdot) : \mathbb{E}[ \delta(Y) ] = \varepsilon} \bar{\mathsf{H}}_{1/(1+\rho)}^{\delta(\cdot)}(X \mid Y) \right) \, \exp\Big( - \rho \log R(\epsilon, \kappa) \Big)
\notag \\
& \overset{\mathclap{\text{(f)}}}{=}
\exp\left( \rho \, H_{1/(1+\rho)}^{\varepsilon}(X \mid Y) \right) \, \exp\Big( - \rho \log R(\epsilon, \kappa) \Big) ,
\label{eq:derivation_unified_converse_avg}
\end{align}
where
\begin{itemize}
\item
(a) follows from \eqref{eq:unified-converse_given-Y},
\item
(b) follows by the definition of $R(\epsilon, \kappa)$ stated in \eqref{def:redundancy-R},
\item
(c) follows by the definition of $\bar{\mathsf{H}}_{\alpha}^{\delta(\cdot)}(X \mid Y)$ stated in \eqref{def:cond-smooth_KN-avg},
\item
(d) follows from \eqref{eq:average-error} and \eqref{eq:delta-y},
\item
(e) follows from the fact that the unsmooth conditional R\'{e}nyi entropy $\varepsilon \mapsto H_{\alpha}^{\varepsilon}$ is nonincreasing in $\varepsilon \in [0, 1)$, and
\item
(f) follows from \lemref{lem:Kuzuoka-formula}.
\end{itemize}
This completes the proof of \lemref{lem:unified_converse_avg}.
\hfill\IEEEQEDhere

\section{Proof of \lemref{lem:unified_converse_max}}
\label{app:unified_converse_max}

A direct calculation shows
\begin{align}
\mathbb{E}[ K(X, Y)^{\rho} ]
& \overset{\mathclap{\text{(a)}}}{\ge}
\exp\left( \rho \, \bar{\mathsf{H}}_{1/(1+\rho)}^{\delta(\cdot)}(X \mid Y) \right) \, \exp\Big( - \rho \log R(\epsilon, \kappa) \Big)
\notag \\
& \overset{\mathclap{\text{(b)}}}{\ge}
\exp\left( \rho \inf_{\delta(\cdot) : \sup_{y \in \mathcal{Y}} \mathbb{E}[\delta(Y) \mid Y = y] \le \varepsilon} \bar{\mathsf{H}}_{1/(1+\rho)}^{\delta(\cdot)}(X \mid Y) \right) \, \exp\Big( - \rho \log R(\epsilon, \kappa) \Big)
\notag \\
& \overset{\mathclap{\text{(c)}}}{=}
\exp\left( \rho \, \bar{\mathsf{H}}_{1/(1+\rho)}^{\varepsilon}(X \mid Y) \right) \, \exp\Big( - \rho \log R(\epsilon, \kappa) \Big)
\notag \\
& \overset{\mathclap{\text{(d)}}}{=}
\exp\left( \rho \, \check{H}_{1/(1+\rho)}^{\varepsilon}(X \mid Y) \right) \, \exp\Big( - \rho \log R(\epsilon, \kappa) \Big)
\end{align}
where
\begin{itemize}
\item
(a) follows as in Steps~(a)--(c) of \eqref{eq:derivation_unified_converse_avg},
\item
(b) follows from \eqref{eq:maximum-error} and \eqref{eq:delta-y},
\item
(c) follows from the fact that the unsmooth conditional R\'{e}nyi entropy $\varepsilon \mapsto H_{\alpha}^{\varepsilon}$ is nonincreasing in $\varepsilon \in [0, 1)$, and
\item
(d) follows from \eqref{eq:check_identity1}.
\end{itemize}
This completes the proof of \lemref{lem:unified_converse_max}.

\section{Proof of \lemref{lem:unified_direct}}
\label{app:unified_direct}

For each $y \in \mathcal{Y}$, choose two real parameters $\eta(y) \ge 1$ and $0 \le \beta(y) < 1$ so that
\begin{align}
\mathbb{P}\{ \kappa(X, Y) > \eta(Y) \mid Y = y \} + \beta(y) \, \mathbb{P}\{ \kappa(X, Y) = \eta(Y) \mid Y = y \}
=
\delta( y ) .
\end{align}
Construct a deterministic map $\epsilon : \mathcal{X} \times \mathcal{Y} \to [0, 1]$ as
\begin{align}
\epsilon(x, y)
=
\begin{cases}
1
& \mathrm{if} \ \kappa(x, y) < \eta( y ) ,
\\
\beta( y )
& \mathrm{if} \ \kappa(x, y) = \eta( y ) ,
\\
0
& \mathrm{if} \ \kappa(x, y) > \eta( y ) .
\end{cases}
\label{eq:achievability-eps}
\end{align}
for each $(x, y) \in \mathcal{X} \times \mathcal{Y}$.
Then, it is clear that
\begin{align}
\mathbb{E}[ \epsilon(X, Y) \mid Y = y ]
=
\delta( y )
\end{align}
for each $y \in \mathcal{Y}$.
Moreover, a direct calculation shows
\begin{align}
\mathbb{E}[ K(X, Y)^{\rho} ]
& \overset{\mathclap{\text{(a)}}}{=}
\sum_{y \in \mathcal{Y}} P_{Y}( y ) \, \left( \sum_{\substack{ x \in \mathcal{X} : \\ \kappa(x, y) < \eta(y) }} P_{X|Y}(x \mid y) \, \kappa(x, y)^{\rho} + \beta(y) \, \eta( y )^{\rho} \right)
\notag \\
& \overset{\mathclap{\text{(b)}}}{\le}
c^{\rho} \, \sum_{y \in \mathcal{Y}} P_{Y}( y ) \, \left( \sum_{x \in \mathcal{A}_{X|Y}^{\delta(y)}(y)} P_{X|Y}(x \mid y)^{1/(1+\rho)} + M(y)^{1/(1+\rho)} \right)^{1 + \rho}
\notag \\
& \overset{\mathclap{\text{(c)}}}{=}
c^{\rho} \, \sum_{y \in \mathcal{Y}} P_{Y}( y ) \, \exp\bigg( \rho \, H_{1/(1+\rho)}^{\delta(y)}( P_{X|Y=y} ) \bigg)
\notag \\
& \overset{\mathclap{\text{(d)}}}{=}
c^{\rho} \, \exp\bigg( \rho \, \bar{\mathsf{H}}_{1/(1+\rho)}^{\delta(\cdot)}(X \mid Y) \bigg) ,
\end{align}
where
\begin{itemize}
\item
(a) follows from \eqref{def:stochastic-K} and \eqref{eq:achievability-eps},
\item
(b) follows from \eqref{eq:kappa-tilted} and the definition of $Q_{X|Y}^{(1/(1+\rho), \delta(\cdot))}$ stated in \eqref{def:tilted-R},
\item
(c) follows from \lemref{lem:majorization}, and
\item
(d) follows from the definition of $\bar{\mathsf{H}}_{1/(1+\rho)}^{\delta(\cdot)}(X \mid Y)$ stated in \eqref{def:cond-smooth_KN-avg}.
\end{itemize}
This completes the proof of \lemref{lem:unified_direct}.
\hfill\IEEEQEDhere

\section{Proof of \lemref{lem:tilde_one-shot}}
\label{app:tilde_one-shot}

Firstly, we shall verify the converse bound of \lemref{lem:tilde_one-shot}, i.e., the left-hand inequality of \eqref{eq:kuzuoka_one-shot}.
Consider a variable-length semi-stochastic code $(F, g)$ such that $\mathcal{C}_{y}(X, Y, F)$ is prefix-free for every $y \in \mathcal{Y}$ and
\begin{align}
\mathbb{P}\{ X \neq g(F(X, Y), Y) \}
\le
\varepsilon .
\label{eq:error_initial-FG}
\end{align}
Construct another stochastic encoder $F_{0} : \mathcal{X} \times \mathcal{Y} \to \{ 0, 1 \}^{\ast}$ as follows:
\begin{align}
F_{0}( x, y )
=
\begin{cases}
\varnothing
& \mathrm{if} \ x \neq g(F( x, y ), y) ,
\\
F( x, y )
& \mathrm{if} \ x = g(F( x, y ), y) 
\end{cases}
\end{align}
for each $x \in \mathcal{X}$.
It is clear that
\begin{align}
X \neq g(F_{0}(X, Y), Y)
\quad \Longrightarrow \quad
X \neq g(F(X, Y), Y) .
\label{eq:competitively-better_g1F0}
\end{align}
Consider a collection $\{ \mathcal{B}( x, y ) \}_{(x, y) \in \mathcal{X} \times \mathcal{Y}}$ of subsets of $\{ 0, 1 \}^{\ast}$ given as
\begin{align}
\mathcal{B}( x, y )
=
\begin{cases}
\{ \varnothing \}
& \mathrm{if} \ x = g( \varnothing, y ) ,
\\
\{ \bvec{b} \in \{ 0, 1 \}^{\ast} \setminus \{ \varnothing \} \mid \mathbb{P}\{ F_{0}( x, y ) = \bvec{b} \} > 0 \} .
& \mathrm{if} \ x \neq g( \varnothing, y ) .
\end{cases}
\end{align}
Furthermore, for each $(x, y) \in \mathcal{X} \times \mathcal{Y}$, choose a binary string $\bvec{b}( x, y ) \in \{ 0, 1 \}^{\ast}$ so that
\begin{align}
\bvec{b}( x, y )
\in
\argmin\limits_{\bvec{b} \in \mathcal{B}( x, y )} \ell( \bvec{b} ) ,
\label{eq:short-string_Set-B}
\end{align}
where suppose that $\bvec{b}( x, y ) = \varnothing$ if $\mathcal{B}( x, y ) = \emptyset$.
Note that this map $\bvec{b}( \cdot, \cdot ) : \mathcal{X} \times \mathcal{Y} \to \{ 0, 1 \}^{\ast}$ is deterministic.
Now, construct another stochastic encoder $F_{1} : \mathcal{X} \times \mathcal{Y} \to \{ 0, 1 \}^{\ast}$ so that
\begin{align}
F_{1}( x, y )
=
\begin{cases}
\varnothing
& \mathrm{if} \ x \neq g(F_{0}( x, y ), y) ,
\\
\bvec{b}( x, y )
& \mathrm{if} \ x = g(F_{0}( x, y ), y)
\end{cases}
\label{eq:construction_F1}
\end{align}
for each $(x, y) \in \mathcal{X} \times \mathcal{Y}$.
Then, it follows from \eqref{eq:competitively-better_g1F0} and \eqref{eq:construction_F1} that
\begin{align}
\exp\Big( \tilde{\Lambda}(X, Y, F, g \, \| \, \rho) \Big)
=
\mathbb{E}[ 2^{\rho \ell(F(X, Y), Y)} \, \bvec{1}_{\{ X = g(F(X, Y), Y) \}} ]
& \ge
\mathbb{E}[ 2^{\rho \ell(\bvec{b}(X, Y))} \, \bvec{1}_{\{ X = g(F(X, Y), Y) \}} ] .
\label{eq:tilde-Lambda_constructive-LB}
\end{align}
Choosing $\epsilon : \mathcal{X} \times \mathcal{Y} \to [0, 1]$ so that
\begin{align}
\epsilon(x, y)
=
\mathbb{P}\{ X \neq g(F(X, Y), Y) \mid (X, Y) = (x, y) \}
\label{eq:eps_each-error}
\end{align}
for each $(x, y) \in \mathcal{X} \times \mathcal{Y}$, we observe that
\begin{align}
2^{\ell( \bvec{b}(x, y) )} \, \bvec{1}_{\{ x = g(F(x, y), y) \}}
=
\begin{cases}
2^{\ell( \bvec{b}(x, y) )}
& \mathrm{with} \ \mathrm{probability} \ 1 - \epsilon( x, y ) ,
\\
0
& \mathrm{with} \ \mathrm{probability} \ \epsilon( x, y ) .
\end{cases}
\end{align}
Therefore, since \eqref{eq:error_initial-FG} implies that \eqref{eq:average-error} holds, it follows from \lemref{lem:unified_converse_avg} and \eqref{eq:tilde-Lambda_constructive-LB} that
\begin{align}
\frac{ \tilde{\Lambda}(X, Y, F, g \, \| \, \rho) }{ \rho }
& \ge
H_{1/(1+\rho)}^{\varepsilon}(X \mid Y) - \log R(\epsilon, 2^{\ell(\bvec{b}(\cdot, \cdot))})
\end{align}
Finally, since $\mathcal{C}_{y}(X, Y, F)$ is prefix-free for every $y \in \mathcal{Y}$, it follows from the Kraft--McMillan inequality that
\begin{align}
\sum_{\substack{ x \in \mathcal{X} : \\ \epsilon( x, y ) < 1}} 2^{- \ell( \bvec{b}( x, y ) )}
\le
\sum_{\bvec{b} \in \mathcal{C}_{y}(X, Y, F)} 2^{- \ell( \bvec{b} )}
\le
1
\end{align}
for every $y \in \mathcal{Y}$, which implies that
\begin{align}
R(\epsilon, 2^{\ell(\bvec{b}(\cdot, \cdot))})
\le
1 .
\label{eq:redundancy_Kraft-McMillan}
\end{align}
This completes the proof of the left-hand inequality of \eqref{eq:kuzuoka_one-shot}.%
\footnote{Here, both constructed stochastic encoders $F_{0}$ and $F_{1}$ do not satisfy the prefix-free constraint, and it does not affect the proof.}

Secondly, we shall verify the achievability bound of \lemref{lem:tilde_one-shot}, i.e., the right-hand inequality of \eqref{eq:kuzuoka_one-shot}.
Fix an arbitrary small positive number $\zeta$.
It follows from \lemref{lem:Kuzuoka-formula} that one can find a deterministic map $\delta^{\ast}( \cdot ) \in \mathcal{E}_{0}( \varepsilon )$ so that
\begin{align}
H_{1/(1+\rho)}^{\varepsilon}(X \mid Y)
\ge
\bar{\mathsf{H}}_{1/(1+\rho)}^{\delta^{\ast}(\cdot)}(X \mid Y) - \zeta .
\label{eq:choice_delta-ast_zeta}
\end{align}
Recall that for each $y \in \mathcal{Y}$, the proper subset $\mathcal{A}_{X|Y=y}^{\delta^{\ast}(y)}$ of $\mathcal{X}$ is defined to satisfy \eqref{eq:largest-A_cond1} and \eqref{eq:largest-A_cond2}, and the element $x^{\ast}(y)$ of $\mathcal{X}$ is chosen as \eqref{def:x-ast}.
Denote by
\begin{align}
\tilde{\mathcal{A}}_{X|Y=y}^{\delta^{\ast}(y)}
\coloneqq
\mathcal{A}_{X|Y=y}^{\delta^{\ast}(y)} \cup \{ x^{\ast}(y) \}
\end{align}
for each $y \in \mathcal{Y}$.
Let $\mathcal{Z}$ be a subset of $\mathcal{X} \times \mathcal{Y}$ given as
\begin{align}
\mathcal{Z}
=
\{ (x, y) \in \mathcal{X} \times \mathcal{Y} \mid x \in \tilde{\mathcal{A}}_{X|Y=y}^{\delta^{\ast}(y)} \} .
\end{align}
Based on the conditional distribution $Q_{X|Y}^{(1/(1+\rho), \delta(\cdot))}$ defined in \eqref{def:tilted-R}, consider the Shannon code $f_{\mathrm{Sh.}} : \mathcal{Z} \to \{ 0, 1 \}^{\ast}$ satisfying
\begin{align}
\ell( f_{\mathrm{Sh.}}( x, y ) )
=
\left\lceil \log \frac{ 1 }{ Q_{X|Y}^{(1/(1+\rho), \delta(\cdot))}( x \mid y ) } \right\rceil
\label{eq:Shannon-length}
\end{align}
for every $(x, y) \in \mathcal{Z}$, where note that $x \mapsto f_{\mathrm{Sh.}}(x, y)$ is prefix-free for every fixed $y$.
It follows from \eqref{eq:Shannon-length} that
\begin{align}
2^{\ell( f_{\mathrm{Sh.}}( x, y ) )} \, Q_{X|Y}^{(1/(1+\rho), \delta(\cdot))}( x \mid y )
<
2
\label{eq:Shannon_code_redundancy}
\end{align}
for every $(x, y) \in \mathcal{Z}$.
Fix a pair $(a, b) \in \mathcal{Z}$ arbitrarily.
Now, construct a stochastic encoder $F : \mathcal{X} \times \mathcal{Y} \to \{ 0, 1 \}^{\ast}$ so that
\begin{align}
F( x, y )
=
\begin{cases}
f_{\mathrm{Sh.}}(a, b)
& \mathrm{if} \ (x, y) \in (\mathcal{X} \times \mathcal{Y}) \setminus \mathcal{Z} ,
\\
B^{\ast}
& \mathrm{if} \ x = x^{\ast}( y ) ,
\\
f_{\mathrm{Sh.}}( x, y )
& \mathrm{if} \ x \in \mathcal{A}_{X|Y=y}^{\delta^{\ast}(y)}
\end{cases}
\label{eq:Shannon-code}
\end{align}
for each $(x, y) \in \mathcal{X} \times \mathcal{Y}$, where $B_{y}^{\ast}$ is a r.v.\ given as
\begin{align}
B_{y}^{\ast}
=
\begin{dcases}
f_{\mathrm{Sh.}}(a, b)
& \mathrm{with} \ \mathrm{probability} \ 1 - \frac{ 1 - \delta(y) - P_{X|Y}(\mathcal{A}_{X|Y=y}^{\delta^{\ast}(y)} \mid y) }{ P_{X|Y}(x^{\ast}(y) \mid y) } ,
\\
f_{\mathrm{Sh.}}(x^{\ast}(y), y)
& \mathrm{with} \ \mathrm{probability} \ \frac{ 1 - \delta(y) - P_{X|Y}(\mathcal{A}_{X|Y=y}^{\delta^{\ast}(y)} \mid y) }{ P_{X|Y}(x^{\ast}(y) \mid y) }
\end{dcases}
\end{align}
for each $y \in \mathcal{Y}$.
Since the Shannon code $x \mapsto f_{\mathrm{Sh.}}(x, y)$ is prefix-free for fixed $y$, it is clear that $\mathcal{C}_{y}(X, Y, F)$ is prefix-free for every $y \in \mathcal{Y}$.
On the other hand, construct a deterministic decoder $g : \{ 0, 1 \}^{\ast} \times \mathcal{Y} \to \mathcal{X}$ so that
\begin{align}
g( \bvec{b}, y )
=
\begin{cases}
x
& \mathrm{if} \ x \in \mathcal{A}_{X|Y=y}^{\delta^{\ast}(y)} \ \mathrm{and} \ \bvec{b} = f_{\mathrm{Sh.}}(x, y) ,
\\
x^{\ast}(y)
& \mathrm{otherwise}
\end{cases}
\label{eq:decoder-g_Shannon-code}
\end{align}
for each $(\bvec{b}, y) \in \{ 0, 1 \}^{\ast} \times \mathcal{Y}$.
We observe that
\begin{align}
\mathbb{P}\{ X \neq g(F(X, Y), Y) \}
& =
1 - \sum_{y \in \mathcal{Y}} P_{Y}( y ) \, \mathbb{P}\{ X = g(F(X, Y), Y) \mid Y = y \}
\notag \\
& \overset{\mathclap{\text{(a)}}}{=}
1 - \sum_{y \in \mathcal{Y}} P_{Y}( y ) \, \left( \mathbb{P}\Big\{ X \in \mathcal{A}_{X|Y=y}^{\delta^{\ast}(y)} \ \Big| \ Y = y \Big\} + \left( \frac{ 1 - \delta(y) - P_{X|Y}(\mathcal{A}_{X|Y=y}^{\delta^{\ast}(y)} \mid y) }{ P_{X|Y}(x^{\ast}(y) \mid y) } \right) \, \mathbb{P}\{ X = x^{\ast}(y) \mid Y = y \} \right)
\notag \\
& =
\sum_{y \in \mathcal{Y}} P_{Y}( y ) \, \delta^{\ast}( y )
\notag \\
& \overset{\mathclap{\text{(b)}}}{=}
\varepsilon ,
\label{eq:error-probab_stochastic-Shannon}
\end{align}
where
\begin{itemize}
\item
(a) follows by the construction of $(F, g)$ stated in \eqref{eq:Shannon-code} and \eqref{eq:decoder-g_Shannon-code},
\item
(b) follows from the fact that $\delta^{\ast}(\cdot) \in \mathcal{E}_{0}( \varepsilon )$; see \lemref{lem:Kuzuoka-formula}.
\end{itemize}
Therefore, we have
\begin{align}
\tilde{\Lambda}_{\mathrm{avg}}^{\ast}(X, Y \, \| \, \rho, \varepsilon)
& \overset{\mathclap{\text{(a)}}}{\le}
\frac{ \tilde{\Lambda}(X, Y, F, g \, \| \, \rho) }{ \rho }
\notag \\
& =
\frac{ 1 }{ \rho } \log \mathbb{E}[ 2^{\rho \ell(F(X, Y))} \, \bvec{1}_{\{ X = g(F(X, Y), Y) \}} ]
\notag \\
& \overset{\mathclap{\text{(b)}}}{<}
\bar{\mathsf{H}}_{1/(1+\rho)}^{\delta^{\ast}(\cdot)}(X \mid Y) + 1
\notag \\
& \overset{\mathclap{\text{(c)}}}{\le}
H_{1/(1+\rho)}^{\varepsilon}(X \mid Y) + 1 + \zeta ,
\label{eq:derivation_tilde_one-shot_achievability}
\end{align}
where
\begin{itemize}
\item
(a) follows from \eqref{eq:error-probab_stochastic-Shannon} and the fact that $\mathcal{C}_{y}(X, Y, F)$ is prefix-free for every $y \in \mathcal{Y}$,
\item
(b) follows from \lemref{lem:unified_direct} and \eqref{eq:Shannon_code_redundancy} with $\kappa(x, y) = 2^{\ell( f_{\mathrm{Sh.}}( x, y ) )}$, and
\item
(c) follows from \eqref{eq:choice_delta-ast_zeta}.
\end{itemize}
Since $\zeta > 0$ is arbitrary, this completes the proof of \lemref{lem:tilde_one-shot}.
\hfill\IEEEQEDhere

\section{Proof of \lemref{lem:no-tilde_tilde}}
\label{app:no-tilde_tilde}

The left-hand inequality of \eqref{eq:no-tilde_tilde} is clear by the definitions of $\Lambda_{\mathrm{avg}}^{\ast}(X, Y \, \| \, \rho, \varepsilon)$ and $\tilde{\Lambda}_{\mathrm{avg}}^{\ast}(X, Y \, \| \, \rho, \varepsilon)$ stated in \eqref{def:Lambda-ast_avg} and \eqref{eq:tilde-Lambda}, respectively.

It remains to verify the right-hand inequality of \eqref{eq:no-tilde_tilde}.
Let $\zeta$ be an arbitrary positive number.
By the definition of $\tilde{\Lambda}_{\mathrm{avg}}^{\ast}(X, Y \, \| \, \rho, \varepsilon)$ stated in \eqref{eq:tilde-Lambda}, one can choose a variable-length semi-stochastic code $(F, g)$ such that $\mathcal{C}_{y}(X, Y, F)$ is prefix-free for every $y \in \mathcal{Y}$ and
\begin{align}
\rho \, \tilde{\Lambda}_{\mathrm{avg}}^{\ast}(X, Y \, \| \, \rho, \varepsilon)
& \ge
\tilde{\Lambda}(X, Y, F, g \, \| \, \rho) - \zeta ,
\label{eq:tilde-Lambda_zeta}
\\
\mathbb{P}\{ X \neq g(F(X, Y), Y) \}
& \le
\varepsilon .
\label{eq:error-probab_tilde-Lambda_zeta}
\end{align}
Now, construct another stochastic encoder $F^{\prime} : \mathcal{X} \times \mathcal{Y} \to \{ 0, 1 \}^{\ast}$ so that
\begin{align}
F^{\prime}(x, y)
=
\begin{cases}
\tilde{\bvec{b}}(x, y)
& \mathrm{if} \ x \neq g(F(x, y), y) ,
\\
F(x, y)
& \mathrm{if} \ x = g(F(x, y), y)
\end{cases}
\label{def:F-prime}
\end{align}
for each $x \in \mathcal{X}$, where $\tilde{\bvec{b}}(x, y)$ is chosen so that
\begin{align}
\tilde{\bvec{b}}(x, y)
\in
\argmin\limits_{\substack{ \bvec{b} \in \mathcal{C}_{y}(X, Y, F) : \\ x = g(\bvec{b}, y) }} \ell( \bvec{b} ) .
\label{def:choice_tilde-b}
\end{align}
Since $\mathcal{C}_{y}(X, Y, F^{\prime}) \subset \mathcal{C}_{y}(X, Y, F)$, it is clear that $\mathcal{C}_{y}(X, Y, F^{\prime})$ is also prefix-free for every $y \in \mathcal{Y}$.
Moreover, we readily see that
\begin{align}
\mathbb{P}\{ X \neq g(F^{\prime}(X, Y), Y) \}
=
\mathbb{P}\{ X \neq g(F(X, Y), Y) \}
\le
\varepsilon .
\label{eq:error-probab_F-prime}
\end{align}
In addition, we see from \eqref{def:F-prime}--\eqref{eq:error-probab_F-prime} that
\begin{align}
\mathbb{E}[ 2^{\rho \ell(F^{\prime}(X, Y))} ]
& =
\mathbb{E}[ 2^{\rho \ell(F(X, Y))} \bvec{1}_{\{ X = g(F(X, Y), Y) \}} ] + \mathbb{E}[ 2^{\rho \ell(\tilde{\bvec{b}}(X, Y))} \bvec{1}_{\{ X \neq g(F(X, Y), Y) \}} ]
\notag \\
& \le
\frac{ \mathbb{E}[ 2^{\rho \ell(F(X, Y))} \bvec{1}_{\{ X = g(F(X, Y), Y) \}} ] }{ 1 - \varepsilon }
\notag \\
& =
\exp\Big( \tilde{\Lambda}(X, Y, F, g \, \| \, \rho) \Big) \, \left( \frac{ 1 }{ 1 - \varepsilon } \right) ,
\label{eq:MGF_F-prime}
\end{align}
where the inequality follows from \eqref{eq:error-probab_tilde-Lambda_zeta} and the fact that
\begin{align}
2^{\rho \, \ell(\tilde{\bvec{b}}(x, y))}
\le
\frac{ \mathbb{E}[ 2^{\rho \ell(F(X, Y))} \bvec{1}_{\{ X = g(F(X, Y), Y) \}} ] }{ 1 - \varepsilon }
\end{align}
for every $(x, y) \in \mathcal{X} \times \mathcal{Y}$.
Therefore, we observe that
\begin{align}
\exp\Big( \rho \, \Lambda_{\mathrm{avg}}^{\ast}(X, Y \, \| \, \rho, \varepsilon) \Big)
& \overset{\mathclap{\text{(a)}}}{\le}
\mathbb{E}[ 2^{\rho \, \ell( F^{\prime}(X, Y) )} ]
\notag \\
& \overset{\mathclap{\text{(b)}}}{\le}
\exp\Big( \tilde{\Lambda}(X, Y, F, g \, \| \, \rho) \Big) \left( \frac{ 1 }{ 1 - \varepsilon } \right)
\notag \\
& \overset{\mathclap{\text{(c)}}}{\le}
\exp\Big( \rho \, \tilde{\Lambda}_{\mathrm{avg}}^{\ast}(X, Y \, \| \, \rho, \varepsilon) + \zeta \Big) \left( \frac{ 1 }{ 1 - \varepsilon } \right)
\label{eq:zeta_no-tilde_tilde}
\end{align}
where
\begin{itemize}
\item
(a) follows from \eqref{eq:error-probab_F-prime} and the definition of $\Lambda_{\mathrm{avg}}^{\ast}(X, Y \, \| \, \rho, \varepsilon)$ stated in \eqref{def:Lambda-ast_avg},
\item
(b) follows from \eqref{eq:MGF_F-prime}, and
\item
(d) follows from \eqref{eq:tilde-Lambda_zeta}.
\end{itemize}
As $\zeta > 0$ is arbitrary, we obtain the right-hand inequality of \eqref{eq:no-tilde_tilde} from \eqref{eq:zeta_no-tilde_tilde}.
This completes the proof of \lemref{lem:no-tilde_tilde}.
\hfill\IEEEQEDhere

\section{Proof of \lemref{lem:tilde_one-shot_max}}
\label{app:tilde_one-shot_max}

Firstly, we shall verify the converse bound of \lemref{lem:tilde_one-shot_max}, i.e., the left-hand inequality of \eqref{eq:kuzuoka_one-shot_max}.
Consider a variable-length semi-stochastic code $(F, g)$ such that $\mathcal{C}_{y}(X, Y, F)$ is prefix-free for every $y \in \mathcal{Y}$ and
\begin{align}
\sup_{y \in \mathcal{Y}} \mathbb{P}\{ X \neq g(F(X, Y), Y) \mid Y = y \}
\le
\varepsilon .
\label{eq:error_initial-FG_max}
\end{align}
Consider the deterministic maps $\bvec{b} : \mathcal{X} \times \mathcal{Y} \to \{ 0, 1 \}^{\ast}$ and $\epsilon : \mathcal{X} \times \mathcal{Y} \to [0, 1]$ as defined in \eqref{eq:short-string_Set-B} and \eqref{eq:eps_each-error}, respectively.
Since \eqref{eq:error_initial-FG_max} implies that \eqref{eq:maximum-error} holds, it follows from \lemref{lem:unified_converse_max} and \eqref{eq:tilde-Lambda_constructive-LB} that
\begin{align}
\frac{ \tilde{\Lambda}(X, Y, F, g \, \| \, \rho) }{ \rho }
& \ge
\check{H}_{1/(1+\rho)}^{\varepsilon}(X \mid Y) - \log R(\epsilon, 2^{\ell(\bvec{b}(\cdot, \cdot))}) ,
\end{align}
which yields the left-hand inequality of \eqref{eq:kuzuoka_one-shot_max} together with \eqref{eq:redundancy_Kraft-McMillan}.

Finally, replacing the map $\delta^{\ast} : \mathcal{Y} \to [0, 1]$ chosen in \eqref{eq:choice_delta-ast_zeta} by the constant $0 \le \varepsilon < 1$, it can be verified by the same way as we did for \eqref{eq:derivation_tilde_one-shot_achievability} that
\begin{align}
\tilde{\Lambda}_{\max}^{\ast}(X, Y \, \| \, \rho, \varepsilon)
& <
\bar{\mathsf{H}}_{1/(1+\rho)}^{\varepsilon}(X \mid Y) + 1
\notag \\
& =
\check{H}_{1/(1+\rho)}^{\varepsilon}(X \mid Y) + 1 ,
\end{align}
where the last inequality follows from \eqref{eq:check_identity1}.
This completes the proof of \lemref{lem:tilde_one-shot_max}.
\hfill\IEEEQEDhere

\section{Proof of \lemref{lem:no-tilde_tilde_max}}
\label{app:no-tilde_tilde_max}

The left-hand inequality of \eqref{eq:no-tilde_tilde_max} is clear by the definitions of $\Lambda_{\max}^{\ast}(X, Y \, \| \, \rho, \varepsilon)$ and $\tilde{\Lambda}_{\max}^{\ast}(X, Y \, \| \, \rho, \varepsilon)$ stated in \eqref{def:Lambda-ast_max} and \eqref{eq:tilde-Lambda_max}, respectively.

It thus remains to verify the right-hand inequality of \eqref{eq:no-tilde_tilde_max}.
Let $\zeta$ be an arbitrary positive real number.
By the definition of $\tilde{\Lambda}_{\max}^{\ast}(X, Y \, \| \, \rho, \varepsilon)$ stated in \eqref{eq:tilde-Lambda_max}, one can choose a variable-length semi-stochastic code $(F, g)$ such that $\mathcal{C}_{y}(X, Y, F)$ is prefix-free for every $y \in \mathcal{Y}$ and
\begin{align}
\rho \, \tilde{\Lambda}_{\max}^{\ast}(X, Y \, \| \, \rho, \varepsilon)
& \ge
\tilde{\Lambda}(X, Y, F, g \, \| \, \rho) - \zeta ,
\label{eq:tilde-Lambda_max_zeta}
\\
\sup_{y \in \mathcal{Y}} \mathbb{P}\{ X \neq g(F(X, Y), Y) \mid Y = y \}
& \le
\varepsilon .
\label{eq:error-probab_tilde-Lambda_max_zeta}
\end{align}
Consider a stochastic encoder $F^{\prime} : \mathcal{X} \times \mathcal{Y} \to \{ 0, 1 \}^{\ast}$ and a deterministic map $\tilde{\bvec{b}} : \mathcal{X} \times \mathcal{Y} \to \{ 0 , 1 \}^{\ast}$ defined as in \eqref{def:F-prime} and \eqref{def:choice_tilde-b}, respectively.
Then, the codeword set $\mathcal{C}_{y}(X, Y, F)$ is prefix-free for every $y \in \mathcal{Y}$, and it follows from \eqref{eq:error-probab_tilde-Lambda_max_zeta} that
\begin{align}
\sup_{y \in \mathcal{Y}} \mathbb{P}\{ X \neq g(F^{\prime}(X, Y), Y) \mid Y = y \}
\le
\varepsilon .
\end{align}
Hence, similar to \eqref{eq:zeta_no-tilde_tilde}, we obtain
\begin{align}
\exp\Big( \rho \, \Lambda_{\max}^{\ast}(X, Y \, \| \, \rho, \varepsilon) \Big)
& \le
\exp\Big( \tilde{\Lambda}(X, Y, F, g \, \| \, \rho) \Big) \left( \frac{ 1 }{ 1 - \varepsilon } \right)
\notag \\
& \le
\exp\Big( \rho \, \tilde{\Lambda}_{\max}^{\ast}(X, Y \, \| \, \rho, \varepsilon) + \zeta \Big) \left( \frac{ 1 }{ 1 - \varepsilon } \right) ,
\end{align}
where the last inequality follows from \eqref{eq:tilde-Lambda_max_zeta}.
This completes the proof of \lemref{lem:no-tilde_tilde_max}.

\section{Proof of \lemref{lem:optimal-strategy_avg}}
\label{app:optimal-strategy_avg}

Equation~\eqref{eq:optimal-guessing_error-probab_avg} can be verified as
\begin{align}
\mathbb{P}\{ \bar{\mathsf{G}}_{\mathrm{avg}}^{\ast}(X, Y) = 0 \}
& =
\sum_{y \in \mathcal{Y}} P_{Y}(y) \, \mathbb{P}\{ \bar{\mathsf{G}}_{\mathrm{avg}}^{\ast}(X, Y) = 0 \mid Y = y \}
\notag \\
& \overset{\mathclap{\text{(a)}}}{=}
\sum_{y \in \mathcal{Y}} P_{Y}( y ) \, \left( 1 - \sum_{k = 1}^{\infty} P_{X|Y}(\varsigma_{y}(k) \mid y) \prod_{j = 1}^{k} \Big( 1 - \pi_{\mathrm{avg}}^{\ast}(j, y) \Big) \right)
\notag \\
& \overset{\mathclap{\text{(b)}}}{=}
\sum_{y \in \mathcal{Y}} P_{Y}( y ) \, \left( 1 - \sum_{k = 1}^{J} P_{X|Y}(\varsigma_{y}(k) \mid y) - \xi \right)
\notag \\
& \overset{\mathclap{\text{(c)}}}{=}
\varepsilon ,
\label{eq:optimal-guessing_error-probab_avg_proof}
\end{align}
where
\begin{itemize}
\item
(a) follows from \eqref{def:giving-up_guessing_function}; see also \cite[Equation~(29)]{kuzuoka_2019},
\item
(b) follows from \eqref{eq:pi-ast_avg}, and
\item
(c) follows from \eqref{def:xi}.
\end{itemize}

Consider a guessing strategy $(\mathsf{g}, \pi)$, and the giving-up guessing function $\bar{G} : \mathcal{X} \times \mathcal{Y} \to \mathbb{N} \cup \{ 0 \}$ induced by $(\mathsf{g}, \pi)$.
Suppose that
\begin{align}
\mathbb{P}\{ \bar{\mathsf{G}}(X, Y) = 0 \}
\le
\varepsilon .
\label{eq:avg-error_guessing}
\end{align}
To prove \eqref{eq:optimal-guessing_moment_avg}, it suffices to show that
\begin{align}
\mathbb{E}[ \bar{\mathsf{G}}(X, Y)^{\rho} ]
\ge
\mathbb{E}[ \bar{\mathsf{G}}_{\mathrm{avg}}^{\ast}(X, Y)^{\rho} ]
\end{align}
for every positive real number $\rho$.
Now, we shall verify that
\begin{align}
\mathbb{P}\{ \bar{\mathsf{G}}(X, Y) \ge k \}
\ge
\mathbb{P}\{ \bar{\mathsf{G}}_{\mathrm{avg}}^{\ast}(X, Y) \ge k \}
\label{eq:majorization_guessing}
\end{align}
for every positive integer $k$.
It follows from \eqref{eq:optimal-guessing_error-probab_avg} and \eqref{eq:avg-error_guessing} that
\begin{align}
\mathbb{P}\{ \bar{\mathsf{G}}(X, Y) = 0 \}
\le
\mathbb{P}\{ \bar{\mathsf{G}}_{\mathrm{avg}}^{\ast}(X, Y) = 0 \} .
\end{align}
In addition, since $x \mapsto \mathsf{g}(x, y)$ is bijective for each $y \in \mathcal{Y}$ and $\varsigma_{y} : \mathbb{N} \to \mathcal{X}$ rearranges the probability masses in $P_{X|Y}(\cdot \mid y)$ in nonincreasing order (see \eqref{def:varsigma}), we see that
\begin{align}
\sum_{l = 1}^{k} \sum_{y \in \mathcal{Y}} P_{Y}( y ) \, \mathbb{P}\{ \mathsf{g}(X, Y) = l \mid Y = y \}
\le
\sum_{l = 1}^{k} \sum_{y \in \mathcal{Y}} P_{Y}( y ) \, P_{X|Y}(\varsigma_{y}(l) \mid y)
\label{eq:rearrange_majorization_avg}
\end{align}
for every positive integer $k$.
Thus, we observe that
\begin{align}
\mathbb{P}\{ \bar{\mathsf{G}}(X, Y) \le k \}
& =
\mathbb{P}\{ \bar{\mathsf{G}}(X, Y) = 0 \} + \sum_{l = 1}^{k} \mathbb{P}\{ \bar{\mathsf{G}}(X, Y) = l \}
\notag \\
& \le
\mathbb{P}\{ \bar{\mathsf{G}}(X, Y) = 0 \} + \sum_{l = 1}^{k} \mathbb{P}\{ \mathsf{g}(X, Y) = l \}
\notag \\
& \overset{\mathclap{\text{(a)}}}{\le}
\mathbb{P}\{ \bar{\mathsf{G}}(X, Y) = 0 \} + \sum_{l = 1}^{k} \sum_{y \in \mathcal{Y}} P_{Y}( y ) \, P_{X|Y}(\varsigma_{y}(l) \mid y)
\notag \\
& \overset{\mathclap{\text{(b)}}}{=}
\mathbb{P}\{ \bar{\mathsf{G}}(X, Y) = 0 \} + \mathbb{P}\{ 1 \le \bar{\mathsf{G}}_{\mathrm{avg}}^{\ast}(X, Y) \le k \}
\notag \\
& \overset{\mathclap{\text{(c)}}}{\le}
\mathbb{P}\{ \bar{\mathsf{G}}_{\mathrm{avg}}^{\ast}(X, Y) \le k \}
\label{eq:CDF_guessing_avg}
\end{align}
for every $0 \le k \le J$, where
\begin{itemize}
\item
(a) follows from \eqref{eq:rearrange_majorization_avg},
\item
(b) follows from \eqref{def:giving-up_guessing_function}, \eqref{eq:optimal-guessing}, and \eqref{eq:pi-ast_avg}, and
\item
(c) follows from \eqref{eq:optimal-guessing_error-probab_avg}.
\end{itemize}
In addition, we get
\begin{align}
\mathbb{P}\{ \bar{\mathsf{G}}_{\mathrm{avg}}^{\ast}(X, Y) \le J + 1 \}
& =
\sum_{l = 0}^{J+1} \mathbb{P}\{ \bar{\mathsf{G}}_{\mathrm{avg}}^{\ast}(X, Y) = l \}
\notag \\
& \overset{\mathclap{\text{(a)}}}{=}
\varepsilon + \sum_{l = 1}^{J+1} \mathbb{P}\{ \bar{\mathsf{G}}_{\mathrm{avg}}^{\ast}(X, Y) = l \}
\notag \\
& \overset{\mathclap{\text{(b)}}}{=}
\varepsilon + \sum_{l = 1}^{J+1} \sum_{y \in \mathcal{Y}} P_{Y}( y ) \, P_{X|Y}(\varsigma_{y}(l) \mid y) \prod_{j = 1}^{l} \Big( 1 - \pi_{\mathrm{avg}}^{\ast}(j, y) \Big)
\notag \\
& \overset{\mathclap{\text{(c)}}}{=}
\varepsilon + \sum_{y \in \mathcal{Y}}  P_{Y}(y) \sum_{l = 1}^{J} P_{X|Y}(\varsigma_{y}(l) \mid y) + \xi
\notag \\
& \overset{\mathclap{\text{(d)}}}{=}
1 ,
\label{eq:CDF_opt-guessing_avg}
\end{align}
where
\begin{itemize}
\item
(a) follows from \eqref{eq:optimal-guessing_error-probab_avg},
\item
(b) follows from \eqref{def:giving-up_guessing_function}, \eqref{eq:optimal-guessing}, and \eqref{eq:pi-ast_avg},
\item
(c) follows from \eqref{eq:pi-ast_avg}, and
\item
(d) follows from \eqref{def:xi}.
\end{itemize}
Combining \eqref{eq:CDF_guessing_avg} and \eqref{eq:CDF_opt-guessing_avg}, we have that \eqref{eq:majorization_guessing} holds.
Therefore, we obtain
\begin{align}
\mathbb{E}[ \bar{\mathsf{G}}(X, Y)^{\rho} ]
& =
\sum_{k = 1}^{\infty} \Big( k^{\rho} - (k-1)^{\rho} \Big) \, \mathbb{P}\{ \bar{\mathsf{G}}(X, Y) \ge k \}
\notag \\
& \ge
\sum_{k = 1}^{\infty} \Big( k^{\rho} - (k-1)^{\rho} \Big) \, \mathbb{P}\{ \bar{\mathsf{G}}_{\mathrm{avg}}^{\ast}(X, Y) \ge k \}
\notag \\
& =
\mathbb{E}[ \bar{\mathsf{G}}_{\mathrm{avg}}^{\ast}(X, Y)^{\rho} ] ,
\end{align}
proving \eqref{eq:optimal-guessing_moment_avg}.
This completes the proof of \lemref{lem:optimal-strategy_avg}.

\section{Proof of \thref{th:one-shot_guess_avg}}
\label{app:one-shot_guess_avg}

\subsection{Converse Part}
\label{app:one-shot_guess_avg_converse}

To prove the left-hand inequality of \eqref{eq:one-shot_guess_avg}, it suffices to consider the optimal guessing strategy $(\mathsf{g}^{\ast}, \pi_{\mathrm{avg}}^{\ast})$ given in \lemref{lem:optimal-strategy_avg}.
Choose the map $\delta : \mathcal{Y} \to [0, 1]$ given by the formula
\begin{align}
\delta( y )
=
\left| - \sum_{k = 1}^{J} P_{X|Y}(\varsigma_{y}(k) \mid y) + \sum_{b \in \mathcal{Y}} P_{Y}( b ) \sum_{l = 1}^{J} P_{X|Y}(\varsigma_{b}(l) \mid b) + \varepsilon \right|_{+}
\label{def:choice_delta_J}
\end{align}
for each $y \in \mathcal{Y}$, where $J$ is given in \eqref{def:J}.
We observe that
\begin{align}
\log( J + 1 )
& \overset{\mathclap{\text{(a)}}}{\le}
\frac{ 1 }{ \varepsilon } \sum_{y \in \mathcal{Y}} P_{Y}(y) \, \delta(y) \log( J + 1 )
\notag \\
& \overset{\mathclap{\text{(b)}}}{\le}
\frac{ 1 }{ \varepsilon } \sum_{\substack{ y \in \mathcal{Y} : \\ \delta(y) > 0 }} P_{Y}( y ) \, \delta(y) \inf\left\{ R > 0 \ \middle| \ \mathbb{P}\left\{ \log \frac{ 1 }{ P_{X|Y}(X \mid Y) } >  R \ \middle| \ Y = y \right\} \le \delta(y) \right\}
\notag \\
& \overset{\mathclap{\text{(c)}}}{\le}
\frac{ 1 }{ \varepsilon } \sum_{\substack{ y \in \mathcal{Y} : \\ \delta(y) > 0 }} P_{Y}( y ) \, \delta(y) \inf\left\{ R > 0 \ \middle| \ \frac{ H( P_{X|Y=y} ) }{ R } \le \delta(y) \right\}
\notag \\
& \le
\frac{ 1 }{ \varepsilon } \sum_{y \in \mathcal{Y}} P_{Y}( y ) \, H( P_{X|Y=y} )
\notag \\
& =
\frac{ H(X \mid Y) }{ \varepsilon } ,
\label{eq:guessing_Markov_avg}
\end{align}
where
\begin{itemize}
\item
(a) follows by the choice of $\delta : \mathcal{Y} \to [0, 1]$ stated in \eqref{def:choice_delta_J},
\item
(b) follows from the fact that
\begin{align}
\delta( y )
<
\sum_{k = J+1}^{\infty} P_{X|Y}(\varsigma_{y}(k) \mid y)
\end{align}
for each $y \in \mathcal{Y}$, and
\item
(c) follows by Markov's inequality.
\end{itemize}
Letting
\begin{align}
\kappa(x, y)
& =
\mathsf{g}^{\ast}(x, y) ,
\\
\epsilon(x, y)
& =
\prod_{k = 1}^{\mathsf{g}^{\ast}(x, y)} \Big( 1 - \pi_{\mathrm{avg}}^{\ast}(k, y) \Big)
\end{align}
for each $(x, y) \in \mathcal{X} \times \mathcal{Y}$, we have
\begin{align}
\frac{ 1 }{ \rho } \log \mathbb{E}[ \bar{\mathsf{G}}_{\mathrm{avg}}^{\ast}(X, Y)^{\rho} ]
& \overset{\mathclap{\text{(a)}}}{\ge}
H_{\alpha}^{\varepsilon}(X \mid Y) - \log \left( \sum_{k = 1}^{J+1} \frac{ 1 }{ k } \right)
\notag \\
& \overset{\mathclap{\text{(b)}}}{\ge}
H_{\alpha}^{\varepsilon}(X \mid Y) - \log (1 + \log(J+1))
\notag \\
& \overset{\mathclap{\text{(c)}}}{\ge}
H_{\alpha}^{\varepsilon}(X \mid Y) - \log \left( 1 + \frac{ H(X \mid Y) }{ \varepsilon } \right) ,
\label{eq:guessing-avg_converse_unified-way}
\end{align}
where
\begin{itemize}
\item
(a) follows from \lemref{lem:unified_converse_avg},
\item
(b) follows from the fact that
\begin{align}
\sum_{k = 1}^{m} \frac{ 1 }{ k }
\le
1 + \log m ,
\label{eq:harmonic_log}
\end{align}
and
\item
(c) follows from \eqref{eq:guessing_Markov_avg}.
\end{itemize}
This completes the proof of the converse bound of \thref{th:one-shot_guess_avg}, i.e., the left-hand inequality of \eqref{eq:one-shot_guess_avg}.
\hfill\IEEEQEDhere

\subsection{Achievability Part}
\label{app:one-shot_guess_avg_direct}

We shall verify the right-hand inequality of \eqref{eq:one-shot_guess_avg}.
Fix a positive real number $\zeta$ arbitrarily, and choose a map $\delta^{\ast} : \mathcal{Y} \to [0, 1]$ by the same manner as \eqref{eq:choice_delta-ast_zeta}.
For each $y \in \mathcal{Y}$, choose an integer $\tilde{J}( y )$ so that
\begin{align}
\tilde{J}( y )
=
\sup\left\{ j \ge 0 \ \middle| \ \sum_{k = 1}^{j} P_{X|Y}(\varsigma_{y}(k) \mid y) < 1 - \delta^{\ast}(y) \right\} ,
\end{align}
and choose a real number $\tilde{M}( y )$ so that
\begin{align}
\tilde{M}( y )
=
1 - \delta^{\ast}( y ) - \sum_{k = 1}^{\tilde{J}(y)} P_{X|Y}(\varsigma_{y}(k) \mid y) .
\end{align}
Consider the optimal guessing function $\mathsf{g}^{\ast} : \mathcal{X} \times \mathcal{Y} \to \mathbb{N}$ given in \eqref{eq:optimal-guessing}.
If $\varsigma_{y}^{-1}(x) \le \tilde{J}( y )$, then
\begin{align}
\mathsf{g}^{\ast}(x, y)
& =
\sum_{k = 1}^{\varsigma_{y}^{-1}(x)} 1
\notag \\
& \le
\sum_{k = 1}^{\varsigma_{y}^{-1}(x)} \left( \frac{ P_{X|Y}(\varsigma_{y}(k) \mid y) }{ P_{X|Y}(x \mid y) } \right)^{1/(1+\rho)}
\notag \\
& \le
\sum_{k = 1}^{J(y)} \left( \frac{ P_{X|Y}(\varsigma_{y}(k) \mid y) }{ P_{X|Y}(x \mid y) } \right)^{1/(1+\rho)} + \left( \frac{ \tilde{M}( y ) }{ P_{X|Y}(x \mid y) } \right)^{1/(1+\rho)}
\notag \\
& =
\frac{ 1 }{ Q_{X|Y}^{(1/(1+\rho), \delta^{\ast}(\cdot))}(x \mid y) } ,
\end{align}
where the last equality follows from \eqref{def:tilted-R}.
In addition, if $\varsigma_{y}^{-1}(x) = \tilde{J}(y) + 1$, then
\begin{align}
\mathsf{g}^{\ast}(x, y)
& =
\sum_{k = 1}^{\tilde{J}(y)+1} 1
\notag \\
& \le
\sum_{k = 1}^{\tilde{J}(y)} \left( \frac{ P_{X|Y}(\varsigma_{y}(k) \mid y) }{ \tilde{M}( y ) } \right)^{1/(1+\rho)} + \left( \frac{ \tilde{M}( y ) }{ \tilde{M}( y ) } \right)^{1/(1+\rho)}
\notag \\
& =
\frac{ 1 }{ Q_{X|Y}^{(1/(1+\rho), \delta^{\ast}(\cdot))}(x \mid y) } .
\end{align}
Therefore, noting that $Q_{X|Y}^{(1/(1+\rho), \delta^{\ast}(\cdot))}(x \mid y) = 0$ if $\varsigma_{y}^{-1}(x) \ge \tilde{J}(y) + 2$, we observe that
\begin{align}
\mathsf{g}^{\ast}(x, y) \, Q_{X|Y}^{(1/(1+\rho), \delta^{\ast}(\cdot))}(x \mid y)
\le
1
\label{eq:guessing_opt-scale_avg}
\end{align}
for every $y \in \mathcal{Y}$.

Given a deterministic map $\pi : \mathcal{X} \times \mathcal{Y} \to [0, 1]$, construct a deterministic map $\epsilon : \mathcal{X} \times \mathcal{Y} \to [0, 1]$ so that
\begin{align}
\epsilon(x, y)
\coloneqq
1 - \prod_{k = 1}^{\varsigma_{y}^{-1}( x )} \Big( 1 - \pi(k, y) \Big) .
\end{align}
Then, the giving-up guessing function $\bar{\mathsf{G}} : \mathcal{X} \times \mathcal{Y} \to \mathbb{N} \cup \{ 0 \}$ induced by $(\mathsf{g}^{\ast}, \pi)$ can be written as
\begin{align}
\bar{\mathsf{G}}(x, y)
=
\begin{cases}
\mathsf{g}(x, y)
& \mathrm{with} \ \mathrm{probability} \ 1 - \epsilon(x, y) ,
\\
0
& \mathrm{with} \ \mathrm{probability} \ \epsilon(x, y) .
\end{cases}
\end{align}
for each $(x, y) \in \mathcal{X} \times \mathcal{Y}$.
Therefore, it follows from \lemref{lem:unified_direct} and \eqref{eq:guessing_opt-scale_avg} that there exists a giving-up policy $\pi^{\ast} : \mathcal{X} \times \mathcal{Y} \to [0, 1]$ such that the guessing strategy $(\mathsf{g}^{\ast}, \pi^{\ast})$ induces $\bar{\mathsf{G}} : \mathcal{X} \times \mathcal{Y} \to \mathbb{N} \cup \{ 0 \}$ that satisfies
\begin{align}
\mathbb{P}\{ \bar{\mathsf{G}}(X, Y) = 0 \}
& =
\sum_{y \in \mathcal{Y}} P_{Y}( y ) \, \mathbb{P}\{ \bar{\mathsf{G}}(X, Y) = 0 \mid Y = y \}
\notag \\
& \overset{\mathclap{\text{(a)}}}{=}
\sum_{y \in \mathcal{Y}} P_{Y}( y ) \, \delta^{\ast}( y )
\notag \\
& \overset{\mathclap{\text{(b)}}}{=}
\varepsilon ,
\\
\frac{ 1 }{ \rho } \log \mathbb{E}[ \bar{\mathsf{G}}(X, Y)^{\rho} ]
& \overset{\mathclap{\text{(c)}}}{\le}
\bar{\mathsf{H}}_{1/(1+\rho)}^{\delta^{\ast}(\cdot)}(X \mid Y)
\notag \\
& \overset{\mathclap{\text{(d)}}}{\le}
H_{1/(1+\rho)}^{\varepsilon}(X \mid Y) + \zeta ,
\end{align}
where
\begin{itemize}
\item
(a) follows from \eqref{eq:cond-error_delta} of \lemref{lem:unified_direct},
\item
(b) follows from the fact that $\delta^{\ast}(\cdot) \in \mathcal{E}_{0}( \varepsilon )$; see \lemref{lem:Kuzuoka-formula},
\item
(c) follows from \eqref{eq:direct-bound_KN} of \lemref{lem:unified_direct}, and
\item
(d) follows by the choice of $\delta^{\ast}(\cdot)$ stated in \eqref{eq:choice_delta-ast_zeta}.
\end{itemize}
As $\zeta > 0$ is arbitrary, this proves the right-hand inequality of \eqref{eq:one-shot_guess_avg}, completing the proof of the achievability bound of \thref{th:one-shot_guess_avg}.
\hfill\IEEEQEDhere

\section{Proof of \lemref{lem:optimal-strategy_max}}
\label{app:optimal-strategy_max}

Equation~\eqref{eq:optimal-guessing_error-probab_max} can be verified by the same way as \eqref{eq:optimal-guessing_error-probab_avg_proof}.
Consider a guessing strategy $(\mathsf{g}, \pi)$, and the giving-up guessing function $\bar{G} : \mathcal{X} \times \mathcal{Y} \to \mathbb{N} \cup \{ 0 \}$ induced by $(\mathsf{g}, \pi)$.
Suppose that
\begin{align}
\mathbb{P}\{ \bar{\mathsf{G}}(X, Y) = 0 \mid Y = y \}
\le
\varepsilon
\label{eq:max-error_guessing}
\end{align}
for every $y \in \mathcal{Y}$.
To prove \eqref{eq:optimal-guessing_moment_max}, it suffices to show that
\begin{align}
\mathbb{E}[ \bar{\mathsf{G}}(X, Y)^{\rho} ]
\ge
\mathbb{E}[ \bar{\mathsf{G}}_{\max}^{\ast}(X, Y)^{\rho} ]
\label{eq:Gmax_optimal}
\end{align}
for every positive real number $\rho$.
Since $x \mapsto \mathsf{g}(x, y)$ is bijective for each $y \in \mathcal{Y}$ and $\varsigma_{y} : \mathbb{N} \to \mathcal{X}$ rearranges the probability masses in $P_{X|Y}(\cdot \mid y)$ in nonincreasing order (see \eqref{def:varsigma}), we see that
\begin{align}
\sum_{l = 1}^{k} \mathbb{P}\{ \mathsf{g}(X, Y) = l \mid Y = y \}
\le
\sum_{l = 1}^{k} P_{X|Y}(\varsigma_{y}(l) \mid y)
\label{eq:rearrange_majorization_max}
\end{align}
for every $y \in \mathcal{Y}$.
Thus, in the same way as we proved \eqref{eq:majorization_guessing}, we may observe that
\begin{align}
\mathbb{P}\{ \bar{\mathsf{G}}(X, Y) \ge k \mid Y = y \}
\ge
\mathbb{P}\{ \bar{\mathsf{G}}_{\max}^{\ast}(X, Y) \ge k \mid Y = y \}
\label{eq:majorization_guessing_max}
\end{align}
for every $y \in \mathcal{Y}$ and every positive integer $k$.
Therefore, we obtain
\begin{align}
\mathbb{E}[ \bar{\mathsf{G}}(X, Y)^{\rho} \mid Y = y ]
& =
\sum_{k = 1}^{\infty} \Big( k^{\rho} - (k-1)^{\rho} \Big) \, \mathbb{P}\{ \bar{\mathsf{G}}(X, Y) \ge k \mid Y = y \}
\notag \\
& \ge
\sum_{k = 1}^{\infty} \Big( k^{\rho} - (k-1)^{\rho} \Big) \, \mathbb{P}\{ \bar{\mathsf{G}}_{\max}^{\ast}(X, Y) \ge k \mid Y = y \}
\notag \\
& =
\mathbb{E}[ \bar{\mathsf{G}}_{\max}^{\ast}(X, Y)^{\rho} \mid Y = y ]
\end{align}
for every $y \in \mathcal{Y}$, proving \eqref{eq:Gmax_optimal}.
Therefore, we have \eqref{eq:optimal-guessing_moment_avg}.
This completes the proof of \lemref{lem:optimal-strategy_max}.
\hfill\IEEEQEDhere

\section{Proof of \thref{th:one-shot_guess_max}}
\label{app:one-shot_guess_max}

\subsection{Converse Part}
\label{app:one-shot_guess_max_converse}

To prove the left-hand inequality of \eqref{eq:one-shot_guess_max}, it suffices to consider the optimal guessing strategy $(\mathsf{g}^{\ast}, \pi_{\max}^{\ast})$ given in \lemref{lem:optimal-strategy_max}.
We observe that
\begin{align}
\log( J(y) + 1 )
& \overset{\mathclap{\text{(a)}}}{\le}
\inf\left\{ R > 0 \ \middle| \ \mathbb{P}\left\{ \log \frac{ 1 }{ P_{X|Y}(X \mid Y) } >  R \ \middle| \ Y = y \right\} \le \varepsilon \right\}
\notag \\
& \overset{\mathclap{\text{(b)}}}{\le}
\inf\left\{ R > 0 \ \middle| \ \frac{ H( P_{X|Y=y} ) }{ R } \le \varepsilon \right\}
\notag \\
& =
\frac{ H( P_{X|Y=y} ) }{ \varepsilon }
\label{eq:guessing_Markov_max}
\end{align}
for every $y \in \mathcal{Y}$, where
\begin{itemize}
\item
(a) follows from the fact that
\begin{align}
\varepsilon
<
\sum_{k = J(y)+1}^{\infty} P_{X|Y}(\varsigma_{y}(k) \mid y)
\end{align}
for each $y \in \mathcal{Y}$, and
\item
(b) follows by Markov's inequality.
\end{itemize}
Letting
\begin{align}
\kappa(x, y)
& =
\mathsf{g}^{\ast}(x, y) ,
\\
\epsilon(x, y)
& =
\prod_{k = 1}^{\mathsf{g}^{\ast}(x, y)} \Big( 1 - \pi_{\max}^{\ast}(k, y) \Big)
\end{align}
for each $(x, y) \in \mathcal{X} \times \mathcal{Y}$, we have
\begin{align}
\frac{ 1 }{ \rho } \log \mathbb{E}[ \bar{\mathsf{G}}_{\max}^{\ast}(X, Y)^{\rho} ]
& \overset{\mathclap{\text{(a)}}}{\ge}
\check{H}_{\alpha}^{\varepsilon}(X \mid Y) - \sup_{y \in \mathcal{Y}} \log \left( \sum_{k = 1}^{J(y)+1} \frac{ 1 }{ k } \right)
\notag \\
& \overset{\mathclap{\text{(b)}}}{\ge}
\check{H}_{\alpha}^{\varepsilon}(X \mid Y) - \sup_{y \in \mathcal{Y}} \log (1 + \log(J(y)+1))
\notag \\
& \overset{\mathclap{\text{(c)}}}{\ge}
\check{H}_{\alpha}^{\varepsilon}(X \mid Y) - \log \left( 1 + \frac{ \sup_{y \in \mathcal{Y}} H( P_{X|Y=y} ) }{ \varepsilon } \right) ,
\end{align}
where
\begin{itemize}
\item
(a) follows from \lemref{lem:unified_converse_max},
\item
(b) follows from \eqref{eq:harmonic_log}, and
\item
(c) follows from \eqref{eq:guessing_Markov_max}.
\end{itemize}
This completes the proof of the converse bound of \thref{th:one-shot_guess_max}, i.e., the left-hand inequality of \eqref{eq:one-shot_guess_max}.
\hfill\IEEEQEDhere

\subsection{Achievability Part}
\label{app:one-shot_guess_max_direct}

We shall verify the right-hand inequality of \eqref{eq:one-shot_guess_max}.
Choose a real number $\check{M}( y )$ so that
\begin{align}
\check{M}( y )
=
1 - \varepsilon - \sum_{k = 1}^{J(y)} P_{X|Y}(\varsigma_{y}(k) \mid y) .
\end{align}
Consider the optimal guessing function $\mathsf{g}^{\ast} : \mathcal{X} \times \mathcal{Y} \to \mathbb{N}$ given in \eqref{eq:optimal-guessing}.
If $\varsigma_{y}^{-1}(x) \le J( y )$, then
\begin{align}
\mathsf{g}^{\ast}(x, y)
& =
\sum_{k = 1}^{\varsigma_{y}^{-1}(x)} 1
\notag \\
& \le
\sum_{k = 1}^{\varsigma_{y}^{-1}(x)} \left( \frac{ P_{X|Y}(\varsigma_{y}(k) \mid y) }{ P_{X|Y}(x \mid y) } \right)^{1/(1+\rho)}
\notag \\
& \le
\sum_{k = 1}^{J(y)} \left( \frac{ P_{X|Y}(\varsigma_{y}(k) \mid y) }{ P_{X|Y}(x \mid y) } \right)^{1/(1+\rho)} + \left( \frac{ \check{M}( y ) }{ P_{X|Y}(x \mid y) } \right)^{1/(1+\rho)}
\notag \\
& =
\frac{ 1 }{ Q_{X|Y}^{(1/(1+\rho), \varepsilon)}(x \mid y) } ,
\end{align}
where the last equality follows from \eqref{def:tilted-R}.
In addition, if $\varsigma_{y}^{-1}(x) = J(y) + 1$, then
\begin{align}
\mathsf{g}^{\ast}(x, y)
& =
\sum_{k = 1}^{J(y)+1} 1
\notag \\
& \le
\sum_{k = 1}^{J(y)} \left( \frac{ P_{X|Y}(\varsigma_{y}(k) \mid y) }{ \check{M}( y ) } \right)^{1/(1+\rho)} + \left( \frac{ \check{M}( y ) }{ \check{M}( y ) } \right)^{1/(1+\rho)}
\notag \\
& =
\frac{ 1 }{ Q_{X|Y}^{(1/(1+\rho), \varepsilon)}(x \mid y) } .
\end{align}
Therefore, noting that $Q_{X|Y}^{(1/(1+\rho), \varepsilon)}(x \mid y) = 0$ if $\varsigma_{y}^{-1}(x) \ge J(y) + 2$, we observe that
\begin{align}
\mathsf{g}^{\ast}(x, y) \, Q_{X|Y}^{(1/(1+\rho), \varepsilon)}(x \mid y)
\le
1
\label{eq:guessing_opt-scale_max}
\end{align}
for every $y \in \mathcal{Y}$.

Given a deterministic map $\pi : \mathcal{X} \times \mathcal{Y} \to [0, 1]$, construct another deterministic map $\epsilon : \mathcal{X} \times \mathcal{Y} \to [0, 1]$ so that
\begin{align}
\epsilon(x, y)
\coloneqq
1 - \prod_{k = 1}^{\varsigma_{y}^{-1}( x )} \Big( 1 - \pi(k, y) \Big) .
\end{align}
Then, the giving-up guessing function $\bar{\mathsf{G}} : \mathcal{X} \times \mathcal{Y} \to \mathbb{N} \cup \{ 0 \}$ induced by $(\mathsf{g}^{\ast}, \pi)$ can be written as
\begin{align}
\bar{\mathsf{G}}(x, y)
=
\begin{cases}
\mathsf{g}(x, y)
& \mathrm{with} \ \mathrm{probability} \ 1 - \epsilon(x, y) ,
\\
0
& \mathrm{with} \ \mathrm{probability} \ \epsilon(x, y) .
\end{cases}
\end{align}
for each $(x, y) \in \mathcal{X} \times \mathcal{Y}$.
Therefore, it follows from \lemref{lem:unified_direct} and \eqref{eq:guessing_opt-scale_max} that there exists a giving-up policy $\pi^{\ast} : \mathcal{X} \times \mathcal{Y} \to [0, 1]$ such that the guessing strategy $(\mathsf{g}^{\ast}, \pi^{\ast})$ induces $\bar{\mathsf{G}} : \mathcal{X} \times \mathcal{Y} \to \mathbb{N} \cup \{ 0 \}$ satisfying
\begin{align}
\mathbb{P}\{ \bar{\mathsf{G}}(X, Y) = 0 \mid Y = y \}
& \overset{\mathclap{\text{(a)}}}{=}
\varepsilon
\end{align}
for every $y \in \mathcal{Y}$, and
\begin{align}
\frac{ 1 }{ \rho } \log \mathbb{E}[ \bar{\mathsf{G}}(X, Y)^{\rho} ]
& \overset{\mathclap{\text{(b)}}}{\le}
\bar{\mathsf{H}}_{1/(1+\rho)}^{\varepsilon}(X \mid Y)
\notag \\
& \overset{\mathclap{\text{(c)}}}{=}
\check{H}_{1/(1+\rho)}^{\varepsilon}(X \mid Y) ,
\end{align}
where
\begin{itemize}
\item
(a) follows from \eqref{eq:cond-error_delta} of \lemref{lem:unified_direct},
\item
(b) follows from \eqref{eq:direct-bound_KN} of \lemref{lem:unified_direct}, and
\item
(d) follows from \eqref{eq:check_identity1}.
\end{itemize}
This proves the right-hand inequality of \eqref{eq:one-shot_guess_avg}, completing the proof of the achievability bound of \thref{th:one-shot_guess_avg}.
\hfill\IEEEQEDhere

\section{Proof of \eqref{eq:one-shot_task_avg_converse}}
\label{app:one-shot_task_avg_converse}

Consider an assignment function $\mathsf{f} : \mathcal{X} \times \mathcal{Y} \to \{ 0, 1, 2, \dots, M \}$ and a stochastic map $\mathsf{E} : 2^{\mathcal{X}} \times \mathcal{Y} \to 2^{\mathcal{X}}$ satisfying \eqref{eq:error-stochas_partition} and
\begin{align}
\mathbb{P}\{ X \notin \mathsf{L}(\mathsf{f}(X, Y), Y) \}
\le
\varepsilon .
\label{eq:error-probab_one-shot_avg_converse}
\end{align}
For each $y \in \mathcal{Y}$, denote by $\{ \mathcal{L}(m, y) \}_{m = 1}^{M}$ the sub-partition of $\mathcal{X}$ induced by the assignment $x \mapsto \mathsf{f}(x, y)$; see \eqref{def:sub-partition}.
Define
\begin{align}
\mathcal{L}( y )
& \coloneqq
\bigcup_{\substack{ m = 1 : \\ \mathbb{P}\{ \mathsf{f}(X, Y) = \mathsf{E}(\mathsf{f}(X, Y), Y) = m  \mid Y = y \} > 0 }}^{M} \mathcal{L}( m, y )
\label{def:union_L-y}
\end{align}
for each $y \in \mathcal{Y}$.

Consider the stochastic sub-partition $\mathsf{L} : \{ 0, 1, 2, \dots, M \} \times \mathcal{Y} \to 2^{\mathcal{X}}$ induced by the pair $(\mathsf{f}, \mathsf{E})$; see \eqref{def:stochastic-sub-partition}.
Since
\begin{align}
m_{1} \neq m_{2}
\quad \Longrightarrow \quad
\mathcal{L}(m_{1}, y) \cap \mathcal{L}(m_{2}, y) = \emptyset
\end{align}
and
\begin{align}
\mathbb{E}[ |\mathsf{L}(\mathsf{f}(X, Y), Y)| \mid Y = y ]
=
\sum_{m = 1}^{M} \mathbb{P}\{ \mathsf{f}(X, Y) = \mathsf{E}(\mathsf{f}(X, Y), Y) = m \mid Y = y \} \, |\mathcal{L}(m, y)|
\end{align}
for every $y \in \mathcal{Y}$, we see that
\begin{align}
|\mathcal{L}( y )| \, \min_{\substack{ m \in \{ 1, \dots, M \} : \\ \mathbb{P}\{ \mathsf{f}(X, Y) = \mathsf{E}(\mathsf{f}(X, Y), Y) = m \mid Y = y \} > 0 }} \mathbb{P}\{ \mathsf{f}(X, Y) = \mathsf{E}(\mathsf{f}(X, Y), Y) = m \mid Y = y \}
\le
\mathbb{E}[ |\mathcal{L}(\mathsf{f}(X, Y), Y)| \mid Y = y ]
\le
|\mathcal{L}( y )|
\label{eq:cond-expectation_partitioning-moment_card-union}
\end{align}
for every $y \in \mathcal{Y}$.
Thus, the task sub-partitioning $\rho$-th moment $\mathbb{E}[ |\mathsf{L}(\mathsf{f}(X, Y), Y)|^{\rho} ]$ is finite if and only if $\mathcal{L}( y )$ is finite for every $y \in \mathcal{Y}$.
Therefore, to prove the converse bound stated in \eqref{eq:one-shot_task_avg_converse}, it suffices to assume that $\mathcal{L}( y )$ is finite for every $y \in \mathcal{Y}$.

Since $\mathcal{L}( y )$ is finite, it follows from \cite[Proposition~III.1]{bunte_lapidoth_2014} that for every $y \in \mathcal{Y}$,
\begin{align}
\sum_{\substack{ x \in \mathcal{L}(y) : \\ \mathbb{P}\{ X \notin \mathsf{L}(\mathsf{f}(X, Y), Y) \mid (X, Y) = (x, y) \} < 1 }} \frac{ 1 }{ |\mathcal{L}(\mathsf{f}(x, y), y)| }
\le
M .
\label{eq:bunte-lapidoth_sum-inverse-cardinality}
\end{align}
Choose two deterministic maps $\epsilon : \mathcal{X} \times \mathcal{Y} \to [0, 1]$ and $\kappa : \mathcal{X} \times \mathcal{Y} \to (0, \infty)$ so that
\begin{align}
\epsilon(x, y)
& =
\mathbb{P}\{ X \notin \mathsf{L}(\mathsf{f}(X, Y), Y) \mid (X, Y) = (x, y) \} ,
\\
\kappa(x, y)
& =
\begin{cases}
1
& \mathrm{if} \ \mathsf{f}(x, y) = 0 ,
\\
|\mathcal{L}(\mathsf{f}(x, y), y)|
& \mathrm{if} \ \mathsf{f}(x, y) \neq 0 ,
\end{cases}
\end{align}
for each $(x, y) \in \mathcal{X} \times \mathcal{Y}$.
Then, it follows from \eqref{eq:error-probab_one-shot_avg_converse} that
\begin{align}
\mathbb{E}[ \epsilon(X, Y) ]
\le
\varepsilon .
\label{eq:error_task_converse_avg}
\end{align}
Moreover, we have
\begin{align}
\frac{ 1 }{ \rho } \log \mathbb{E}[ |\mathsf{L}(\mathsf{f}(X, Y), Y)|^{\rho} ]
& \overset{\mathclap{\text{(a)}}}{=}
\frac{ 1 }{ \rho } \log \mathbb{E}[ K(X, Y)^{\rho} ]
\notag \\
& \overset{\mathclap{\text{(b)}}}{\ge}
H_{1/(1+\rho)}^{\varepsilon}(X \mid Y) - \sup_{y \in \mathcal{Y}} \log \left( \sum_{\substack{ x \in \mathcal{L}(y) : \\ \mathbb{P}\{ X \notin \mathsf{L}(\mathsf{f}(X, Y), Y) \mid (X, Y) = (x, y) \} < 1 }} \frac{ 1 }{ |\mathcal{L}(\mathsf{f}(x, y), y)| } \right)
\notag \\
& \overset{\mathclap{\text{(c)}}}{\ge}
H_{1/(1+\rho)}^{\varepsilon}(X \mid Y) - \log M
\label{eq:one-shot_task_avg_converse_proof}
\end{align}
\begin{itemize}
\item
(a) follows by the definition of $K : \mathcal{X} \times \mathcal{Y} \to [0, \infty)$ stated in \eqref{def:stochastic-K},
\item
(b) follows from \lemref{lem:unified_converse_avg}, and
\item
(c) follows from \eqref{eq:bunte-lapidoth_sum-inverse-cardinality}.
\end{itemize}
This completes the proof of \eqref{eq:one-shot_task_avg_converse}.
\hfill\IEEEQEDhere

\section{Proof of \lemref{lem:task_one-shot_direct_avg_tilde}}
\label{app:task_one-shot_direct_avg_tilde}

In the proof, we employ the following technical result:

\begin{lemma}[{Bunte and Lapidoth \cite[Proposition~III.2]{bunte_lapidoth_2014}}]
\label{lem:Bunte-Lapidoth}
Let $\mathcal{S}$ be a finite set, $\lambda : \mathcal{S} \to \mathbb{N} \cup \{ \infty \}$ a function, and $M$ a positive integer.
If
\begin{align}
M
\ge
2 \sum_{\substack{ s \in \mathcal{S} : \\ \lambda( s ) < \infty }} \frac{ 1 }{ \lambda( s ) } + \log |\mathcal{S}| + 2 ,
\end{align}
then there exists a partition $\{ \mathcal{L}_{m} \}_{m = 1}^{M}$ of $\mathcal{S}$ such that
\begin{align}
s \in \mathcal{L}_{m}
\quad \Longrightarrow \quad
|\mathcal{L}_{m}| \le \lambda( s ) .
\end{align}
\end{lemma}

For each $y \in \mathcal{Y}$, define
\begin{align}
\mathcal{S}_{y}
\coloneqq
\{ \varsigma_{y}( k ) \mid 1 \le k \le J + 1 \} ,
\end{align}
where the bijection $\varsigma_{y} : \mathbb{N} \to \mathcal{X}$ and the number $J$ are defined in \eqref{def:varsigma} and \eqref{def:J}, respectively.
Since $|\mathcal{S}_{y}| = J+1$, it follows from \eqref{eq:guessing_Markov_avg} that
\begin{align}
\log |\mathcal{S}_{y}|
\le
\frac{ H(X \mid Y) }{ \varepsilon } 
\label{eq:card-Sy}
\end{align}
for every $y \in \mathcal{Y}$.
Letting $\delta : \mathcal{Y} \to [0, 1]$ be a deterministic map given as
\begin{align}
\delta( y )
=
1 - \left( \sum_{k = 1}^{J} P_{X|Y}(\varsigma_{y}^{-1}(k) \mid y) + \upsilon \, P_{X|Y}(\varsigma_{y}^{-1}(J+1) \mid y) \right)
\label{def:delta-tilde}
\end{align}
for each $y \in \mathcal{Y}$, it follows by the definition of $Q_{X|Y}^{(1/(1+\rho), \delta(\cdot))}$ stated in \eqref{def:tilted-R} that
\begin{align}
Q_{X|Y}^{(1/(1+\rho), \delta(\cdot))}(x \mid y)
=
\begin{dcases}
\frac{ P_{X|Y}(x \mid y)^{1/(1+\rho)} }{ \sum_{k = 1}^{J} P_{X|Y}(\varsigma_{y}^{-1}(k) \mid y)^{1/(1+\rho)} + \upsilon^{1/(1+\rho)} \, P_{X|Y}(\varsigma_{y}^{-1}(J+1) \mid y)^{1/(1+\rho)} }
& \mathrm{if} \ 1 \le \varsigma_{y}^{-1}(x) \le J ,
\\
\frac{ \upsilon^{1/(1+\rho)} \, P_{X|Y}(\varsigma_{y}^{-1}(J+1) \mid y)^{1/(1+\rho)} }{ \sum_{k = 1}^{J} P_{X|Y}(\varsigma_{y}^{-1}(k) \mid y)^{1/(1+\rho)} + \upsilon^{1/(1+\rho)} \, P_{X|Y}(\varsigma_{y}^{-1}(J+1) \mid y)^{1/(1+\rho)} }
& \mathrm{if} \ \varsigma_{y}^{-1}(x) = J + 1 ,
\\
0
& \mathrm{if} \ J + 2 \le \varsigma_{y}^{-1}(x) < \infty
\end{dcases}
\label{eq:task_direct_avg_tilted-Q}
\end{align}
for every $(x, y) \in \mathcal{X} \times \mathcal{Y}$.
Note that
\begin{align}
Q_{X|Y}^{(1/(1+\rho), \delta(\cdot))}(x \mid y) > 0
\quad \Longrightarrow \quad
x \in \mathcal{S}_{y} .
\end{align}
In addition, for each $y \in \mathcal{Y}$, define the function $\lambda_{y} : \mathcal{X} \to \mathbb{N} \cup \{ \infty \}$ by
\begin{align}
\lambda_{y}( x )
\coloneqq
\begin{dcases}
\left\lceil \frac{ 2 \, \varepsilon }{ (\varepsilon \, (M - 2) - H(X \mid Y) ) \, Q_{X|Y}^{(1/(1+\rho), \delta(\cdot))}(x \mid y) } \right\rceil
& \mathrm{if} \ Q_{X|Y}^{(1/(1+\rho), \delta(\cdot))}(x \mid y) > 0 ,
\\
\infty
& \mathrm{if} \ Q_{X|Y}^{(1/(1+\rho), \delta(\cdot))}(x \mid y) = 0 ,
\end{dcases}
\end{align}
where $M$ is a positive integer satisfying \eqref{eq:at-most_M_avg}.
Then, a direct calculation shows
\begin{align}
\sum_{\substack{ x \in \mathcal{X} : \\ \lambda_{y}( x ) < \infty }} \frac{ 1 }{ \lambda_{y}( x ) }
& =
\sum_{\substack{ x \in \mathcal{S}_{y} : \\ \lambda_{y}( x ) < \infty }} \frac{ 1 }{ \lambda_{y}( x ) }
\notag \\
& \le
\frac{ \varepsilon \, (M - 2) - H(X \mid Y) }{ 2 \, \varepsilon } \sum_{x \in \mathcal{S}_{y}} Q_{X|Y}^{(1/(1+\rho), \delta(\cdot))}(x \mid y)
\notag \\
& =
\frac{ \varepsilon \, (M - 2) - H(X \mid Y) }{ 2 \, \varepsilon }
\label{eq:sum_inverse-lambda-y}
\end{align}
for every $y \in \mathcal{Y}$.
Therefore, it follows from \lemref{lem:Bunte-Lapidoth} and \eqref{eq:card-Sy} that there exists an assignment function $\mathsf{f} : \mathcal{X} \times \mathcal{Y} \to \{ 0, 1, 2, \dots, M \}$ satisfying
\begin{align}
1 \le \varsigma_{y}^{-1}( x ) \le J + 1
\iff
\mathsf{f}(x, y) \neq 0
\label{eq:task_direct_avg_domain}
\end{align}
and
\begin{align}
|\mathcal{L}(\mathsf{f}(x, y), y)|
\le
\lambda_{y}( x )
\label{eq:bound-lambda}
\end{align}
for every $(x, y) \in \mathcal{X} \times \mathcal{Y}$, provided that \eqref{eq:at-most_M_avg} holds, where $\mathcal{L} : \{ 0, 1, 2, \dots, M \} \times \mathcal{Y} \to 2^{\mathcal{X}}$ is the sub-partition induced by $\mathsf{f}$; see \eqref{def:sub-partition}.

On the other hand, consider a stochastic map $\mathsf{E} : 2^{\mathcal{X}} \times \mathcal{Y} \to 2^{\mathcal{X}}$ satisfying \eqref{eq:error-stochas_partition} and
\begin{align}
\mathbb{P}\{ \mathsf{E}(\mathcal{L}(\mathsf{f}( \varsigma_{y}( k ), y ), y)) = \emptyset \}
=
\begin{cases}
0
& \mathrm{if} \ 0 \le k \le J ,
\\
1 - \upsilon
& \mathrm{if} \ k = J + 1 ,
\\
1
& \mathrm{if} \ J + 2 \le k < \infty
\end{cases}
\label{eq:task_direct_avg_stochastic-E}
\end{align}
for each $(k, y) \in \mathbb{N} \times \mathcal{Y}$, where $\upsilon$ is defined in \eqref{def:upsilon}.
We see from \eqref{eq:task_direct_avg_stochastic-E} that
\begin{align}
\mathbb{P}\{ X \notin \mathsf{L}(\mathsf{f}(X, Y), Y) \mid (X, Y) = (x, y) \}
=
\begin{dcases}
0
& \mathrm{if} \ 0 \le \varsigma_{y}^{-1}( x ) \le J ,
\\
1 - \upsilon
& \mathrm{if} \ \varsigma_{y}^{-1}( x ) = J + 1 ,
\\
1
& \mathrm{if} \ J + 2 \le \varsigma_{y}^{-1}( x ) < \infty
\end{dcases}
\label{eq:task_direct_avg_error-xy}
\end{align}
for every $(x, y) \in \mathcal{X} \times \mathcal{Y}$, where the stochastic sub-partition $\mathsf{L} : \{ 0, 1, 2, \dots, M \} \times \mathcal{Y} \to 2^{\mathcal{X}}$ induced by the pair $(\mathsf{f}, \mathsf{E})$ is defined in \eqref{def:stochastic-sub-partition}.
Therefore, we have
\begin{align}
\mathbb{E}[ |\mathsf{L}(\mathsf{f}(X, Y), Y)|^{\rho} ]
& \overset{\mathclap{\text{(a)}}}{=}
\sum_{y \in \mathcal{Y}} P_{Y}( y ) \left( \sum_{k = 1}^{J} P_{X|Y}(\varsigma_{y}(k) \mid y) \, |\mathcal{L}(\mathsf{f}(\varsigma_{y}(k), y), y)|^{\rho} + \upsilon \, P_{X|Y}(\varsigma_{y}(J+1) \mid y) \, |\mathcal{L}(\mathsf{f}(\varsigma_{y}(J+1), y), y)|^{\rho} \right)
\notag \\
& \overset{\mathclap{\text{(b)}}}{\le}
\sum_{y \in \mathcal{Y}} P_{Y}( y ) \left( \sum_{k = 1}^{J} P_{X|Y}(\varsigma_{y}(k) \mid y) \, \left\lceil \frac{ 2 \, \varepsilon }{ (\varepsilon \, (M - 2) - H(X \mid Y) ) \, Q_{X|Y}^{(1/(1+\rho), \delta(\cdot))}(\varsigma_{y}(k) \mid y) } \right\rceil^{\rho}
\right.\notag \\
& \left. \qquad \qquad \qquad
{} + \upsilon \, P_{X|Y}(\varsigma_{y}(J+1) \mid y) \, \left\lceil \frac{ 2 \, \varepsilon }{ (\varepsilon \, (M - 2) - H(X \mid Y) ) \, Q_{X|Y}^{(1/(1+\rho), \delta(\cdot))}(\varsigma_{y}(J+1) \mid y) } \right\rceil^{\rho} \right)
\notag \\
& \overset{\mathclap{\text{(c)}}}{\le}
\left( \frac{ 4 \, \varepsilon }{ \varepsilon \, (M - 2) - H(X \mid Y) } \right)^{\rho} \sum_{y \in \mathcal{Y}} P_{Y}( y ) \left( \sum_{k = 1}^{J} P_{X|Y}(\varsigma_{y}(k) \mid y) \, \left( \frac{ 1 }{ Q_{X|Y}^{(1/(1+\rho), \delta(\cdot))}(\varsigma_{y}(k) \mid y) } \right)^{\rho}
\right.\notag \\
& \left. \qquad \qquad \qquad \qquad \qquad \qquad \qquad
{} + \upsilon \, P_{X|Y}(\varsigma_{y}(J+1) \mid y) \, \left( \frac{ 1 }{ Q_{X|Y}^{(1/(1+\rho), \delta(\cdot))}(\varsigma_{y}(J+1) \mid y) } \right)^{\rho} \right) + (1 - \varepsilon)
\notag \\
& \overset{\mathclap{\text{(d)}}}{=}
\left( \frac{ 4 \, \varepsilon }{ \varepsilon \, (M - 2) - H(X \mid Y) } \right)^{\rho} \sum_{y \in \mathcal{Y}} P_{Y}( y ) \left( \sum_{k = 1}^{J} P_{X|Y}(\varsigma_{y}(k) \mid y)^{1/(1+\rho)} + \upsilon^{1/(1+\rho)} \, P_{X|Y}(\varsigma_{y}(J+1) \mid y)^{1/(1+\rho)} \right)^{1+\rho}
\notag \\
& \qquad
{} + (1 - \varepsilon)
\notag \\
& \overset{\mathclap{\text{(e)}}}{=}
\left( \frac{ 4 \, \varepsilon }{ \varepsilon \, (M - 2) - H(X \mid Y) } \right)^{\rho} \exp\Big( \rho \, \tilde{H}_{1/(1+\rho)}^{\varepsilon}(X \mid Y) \Big) + (1 - \varepsilon)
\notag \\
& \le
\exp\left( \rho \, \tilde{H}_{1/(1+\rho)}^{\varepsilon}(X \mid Y) - \rho \log \left( \frac{ \varepsilon \, (M - 2) - H(X \mid Y) }{ 4 \, \varepsilon } \right) \right) + 1 ,
\label{eq:bunte-lapidoth_avg}
\end{align}
where
\begin{itemize}
\item
(a) follows from \eqref{eq:task_direct_avg_error-xy},
\item
(b) follows from \eqref{eq:bound-lambda},
\item
(c) follows from the fact that
\begin{align}
\lceil u \rceil^{\rho}
<
1 + 2^{\rho} \, u^{\rho}
\label{eq:bonte-lapidoth_ceiling}
\end{align}
for every $u \ge 0$; cf.\ \cite[Equation~(26)]{bunte_lapidoth_2014},
\item
(d) follows from \eqref{eq:task_direct_avg_tilted-Q}, and
\item
(e) follows by the definition of $\tilde{H}_{1/(1+\rho)}^{\varepsilon}(X \mid Y)$ stated in \eqref{def:yet}.
\end{itemize}
Finally, noting that
\begin{align}
\tilde{H}_{1/(1+\rho)}^{\varepsilon}(X \mid Y)
\le
\log \left( \frac{ \varepsilon \, (M - 2) - H(X \mid Y) }{ 4 \, \varepsilon } \right)
\iff
\exp\left( \rho \, \tilde{H}_{1/(1+\rho)}^{\varepsilon}(X \mid Y) - \rho \log \left( \frac{ \varepsilon \, (M - 2) - H(X \mid Y) }{ 4 \, \varepsilon } \right) \right)
\le
1 ,
\end{align}
we obtain \eqref{eq:task_one-shot_direct_avg_tilde} from \eqref{eq:bunte-lapidoth_avg}.
This completes the proof of \lemref{lem:task_one-shot_direct_avg_tilde}.
\hfill\IEEEQEDhere

\section{Proof of \lemref{lem:ineq-Kuzuoka_tilde}}
\label{app:ineq-Kuzuoka_tilde}

Let $\delta : \mathcal{Y} \to [0, 1]$ be given by \eqref{def:delta-tilde}.
After some algebra, we observe that
\begin{align}
\tilde{H}_{\alpha}^{\varepsilon}(X \mid Y)
=
\bar{\mathsf{H}}_{\alpha}^{\delta(\cdot)}(X \mid Y) ,
\label{eq:tilde_KN}
\end{align}
where the right-hand side is defined in \eqref{def:cond-smooth_KN-avg}.
Moreover, it follows by the definitions of $J$, $\xi$, and $\upsilon$ stated in \eqref{def:J}, \eqref{def:xi}, and \eqref{def:upsilon} that
\begin{align}
\sum_{y \in \mathcal{Y}} P_{Y}( y ) \, \delta( y )
=
\varepsilon ,
\end{align}
implying that $\delta( \cdot ) \in \mathcal{E}_{0}( \varepsilon )$.
Therefore, it follows from \lemref{lem:Kuzuoka-formula} that the left-hand inequality of \eqref{eq:ineq-Kuzuoka_tilde} holds.

To prove the right-hand inequality of \eqref{eq:ineq-Kuzuoka_tilde}, we shall revisit the guessing problem discussed in \appref{app:one-shot_guess_avg_converse}.
Consider the giving-up guessing function $\bar{\mathsf{G}}_{\mathrm{avg}}^{\ast} : \mathcal{X} \times \mathcal{Y} \to \mathbb{N} \cup \{ 0 \}$ induced by the optimal guessing strategy $(\mathsf{g}^{\ast}, \pi_{\mathrm{avg}}^{\ast})$ given in \lemref{lem:optimal-strategy_avg}.
Similar to \eqref{eq:unified-converse_given-Y}, it follows from H\"{o}lder's inequality that
\begin{align}
\mathbb{E}[ \mathsf{G}_{\mathrm{avg}}^{\ast}(X, Y)^{\rho} \mid Y = y ]
\ge
\exp\left( \rho \, H_{1/(1+\rho)}^{\delta(y)}( P_{X|Y=y} ) - \rho \log \left( \sum_{k = 1}^{J} \frac{ 1 }{ k } \right) \right)
\end{align}
where $\delta : \mathcal{Y} \to [0, 1]$ is given by \eqref{def:delta-tilde}.
Hence, it follows from \eqref{eq:tilde_KN} that
\begin{align}
\mathbb{E}[ \mathsf{G}_{\mathrm{avg}}^{\ast}(X, Y)^{\rho} ]
\ge
\exp\left( \rho \, \tilde{H}_{\alpha}^{\varepsilon}(X \mid Y) - \rho \log \left( \sum_{k = 1}^{J} \frac{ 1 }{ k } \right) \right) .
\label{eq:unified-converse_avg_guessing_non-inf}
\end{align}
Therefore, it holds that
\begin{align}
\mathsf{G}_{\mathrm{avg}}^{\ast}(X, Y \, \| \, \rho, \varepsilon)
& \overset{\mathclap{\text{(a)}}}{=}
\mathsf{G}_{\mathrm{avg}}^{\ast}(X, Y \, \| \, \rho, \varepsilon)
\notag \\
& \overset{\mathclap{\text{(b)}}}{\ge}
\tilde{H}_{\alpha}^{\varepsilon}(X \mid Y) - \log \left( \sum_{k = 1}^{J} \frac{ 1 }{ k } \right)
\notag \\
& \overset{\mathclap{\text{(c)}}}{\ge}
\tilde{H}_{\alpha}^{\varepsilon}(X \mid Y) - \log \left( 1 + \frac{ H(X \mid Y) }{ \varepsilon } \right) ,
\label{eq:one-shot_guess_avg_strengthened-converse}
\end{align}
where
\begin{itemize}
\item
(a) follows from \lemref{lem:optimal-strategy_avg},
\item
(b) follows from \eqref{eq:unified-converse_avg_guessing_non-inf}, and
\item
(c) follows as in Steps~(b) and~(c) of \eqref{eq:guessing-avg_converse_unified-way}.
\end{itemize}
Combining the right-hand inequality of \eqref{eq:one-shot_guess_avg} and \eqref{eq:one-shot_guess_avg_strengthened-converse}, we obtain the right-hand inequality of \eqref{eq:ineq-Kuzuoka_tilde}.
This completes the proof of \lemref{lem:ineq-Kuzuoka_tilde}.

\section{Proof of \thref{th:one-shot_task_max}}
\label{app:one-shot_task_max}

\subsection{Proof of \eqref{eq:one-shot_task_max_converse}}
\label{app:one-shot_task_max_converse}

We can prove \eqref{eq:one-shot_task_max_converse} in the same way as we did in \appref{app:one-shot_task_avg_converse}.
Replacing \eqref{eq:error-probab_one-shot_avg_converse} by
\begin{align}
\sup_{y \in \mathcal{Y}} \mathbb{P}\{ X \notin \mathsf{L}(\mathsf{f}(X, Y), Y) \mid Y = y \}
\le
\varepsilon ,
\label{eq:error-probab_one-shot_max_converse}
\end{align}
Equation~\eqref{eq:error_task_converse_avg} can be strengthened to
\begin{align}
\mathbb{E}[ \epsilon(X, Y) \mid Y = y ]
\le
\varepsilon .
\end{align}
Thus, by using \lemref{lem:unified_converse_max} in Step~(b) of \eqref{eq:one-shot_task_avg_converse_proof} instead on \lemref{lem:unified_converse_avg}, we obtain
\begin{align}
\frac{ 1 }{ \rho } \log \mathbb{E}[ |\mathsf{L}(\mathsf{f}(X, Y), Y)|^{\rho} ]
& \ge
\check{H}_{1/(1+\rho)}^{\varepsilon}(X \mid Y) - \log M ,
\end{align}
as desired.
\hfill\IEEEQEDhere

\subsection{Proof of \eqref{eq:one-shot_task_max_direct}}
\label{app:one-shot_task_max_direct}

Recall that the numbers $J( y )$ and $\xi( y )$ are defined in \eqref{def:J-y} and \eqref{def:xi-y}, respectively, for each $y \in \mathcal{Y}$.
In addition, define
\begin{align}
\upsilon( y )
& \coloneqq
\frac{ \xi( y ) }{ P_{X|Y}(\varsigma_{y}(J(y) + 1) \mid y) } ,
\end{align}
for each $y \in \mathcal{Y}$, where the bijection $\varsigma_{y} : \mathbb{N} \to \mathcal{X}$ is defined in \eqref{def:varsigma}.

For each $y \in \mathcal{Y}$, define
\begin{align}
\check{\mathcal{S}}_{y}
\coloneqq
\{ \varsigma_{y}( k ) \mid 1 \le k \le J( y ) + 1 \} .
\end{align}
Since $|\check{S}_{y}| = J( y ) + 1$, it follows from \eqref{eq:guessing_Markov_max} that
\begin{align}
\log |\check{S}_{y}|
\le
\frac{ \sup_{y \in \mathcal{Y}} H( P_{X|Y=y} ) }{ \varepsilon } .
\label{eq:card-checkSy}
\end{align}
Letting $\delta( y ) = \varepsilon$ for each $y \in \mathcal{Y}$, i.e., the deterministic map $\delta : \mathcal{Y} \to [0, 1]$ is constant, it follows by the definition of $Q_{X|Y}^{(1/(1+\rho), \varepsilon)}$ stated in \eqref{def:tilted-R} that
\begin{align}
Q_{X|Y}^{(1/(1+\rho), \varepsilon)}(x, y)
=
\begin{dcases}
\frac{ P_{X|Y}(x \mid y)^{1/(1+\rho)} }{ \sum_{k = 1}^{J(y)} P_{X|Y}(\varsigma_{y}^{-1}(k) \mid y)^{1/(1+\rho)} + \upsilon( y )^{1/(1+\rho)} \, P_{X|Y}(\varsigma_{y}^{-1}(J(y)+1 \mid y)^{1/(1+\rho)} }
& \mathrm{if} \ 1 \le \varsigma_{y}^{-1}( x ) \le J( y ) ,
\\
\frac{ \upsilon( y )^{1/(1+\rho)} \, P_{X|Y}(\varsigma_{y}^{-1}(J(y)+1 \mid y)^{1/(1+\rho)} }{ \sum_{k = 1}^{J(y)} P_{X|Y}(\varsigma_{y}^{-1}(k) \mid y)^{1/(1+\rho)} + \upsilon( y )^{1/(1+\rho)} \, P_{X|Y}(\varsigma_{y}^{-1}(J(y)+1 \mid y)^{1/(1+\rho)} }
& \mathrm{if} \ \varsigma_{y}^{-1}( x ) = J( y ) + 1 ,
\\
0
& \mathrm{if} \ J( y ) + 2 \le \varsigma_{y}^{-1}( x ) < \infty
\end{dcases}
\label{eq:task_direct_max_tilted-Q}
\end{align}
for every $(x, y) \in \mathcal{X} \times \mathcal{Y}$.
In addition, for each $y \in \mathcal{Y}$, define the function $\check{\lambda}_{y} : \mathcal{X} \to \mathbb{N} \cup \{ \infty \}$ by
\begin{align}
\check{\lambda}_{y}( x )
\coloneqq
\begin{dcases}
\left\lceil \frac{ 2 \, \varepsilon }{ (\varepsilon \, (M - 2) - \sup_{y \in \mathcal{Y}} H( P_{X|Y=y} ) ) \, Q_{X|Y}^{(1/(1+\rho), \varepsilon)}(x \mid y) } \right\rceil
& \mathrm{if} \ Q_{X|Y}^{(1/(1+\rho), \varepsilon)}(x \mid y) > 0 ,
\\
\infty
& \mathrm{if} \ Q_{X|Y}^{(1/(1+\rho), \varepsilon)}(x \mid y) = 0 ,
\end{dcases}
\end{align}
where $M$ is a positive integer satisfying
\begin{align}
M
>
2 + \frac{ \sup_{y \in \mathcal{Y}} H( P_{X|Y=y} ) }{ \varepsilon } .
\label{eq:at-most_M_max}
\end{align}
Then, a similar calculation to \eqref{eq:sum_inverse-lambda-y} yields
\begin{align}
\sum_{\substack{ x \in \mathcal{X} : \\ \check{\lambda}_{y}( x ) < \infty }} \frac{ 1 }{ \check{\lambda}_{y}( x ) }
& \le
\frac{ \varepsilon \, (M - 2) - \sup_{y \in \mathcal{Y}} H( P_{X|Y=y} ) }{ 2 \, \varepsilon }
\end{align}
for every $y \in \mathcal{Y}$.
Therefore, it follows from \lemref{lem:Bunte-Lapidoth} and \eqref{eq:card-checkSy} that there exists an assignment function $\mathsf{f} : \mathcal{X} \times \mathcal{Y} \to \{ 0, 1, 2, \dots, M \}$ satisfying
\begin{align}
1 \le \varsigma_{y}^{-1}( x ) \le J(y) + 1
\iff
\mathsf{f}(x, y) \neq 0
\label{eq:task_direct_max_domain}
\end{align}
and
\begin{align}
|\mathcal{L}(\mathsf{f}(x, y), y)|
\le
\check{\lambda}_{y}( x )
\label{eq:bound-check-lambda}
\end{align}
for every $(x, y) \in \mathcal{X} \times \mathcal{Y}$, provided that \eqref{eq:at-most_M_avg} holds, where $\mathcal{L} : \{ 0, 1, 2, \dots, M \} \times \mathcal{Y} \to 2^{\mathcal{X}}$ is the sub-partition induced by $\mathsf{f}$; see \eqref{def:sub-partition}.

On the other hand, consider a stochastic map $\mathsf{E} : 2^{\mathcal{X}} \times \mathcal{Y} \to 2^{\mathcal{X}}$ satisfying \eqref{eq:error-stochas_partition} and
\begin{align}
\mathbb{P}\{ \mathsf{E}(\mathcal{L}(\mathsf{f}( \varsigma_{y}( k ), y ), y)) = \emptyset \}
=
\begin{cases}
0
& \mathrm{if} \ 0 \le k \le J(y) ,
\\
1 - \upsilon(y)
& \mathrm{if} \ k = J(y) + 1 ,
\\
1
& \mathrm{if} \ J(y) + 2 \le k < \infty
\end{cases}
\label{eq:task_direct_max_stochastic-E}
\end{align}
for each $(k, y) \in \mathbb{N} \times \mathcal{Y}$.
We see from \eqref{eq:task_direct_max_stochastic-E} that
\begin{align}
\mathbb{P}\{ X \notin \mathsf{L}(\mathsf{f}(X, Y), Y) \mid (X, Y) = (x, y) \}
=
\begin{dcases}
0
& \mathrm{if} \ 0 \le \varsigma_{y}^{-1}( x ) \le J(y) ,
\\
1 - \upsilon(y)
& \mathrm{if} \ \varsigma_{y}^{-1}( x ) = J(y) + 1 ,
\\
1
& \mathrm{if} \ J(y) + 2 \le \varsigma_{y}^{-1}( x ) < \infty
\end{dcases}
\label{eq:task_direct_max_error-xy}
\end{align}
for every $(x, y) \in \mathcal{X} \times \mathcal{Y}$, where the stochastic sub-partition $\mathsf{L} : \{ 0, 1, 2, \dots, M \} \times \mathcal{Y} \to 2^{\mathcal{X}}$ induced by the pair $(\mathsf{f}, \mathsf{E})$ is defined in \eqref{def:stochastic-sub-partition}.
Therefore, we have
\begin{align}
\mathbb{E}[ |\mathsf{L}(\mathsf{f}(X, Y), Y)|^{\rho} ]
& \overset{\mathclap{\text{(a)}}}{=}
\sum_{y \in \mathcal{Y}} P_{Y}( y ) \left( \sum_{k = 1}^{J(y)} P_{X|Y}(\varsigma_{y}(k) \mid y) \, |\mathcal{L}(\mathsf{f}(\varsigma_{y}(k), y), y)|^{\rho}
\right.\notag \\
& \left. \vphantom{\sum_{k = 1}^{J(y)}} \qquad \qquad \qquad
{} + \upsilon(y) \, P_{X|Y}(\varsigma_{y}(J(y)+1) \mid y) \, |\mathcal{L}(\mathsf{f}(\varsigma_{y}(J(y)+1), y), y)|^{\rho} \right)
\notag \\
& \overset{\mathclap{\text{(b)}}}{\le}
\sum_{y \in \mathcal{Y}} P_{Y}( y ) \left( \sum_{k = 1}^{J(y)} P_{X|Y}(\varsigma_{y}(k) \mid y) \, \left\lceil \frac{ 2 \, \varepsilon }{ (\varepsilon \, (M - 2) - \sup_{y \in \mathcal{Y}} H(P_{X|Y=y}) ) \, Q_{X|Y}^{(1/(1+\rho), \varepsilon)}(\varsigma_{y}(k) \mid y) } \right\rceil^{\rho}
\right.\notag \\
& \left. \qquad
{} + \upsilon(y) \, P_{X|Y}(\varsigma_{y}(J(y)+1) \mid y) \, \left\lceil \frac{ 2 \, \varepsilon }{ (\varepsilon \, (M - 2) - \sup_{y \in \mathcal{Y}} H(P_{X|Y=y}) ) \, Q_{X|Y}^{(1/(1+\rho), \varepsilon)}(\varsigma_{y}(J(y)+1) \mid y) } \right\rceil^{\rho} \right)
\notag \\
& \overset{\mathclap{\text{(c)}}}{\le}
\left( \frac{ 4 \, \varepsilon }{ \varepsilon \, (M - 2) - \sup_{y \in \mathcal{Y}} H(P_{X|Y=y}) } \right)^{\rho} \sum_{y \in \mathcal{Y}} P_{Y}( y ) \left( \sum_{k = 1}^{J(y)} P_{X|Y}(\varsigma_{y}(k) \mid y) \, \left( \frac{ 1 }{ Q_{X|Y}^{(1/(1+\rho), \varepsilon)}(\varsigma_{y}(k) \mid y) } \right)^{\rho}
\right.\notag \\
& \left. \qquad \qquad \qquad \qquad \qquad \qquad
{} + \upsilon(y) \, P_{X|Y}(\varsigma_{y}(J(y)+1) \mid y) \, \left( \frac{ 1 }{ Q_{X|Y}^{(1/(1+\rho), \varepsilon)}(\varsigma_{y}(J(y)+1) \mid y) } \right)^{\rho} \right) + (1 - \varepsilon)
\notag \\
& \overset{\mathclap{\text{(d)}}}{=}
\left( \frac{ 4 \, \varepsilon }{ \varepsilon \, (M - 2) - \sup_{y \in \mathcal{Y}} H(P_{X|Y=y}) } \right)^{\rho}
\notag \\
& \qquad \quad
{} \times \sum_{y \in \mathcal{Y}} P_{Y}( y ) \left( \sum_{k = 1}^{J(y)} P_{X|Y}(\varsigma_{y}(k) \mid y)^{1/(1+\rho)} + \upsilon(y)^{1/(1+\rho)} \, P_{X|Y}(\varsigma_{y}(J(y)+1) \mid y)^{1/(1+\rho)} \right)^{1+\rho} + (1 - \varepsilon)
\notag \\
& \overset{\mathclap{\text{(e)}}}{=}
\left( \frac{ 4 \, \varepsilon }{ \varepsilon \, (M - 2) - \sup_{y \in \mathcal{Y}} H(P_{X|Y=y}) } \right)^{\rho} \exp\Big( \rho \, \check{H}_{1/(1+\rho)}^{\varepsilon}(X \mid Y) \Big) + (1 - \varepsilon)
\notag \\
& \le
\exp\left( \rho \, \check{H}_{1/(1+\rho)}^{\varepsilon}(X \mid Y) - \rho \log \left( \frac{ \varepsilon \, (M - 2) - \sup_{y \in \mathcal{Y}} H(P_{X|Y=y}) }{ 4 \, \varepsilon } \right) \right) + 1 ,
\label{eq:bunte-lapidoth_max}
\end{align}
where
\begin{itemize}
\item
(a) follows from \eqref{eq:task_direct_max_error-xy},
\item
(b) follows from \eqref{eq:bound-check-lambda},
\item
(c) follows from \eqref{eq:bonte-lapidoth_ceiling},
\item
(d) follows from \eqref{eq:task_direct_max_tilted-Q}, and
\item
(e) follows from \eqref{eq:check_identity1}.
\end{itemize}
Finally, noting that
\begin{align}
&
\check{H}_{1/(1+\rho)}^{\varepsilon}(X \mid Y)
\le
\log \left( \frac{ \varepsilon \, (M - 2) - \sup_{y \in \mathcal{Y}} H(P_{X|Y=y}) }{ 4 \, \varepsilon } \right)
\notag \\
& \qquad \qquad \qquad
\iff
\exp\left( \rho \, \tilde{H}_{1/(1+\rho)}^{\varepsilon}(X \mid Y) - \rho \log \left( \frac{ \varepsilon \, (M - 2) - \sup_{y \in \mathcal{Y}} H(P_{X|Y=y}) }{ 4 \, \varepsilon } \right) \right)
\le
1 ,
\end{align}
we obtain \eqref{eq:one-shot_task_max_direct} from \eqref{eq:bunte-lapidoth_max}.
\hfill\IEEEQEDhere

\bibliographystyle{IEEEtran}
\bibliography{IEEEabrv,mybib}

\begin{thebibliography}{10}
\providecommand{\url}[1]{#1}
\csname url@samestyle\endcsname
\providecommand{\newblock}{\relax}
\providecommand{\bibinfo}[2]{#2}
\providecommand{\BIBentrySTDinterwordspacing}{\spaceskip=0pt\relax}
\providecommand{\BIBentryALTinterwordstretchfactor}{4}
\providecommand{\BIBentryALTinterwordspacing}{\spaceskip=\fontdimen2\font plus
\BIBentryALTinterwordstretchfactor\fontdimen3\font minus
  \fontdimen4\font\relax}
\providecommand{\BIBforeignlanguage}[2]{{%
\expandafter\ifx\csname l@#1\endcsname\relax
\typeout{** WARNING: IEEEtran.bst: No hyphenation pattern has been}%
\typeout{** loaded for the language `#1'. Using the pattern for}%
\typeout{** the default language instead.}%
\else
\language=\csname l@#1\endcsname
\fi
#2}}
\providecommand{\BIBdecl}{\relax}
\BIBdecl

\bibitem{renyi_1961}
A.~R\'{e}nyi, ``On measures of entropy and information,'' in \emph{Proc.\ 4th
  Berkeley Symp.\ Math.\ Statist.\ Probab.}, 1961, pp. 574--561.

\bibitem{campbell_1965}
L.~L. Campbell, ``A coding theorem and {R\'{e}nyi's} entropy,'' \emph{Inf.\
  Control}, vol.~8, no.~4, pp. 423--429, 1965.

\bibitem{courtade_verdu_isit2014_lossless}
T.~A. Courtade and S.~Verd\'{u}, ``Cumulant generating function of codeword
  lengths in optimal lossless compression,'' in \emph{Proc.\ {IEEE}\ Int.\
  Symp.\ Inf.\ Theory}, Honolulu, HI, USA, June--July 2014, pp. 2494--2498.

\bibitem{massey_isit1994}
J.~L. Massey, ``Guessing and entropy,'' in \emph{Proc.\ {IEEE}\ Int.\ Symp.\
  Inf.\ Theory}, Chicago, IL, USA, Jun. 1994, p. 204.

\bibitem{arikan_1996}
E.~Ar{\i}kan, ``An inequality on guessing and its application to sequential
  decoding,'' \emph{{IEEE} Trans. Inf. Theory}, vol.~42, no.~1, pp. 99--105,
  Jan. 1996.

\bibitem{bunte_lapidoth_2014}
C.~Bunte and A.~Lapidoth, ``Encoding tasks and {R\'{e}nyi} entropy,''
  \emph{{IEEE} Trans. Inf. Theory}, vol.~60, no.~9, pp. 5065--5076, Sept. 2014.

\bibitem{renner_wolf_isit2004}
R.~Renner and S.~Wolf, ``Smooth {R\'{e}nyi} entropy and applications,'' in
  \emph{Proc.\ {IEEE}\ Int.\ Symp.\ Inf.\ Theory}, Chicago, USA, June--July
  2004, p. 232.

\bibitem{renner_wolf_asiacrypt2005}
------, ``Simple and tight bounds for information reconciliation and privacy
  amplification,'' in \emph{Advances in Cryptology---ASIACRYPT 2005}, Chennai,
  India, Dec. 2005, pp. 199--216.

\bibitem{konig_renner_schaffner_2009}
R.~K\"{o}nig, R.~Renner, and C.~Schaffner, ``The operational meaning of min-
  and max-entropy,'' \emph{{IEEE} Trans. Inf. Theory}, vol.~55, no.~9, pp.
  4337--4347, Sept. 2009.

\bibitem{koga_itw2013}
H.~Koga, ``Characterization of the smooth {R\'{e}nyi} entropy using
  majorization,'' in \emph{Proc.\ {IEEE}\ Inf.\ Theory\ Workshop}, Sevilla,
  Spain, Sept. 2013.

\bibitem{kuzuoka_2019}
S.~Kuzuoka, ``On the conditional smooth {R\'{e}nyi} entropy and its
  applications in guessing and source coding,'' \emph{{IEEE} Trans. Inf.
  Theory}, vol.~66, no.~3, pp. 1674--1690, Mar. 2020.

\bibitem{arimoto_1977}
S.~Arimoto, ``Information measures and capacity of order for discrete
  memoryless channels,'' in \emph{Topics Inf.\ Theory, 2nd Colloq.\ Math.\
  Soc.\ J.\ Bolyai}, vol.~16, Keszthely, Hungary, 1977, pp. 41--52.

\bibitem{tan_2014}
V.~Y.~F. Tan, ``Asymptotic estimates in information theory with non-vanishing
  error probabilities,'' \emph{Found.\ Trends\ Commun.\ Inf.\ Theory}, vol.~11,
  no. 1--2, pp. 1--184, Sept. 2014.

\bibitem{polyanskiy_poor_verdu_2010}
Y.~Polyanskiy, H.~V. Poor, and S.~Verd\'{u}, ``Channel coding rate in the
  finite blocklength regime,'' \emph{{IEEE} Trans. Inf. Theory}, vol.~56,
  no.~5, pp. 2307--2359, May 2010.

\bibitem{kumar_sunny_thakre_kumar_2019}
M.~A. Kumar, A.~Sunny, A.~Thakre, and A.~Kumar, ``A unified framework for
  problems on guessing, source coding and task partitioning,'' Jul. 2019,
  [Online]. Available at \url{https://arxiv.org/abs/1907.06889}.

\bibitem{sakai_tan_2019_VL}
Y.~Sakai and V.~Y.~F. Tan, ``Variable-length source dispersions differ under
  maximum and average error criteria,'' \emph{\emph{submitted to} IEEE Trans.\
  Inf.\ Theory}, Oct. 2019, {Available} at
  \url{https://arxiv.org/abs/1910.05724}.

\bibitem{strassen_1964}
V.~Strassen, ``Asymptotische absch\"{a}tzungen in shannon's
  informationstheorie,'' in \emph{Trans.\ 3rd Prague Conf.\ Inf.\
  Theory}.\hskip 1em plus 0.5em minus 0.4em\relax Prague: Academia, 1962, pp.
  689--723.

\bibitem{vembu_verdu_1995}
S.~Vembu and A.~Verd\'{u}, ``Generating random bits from an arbitrary sources:
  {Fundamental} limits,'' \emph{{IEEE} Trans. Inf. Theory}, vol.~41, no.~5, pp.
  1322--1332, May 1995.

\bibitem{han_2003}
T.~S. Han, \emph{Information Spectrum Methods in Information Theory}.\hskip 1em
  plus 0.5em minus 0.4em\relax New~York: Springer-Verlag, 2003.

\bibitem{tomamichel_colbeck_renner_2009}
M.~Tomamichel, R.~Colbeck, and R.~Renner, ``A fully quantum asymptotic
  equipartition property,'' \emph{{IEEE} Trans. Inf. Theory}, vol.~55, no.~12,
  pp. 5840--5847, Dec. 2009.

\bibitem{tomamichel_colbeck_renner_2010}
------, ``Duality between smooth min- and max-eentropies,'' \emph{{IEEE} Trans.
  Inf. Theory}, vol.~56, no.~9, pp. 4674--4681, 2010.

\bibitem{uyematsu_2010}
T.~Uyematsu, ``A new unified method for fixed-length source coding problems of
  general sources,'' \emph{{IEICE} Trans.\ Fundamentals}, vol. E93-A, no.~11,
  pp. 1868--1877, Nov. 2010.

\bibitem{uyematsu_isit2010}
------, ``Relating source coding and resolvability: A direct approach,'' in
  \emph{Proc.\ {IEEE}\ Int.\ Symp.\ Inf.\ Theory}, Austin, TX, USA, Jun. 2010,
  pp. 1350--1354.

\bibitem{uyematsu_kunimatsu_itw2013}
T.~Uyematsu and S.~Kunimatsu, ``A new unified method for intrinsic randomness
  problems of general sources,'' in \emph{Proc.\ {IEEE}\ Inf.\ Theory\
  Workshop}, Sevilla, Spain, Sept. 2013, pp. 624--628.

\bibitem{saito_matsushima_2016}
S.~Saito and T.~Matsushima, ``Threshold of overflow probability using smooth
  max-entropy in lossless fixed-to-variable length source coding for general
  sources,'' \emph{{IEICE} Trans.\ Fundamentals}, vol. E99-A, no.~12, pp.
  2286--2290, Dec. 2016.

\bibitem{kuzuoka_isit2016}
S.~Kuzuoka, ``On the smooth {R\'{e}nyi} entropy and variable-length source
  coding allowing errors,'' in \emph{Proc.\ {IEEE}\ Int.\ Symp.\ Inf.\ Theory},
  Barcelona, Spain, Jul. 2016, pp. 745--749, [Online]. Available at
  \url{https://arxiv.org/abs/1512.06499}.

\bibitem{sason_verdu_2018}
I.~Sason and S.~Verd\'{u}, ``Improved bounds on lossless source coding and
  guessing moments via {R\'{e}nyi} measures,'' \emph{{IEEE} Trans. Inf.
  Theory}, vol.~64, no.~6, pp. 4323--4346, Jun. 2018.

\bibitem{yagi_han_isit2017}
H.~Yagi and T.~S. Han, ``Variable-length resolvability for general sources,''
  in \emph{Proc.\ {IEEE}\ Int.\ Symp.\ Inf.\ Theory}, Aachen, Germany, Jun.
  2017, pp. 1748--1752.

\bibitem{han_2000}
T.~S. Han, ``Weak variable-length source coding,'' \emph{{IEEE} Trans. Inf.
  Theory}, vol.~46, no.~4, pp. 1217--1226, Jul. 2000.

\bibitem{koga_yamamoto_2005}
H.~Koga and H.~Yamamoto, ``Asymptotic properties on codeword lengths of an
  optimal {FV} code for general sources,'' \emph{{IEEE} Trans. Inf. Theory},
  vol.~51, no.~4, pp. 1546--1555, Apr. 2005.

\bibitem{kostina_polyanskiy_verdu_2015}
V.~Kostina, Y.~Polyanskiy, and S.~Verd\'{u}, ``Variable-length compression
  allowing errors,'' \emph{{IEEE} Trans. Inf. Theory}, vol.~61, no.~8, pp.
  4316--4330, Aug. 2015.

\bibitem{feller_1971}
W.~Feller, \emph{An Introduction to Probability Theory and Its Applications},
  2nd~ed.\hskip 1em plus 0.5em minus 0.4em\relax New~York: Wiley, 1971, vol.~2.

\bibitem{gallager_1968}
R.~G. Galleger, \emph{Information Theory and Reliable Communications}.\hskip
  1em plus 0.5em minus 0.4em\relax New~York: John Wiley \& Sons, 1968.

\end{thebibliography}

\end{document}